# Braneworld Black Holes

## Richard Whisker

A Thesis presented for the degree of

Doctor of Philosophy

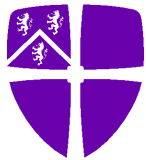

Institute for Particle Physics Phenomenology

Department of Physics

University of Durham

England

November 2006

# Braneworld Black Holes

## Richard Whisker

Submitted for the degree of Doctor of Philosophy

November 2006

## Abstract

The braneworld paradigm provides an interesting framework within which to explore the possibility that our Universe lives in a fundamentally higher dimensional spacetime. In this thesis we investigate black holes in the Randall-Sundrum braneworld scenario. We begin with an overview of extra-dimensional physics, from the original proposal of Kaluza and Klein up to the modern braneworld picture of extra dimensions. A detailed description of braneworld gravity is given, with particular emphasis on its compatibility with experimental tests of gravity. We then move on to a discussion of static, spherically symmetric braneworld black hole solutions. Assuming an equation of state for the "Weyl term", which encodes the effects of the extra dimension, we are able to classify the general behaviour of these solutions. We then use the strong field limit approach to investigate the gravitational lensing properties of some candidate braneworld black hole solutions. It is found that braneworld black holes could have significantly different observational signatures to the Schwarzschild black hole of standard general relativity. Rotating braneworld black hole solutions are also discussed, and we attempt to generate rotating solutions from known static solutions using the Newman-Janis complexification "trick".

# Declaration

The work in this thesis is based on research carried out at the Institute for Particle Physics Phenomenology, Department of Physics, Durham University, England. No part of this thesis has been submitted here or elsewhere for any other degree or qualification and is all my own work unless referenced to the contrary in the text.

Chapter 2 of this thesis is a review of necessary background material. The beginning of Chapter 3 is also a review, although Section 3.4 onwards is original work done in collaboration with my supervisor Ruth Gregory, Kris Beckwith and Chris Done [1]. Chapter 4 is all my own work, based on [2]. The beginning of Chapter 5 contains some review material, but Section 5.4 onwards is all my own work.

# Acknowledgements

Firstly I would like to thank my Mum and Dad for their love, support and encouragement throughout my life, and in all that I have chosen to do. You have contributed far more to this thesis than you probably realise, thankyou. Thanks also to Robert and Alison, I know you've always been there for me.

Ruth, thanks for all your supervision and guidance, even though I was only technically supposed to be half your student. Thanks to all my friends in the department for making the Ph.D. so much fun: Angelique for keeping us all in line, Mark for being Mark, Tom for the banter, and Paul (and Julie) for their great cooking, friendship and wine. Additional thanks to Tom and Paul for proof-reading this thesis so carefully and making sure all of my equations were correctly punctuated! Any mistakes remaining in this thesis are entirely their fault. Thanks to Mike, Owen, James, Jose and Wadey for being such a good laugh, and to all the guys at football on Fridays for preventing me becoming fat. Speaking of football, thanks to Jose Mourinho for putting Chelsea where they belong, at the top.

Thanks to Lorraine and all the staff at Gosforth High School for making my year on the teaching fellowship scheme such an enjoyable one, and to Mike and the Ogden Trust for making it possible. I also gratefully acknowledge financial support from PPARC, without which I could not have contemplated undertaking a Ph.D.

The final word goes to Sarah, for her love, support and patience over the last four years. Thank you for following me to Durham and making my time here so happy, it wouldn't have been the same without you.

# Contents

# List of Figures

# List of Tables

# Notation

We use early Latin indices $a, b, \cdots$ to denote bulk (5-D) coordinates, Greek indices $\mu, \nu, \cdots$ for brane (4-D) coordinates, and later Latin indices $i, j, \cdots$ for spatial 3-D quantities. We use natural units with $c = \hbar = 1$, and sometimes also set the 4-D Newton's constant $G_4 = 1$. The sign conventions we adopt are those of Misner, Thorne and Wheeler [3]:

- Metric signature: $(-, +, \cdots, +)$

- Einstein equations: $G_{ab} = 8\pi G\, T_{ab} - \Lambda g_{ab}$

- Riemann tensor: $R^a{}_{bcd} = \partial_c \Gamma^a{}_{bd} - \partial_d \Gamma^a{}_{bc} + \Gamma^a{}_{ec}\Gamma^e{}_{db} - \Gamma^a{}_{ed}\Gamma^e{}_{cb}$

- Ricci tensor: $R_{ab} = R^c{}_{acb}$

# Chapter 1

# Introduction

Why is it that we appear to live in a universe with four dimensions – three of space and one of time? This is undoubtedly one of the most fundamental questions we can ask about the nature of the universe in which we live, and also one of the most elusive. Intriguingly, four dimensions seems to be *special* in that it is the minimum number of dimensions in which a gravitational field can exist in empty space, and therefore that we can have interesting physics and, indeed, life as we know it.

It is natural to ask if there might exist additional dimensions that are somehow *hidden* from us, creating the illusion that we live in just four dimensions (4-D). This might seem like a question for philosophers rather than physicists. On the contrary, extra dimensions are often postulated by theories that attempt to solve two of the biggest problems in theoretical physics: quantum gravity and unification of the forces.

General relativity (GR) [3–7] is a beautifully elegant theory of gravity that successfully accounts for a plethora of cosmological and astrophysical observations. This is all the more remarkable considering that most of these observations were not even thought of at the time GR was conceived. However, it also predicts its own demise in the form of singularities, as was recognised by Einstein himself soon after the formulation of the theory [8]. GR must therefore breakdown at high energies, where quantum effects begin to become important. Quantum mechanics (QM) [9–12], at the other extreme, successfully accounts for all known small scale physics, including the structure of matter itself. However, it suffers from conceptual problems – the measurement problem in particular – which it might be hoped will be resolved either by a more fundamental theory, or by a deeper understanding of the existing theory. We therefore seek a *quantum gravity* theory that can solve these problems and successfully describe physics in the overlap between GR and QM. A priori, this need not be based either on QM *or* GR and might supercede both,[1] with the only requirement being that it reproduces the results of both in an appropriate limit.

It is possible that the forces of the standard model, and perhaps also gravity, are in fact different manifestations of the *same* underlying force. This is the hope of *unification*, where the goal is to unify the known forces into a single force describing all the interactions in nature. Unification is an appealing principle and has been successful in the past at providing a deeper understanding of seemingly disparate phenomena; the most obvious example being Maxwell's unification of electricity and magnetism into electromagnetism. However, it is also possible that the forces of nature can not be unified in this way, and we should bear in mind another feature of the historical development of physics: the discovery of *new* particles and forces as we probe higher energy scales. It should also be emphasised that the issues of quantum gravity and unification of the forces are not necessarily connected, although a theory that can address both is clearly desirable.

Extra dimensions were first considered in connection to these issues by the pioneering works of Kaluza [13] and Klein [14] in the 1920s. Kaluza's brilliant idea was to use the extra components of the metric tensor in higher dimensions to unify electromagnetism with gravity, as described in the next section. More recently, a popular potential theory of both unification and quantum gravity is *string theory* [15], in which point particles are replaced by strings as the fundamental "building blocks" of nature. A huge amount of research effort has been expended on string theory, which requires us to live in 10-D for consistency (or 11-D for "M-theory"), although it is still not clear if it can correctly describe nature. Even if string theory turns out to be incorrect however, it is possible that extra dimensions might play a role in addressing these issues.

---

[1]This is why I prefer the phrase "theory of quantum gravity" rather than "quantum theory of gravity", since the latter seems already to elevate quantum theory to a more fundamental status.

Another area of theoretical physics in which extra dimensions might be invoked to resolve outstanding problems is cosmology. Increasingly precise measurements have led to the emergence of a "concordance model" in cosmology, which agrees with most cosmological data including the cosmic microwave background anisotropy [16], large scale structure formation [17], supernova IA distance measurements [18], and the abundance of light elements [19]. However, the model requires that the content of the universe is dominated by *dark energy* and *dark matter*, the true nature of which are not understood. It is quite possible that it is the theory of gravity that should be altered to fit these data instead, and extra dimensions provide an interesting framework for exploring this possibility.

If extra dimensions exist, we clearly require some mechanism to explain why it is that we do not experience them! Traditionally, this is achieved by supposing that they are *compactified* on a small scale, as suggested by Klein. Recently however, an alternative *braneworld* picture has emerged, in which we are confined to live on a 4-D *brane* embedded in a higher dimensional *bulk*. This allows extra dimensions which are large, or even infinite, thus providing an interesting alternative to compactification.

All this talk of extra dimensions is purely conjecture unless experimental evidence for them can be found. It should be borne in mind that an extra dimension can be thought of as providing us with an extra "degree of freedom" and so it should be no surprise if a theory with extra dimensions is better able to fit a particular set of data than standard 4-D theory can. What we therefore require is a "smoking gun"

– an unmistakable signature of extra dimensions that can not sensibly be accounted for conventionally. Such a discovery would profoundly alter our view of the universe in which we live.

## 1.1  Kaluza-Klein theory

In 1921, after a delay of two years, Kaluza published his seminal paper attempting to unify electromagnetism with gravity by considering GR in five dimensions [13]. Earlier work on the unification of electromagnetism with gravity through the use of an extra dimension was performed by Nordström [20], in the context of his pre-Einsteinian theory of gravity.

The vacuum Einstein equations in 4-D can be derived from the Einstein-Hilbert action

$$S_{EH} = \frac{1}{16\pi G} \int \mathrm{d}^4 x \sqrt{g}\, R, \tag{1.1}$$

where $g = \det(g_{\mu\nu})$, while the sourceless Maxwell equations follow from the action

$$S_{EM} = -\frac{1}{4} \int \mathrm{d}^4 x \sqrt{g}\, F_{\mu\nu} F^{\mu\nu}, \tag{1.2}$$

where $F_{\mu\nu}$ is the electromagnetic field strength tensor. Adding together (1.1) and (1.2) we obtain the Einstein-Maxwell equations for gravity coupled to an electromagnetic field. Instead, Kaluza's idea was to consider pure gravity in five dimensions, described by the action

$$S = \frac{1}{16\pi \hat{G}} \int \mathrm{d}^4 x \mathrm{d}y \sqrt{\hat{g}}\, \hat{R}, \tag{1.3}$$

where $y$ is the coordinate of the extra dimension and hats denote 5-D quantities. In order that the resulting theory is independent of the extra dimension, Kaluza simply imposed the "cylinder condition" that the metric components be independent of $y$:

$$\frac{\partial \hat{g}_{ab}}{\partial y} = 0. \tag{1.4}$$

Then, writing the metric in the suggestive form

$$\hat{g}_{ab} = \phi^{-1/3} \begin{pmatrix} g_{\mu\nu} + \phi A_\mu A_\nu & \phi A_\mu \\ \phi A_\nu & \phi \end{pmatrix}, \quad (1.5)$$

(1.3) becomes

$$S = \frac{1}{16\pi G} \int d^4x \sqrt{g} \left( R - \frac{1}{4}\phi F_{\mu\nu} F^{\mu\nu} - \frac{1}{6\phi^2}\partial_\mu\phi\, \partial^\mu\phi \right), \quad (1.6)$$

where $F_{\mu\nu} = \partial_\mu A_\nu - \partial_\nu A_\mu$ and $G = \hat{G}/\int dy$.

The action (1.6) describes 4-D gravity together with electromagnetism and a massless Klein-Gordon scalar field $\phi$. The appearance of this scalar troubled Kaluza, who simply removed it by setting $\phi = 1$. Equation (1.6) is then precisely the Einstein-Maxwell action for gravity plus electromagnetism. However, the condition $\phi =$ constant is only consistent with the 5-D Einstein equations if $F_{\mu\nu} = 0$. Hence, we must learn to live with the scalar field $\phi$, which is called the *dilaton*. Today, we are more comfortable with the existence of fundamental scalar fields. Indeed, the scalar Higgs and inflaton fields underpin modern theories of particle physics and cosmology! Even so, the dilaton is a serious drawback to the theory since the action (1.6) actually describes a phenomenologically unacceptable scalar-tensor theory of gravity. It is possible to avoid this problem, for example by giving the dilaton a potential $V(\phi)$, however the original appeal and beauty of Kaluza's idea – that electromagnetism can be regarded as a consequence of pure gravity in an *empty* spacetime – is then lost.

Kaluza's assumption that a fifth dimension exists but that no physical quantities depend on it seems somewhat contrived and unsatisfactory. In 1926 Klein [14] instead suggested that the extra dimension be *compactified* on a circle by identifying $y$ with $y + 2\pi R$, as depicted in Figure 1.1. This implies that any quantity $f(x^\mu, y)$ is periodic in $y$, $f(x^\mu, y) = f(x^\mu, y + 2\pi R)$, and can therefore be Fourier expanded:

$$f(x^\mu, y) = \sum_{n=-\infty}^{n=\infty} f_n(x^\mu) e^{iny/R}. \tag{1.7}$$

Now take $f(x^\mu, y)$ to be a massless scalar field satisfying the wave equation $\Box^{(5)} f = 0$. It follows that the fourier modes $f_n$ satisfy a 4-D Klein-Gordon equation with an effective mass $m_n$:

$$\Box^{(4)} f_n = m_n^2 f_n, \qquad m_n = \frac{n}{R}. \tag{1.8}$$

From a 4-D point of view $f$ therefore consists of a massless *zero mode* $f_0$ together with a tower of massive Kaluza-Klein (KK) modes $f_{n \neq 0}$. Klein realised that if the extra dimension was sufficiently small, the KK modes would be so massive that they will not appear in the world of *low energy* physics. Hence only the zero mode, which is independent of $y$, will be observable and physics will appear four-dimensional, as required. This argument applies to all 5-D fields, including the metric $\hat{g}_{ab}$ itself, as required for the action (1.6) to follow from (1.3).

The fact that KK modes of known particles have not been observed in collider experiments provides an upper bound on the size of the extra dimension:

$$R \lesssim O(\text{TeV}^{-1}) \sim 10^{-18} \text{m}, \tag{1.9}$$

with the standard "lore" being that extra dimensions must be compactified at the Planck length $R \approx l_p \sim 10^{-35}$m, which is the only natural length scale in the theory.

The remarkable fact that electromagnetism is contained in 5-D GR now receives a clear explanation: the *internal* $U(1)$ gauge invariance of electromagnetism is a manifestation of the local *coordinate* invariance under rotations around the small circle in $y$. This idea can be extended to incorporate other gauge groups by considering larger numbers of extra dimensions, all compactified on a small scale (although not necessarily with the topology of a circle), whereby it might be hoped that all the internal gauge symmetries of the Standard Model can be understood as local coordinate invariances of spacetime. Unfortunately, however, the theory can not be made consistent with all the observed features of the Standard Model (see, e.g., the discussions in [21, 22]). Nevertheless, the principles of KK theory, in particular the idea that extra dimensions are *compactified* on such a small scale that we do not experience them, pervades most subsequent discussion of extra-dimensional theories.

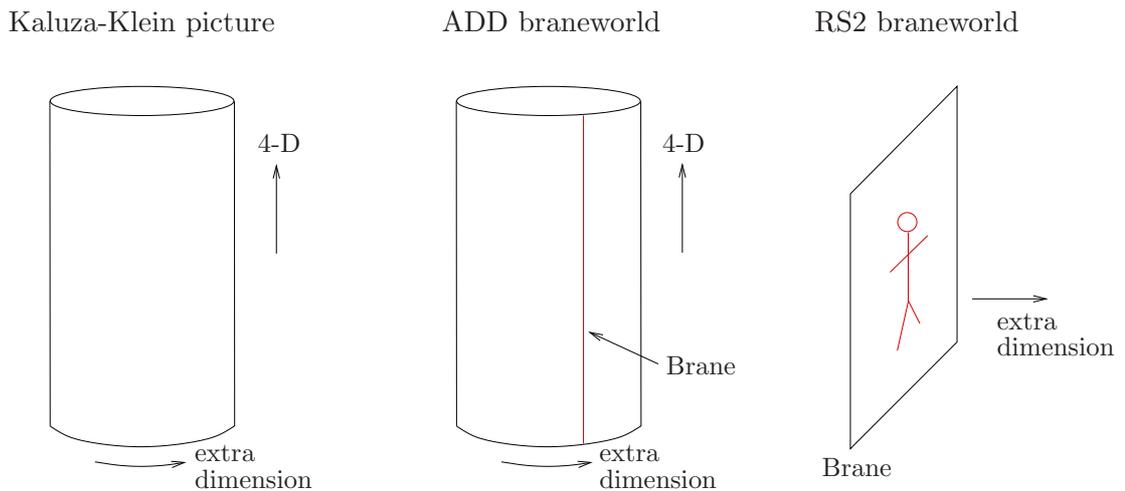

Figure 1.1: Cartoon showing three different pictures of an extra dimension.

## 1.2 ADD braneworlds

In 1998, theories of extra dimensions took a change in direction with the proposal by Arkani-Hamed, Dimopoulos and Dvali (ADD) [23–25] of an interesting model in which the extra dimensions can be *large*, in contrast with the conventional "wisdom" that they must be compactified at the Planck scale. Essentially a generalisation of the KK model, the ADD model has $n$ flat, compact extra dimensions of size $R$, but postulates that the Standard Model fields are *confined* to a 4-D *brane*, with only gravity propagating in the *bulk* (see Figure 1.1).

The assumption of confinement to a brane might at first seem counter-intuitive and no less ad-hoc than Kaluza's cylinder condition, however there are in fact several mechanisms that can produce such confinement. In string theory, for example, D-branes have confined gauge theories on their worldvolumes, while earlier incarnations of the braneworld idea [26–29] used topological defects to model the brane, and zero modes on the defect to produce confinement. Even so, in most braneworld models the precise mechanism remains unspecified and confinement is treated simply as an hypothesis. Confinement to a brane automatically ensures that there are no KK modes of the Standard Model fields. Therefore the bound (1.9) is evaded and the extra dimensions are allowed to be as large as

$$R \sim 0.1\text{mm}, \tag{1.10}$$

the scale down to which Newton's law has been experimentally tested [30].

A primary motivation of the ADD model is a possible resolution to the *hierarchy problem*, that is the unexplained large discrepancy between the Planck scale $M_p \sim 10^{16}$ TeV associated with gravity and the electroweak scale $M_{EW} \sim 1$ TeV of elementary particle physics. Consider the Newtonian potential between two test masses on the brane. For small separations $r \ll R$ the potential is that of higher-dimensional gravity, but for large separations $r \gg R$ gravity is insensitive to the extra dimension and the potential behaves like usual 4-D gravity:

$$V(r) \approx \frac{m_1 m_2}{M_f^{2+n}} \frac{1}{r^{n+1}}, \quad r \ll R, \tag{1.11}$$

$$V(r) \approx \frac{m_1 m_2}{M_f^{2+n}} \frac{1}{R^n r}, \quad r \gg R, \tag{1.12}$$

where $M_f$ is the fundamental mass scale of gravity in the full $(4+n)$-D spacetime. Therefore an observer on the brane experiences an *effective* 4-D Planck scale given by

$$M_p^2 = M_f^{2+n} R^n. \tag{1.13}$$

Hence the fundamental scale $M_f$ can be much lower than the Planck mass, while still giving rise to a large effective Planck mass $M_p \sim 10^{16}$ TeV on the brane due to the large volume of the extra dimensions. Essentially, gravity is understood as being weaker than the Standard Model forces because it "spreads" over the extra dimensions, whereas the other forces do not. If we wish to take the fundamental scale to be comparable to the weak scale, $M_f \sim M_{EW} \sim 1\,\text{TeV}$, equation (1.13) together with the bound (1.10) requires us to have $n \geq 2$ extra dimensions.

Of course, this does not really solve the hierarchy problem but merely reformulates it, since we now have to explain why $R$ should be so much larger than the length scale $10^{-18}\,\text{m}$ associated with $M_f$! Nevertheless, the possibility that gravity can have a fundamental mass scale much lower than the observed Planck mass has generated much excitement in the particle physics community, since it opens the door to the possibility of observing extra-dimensional effects in upcoming collider experiments [31]. In particular, black holes could be formed as a result of high energy particle collisions on the brane [32, 33].

## 1.3 Randall-Sundrum braneworlds

Unlike the flat extra dimensions of the ADD scenario, the Randall-Sundrum (RS) models [34, 35] proposed shortly afterwards allowed the bulk geometry to be curved, and endowed the brane with a tension. Hence the brane becomes a gravitating object, interacting dynamically with the bulk, making these models more interesting from the GR viewpoint. In their first model (RS1), which attempts to address the hierarchy problem, the extra dimension is still compact. However, they showed in their second paper (RS2) that, remarkably, 4-D gravity is recovered on the brane even for an extra dimension that is *infinite* in size! The RS2 model thus provides an interesting alternative to compactification, with gravity being effectively localised to the brane by the *curvature* of the bulk rather than straightforward compactification. Let us now review the models in turn.

### 1.3.1 Randall-Sundrum I (RS1)

The RS1 model consists of two branes of tension $\sigma_1$ and $\sigma_2$ bounding a slice of anti-de Sitter (AdS) space. Since the extra dimension is compact we need to specify conditions at the boundaries of the spacetime. We assume $\mathbb{Z}_2$ symmetry across both branes by identifying points $(x^\mu, y)$ with $(x^\mu, -y)$, and taking $y$ to be periodic with period $2L$. This describes an $S^1/\mathbb{Z}_2$ orbifold topology (i.e. a circle folded across a diameter), and we locate the two branes at the orbifold fixed points $y = 0$ and $y = L$, which form the boundaries of the spacetime. The assumption of $\mathbb{Z}_2$ symmetry can be loosely motivated from string theory constructions of braneworld models [36,37], although a more pragmatic motivation is that of simplicity.

The bulk Einstein equations to be solved are

$$R_{ab} - \frac{1}{2}R g_{ab} = -\Lambda_5 g_{ab}, \tag{1.14}$$

where the bulk cosmological constant $\Lambda_5$ can be expressed in terms of the AdS$_5$ curvature length $l$:

$$\Lambda_5 = -\frac{6}{l^2}. \tag{1.15}$$

Now, requiring that the branes preserve 4-D Lorentz invariance, we choose the metric ansatz

$$\mathrm{d}s^2 = a^2(y)\eta_{\mu\nu}\mathrm{d}x^\mu \mathrm{d}x^\nu + \mathrm{d}y^2. \tag{1.16}$$

Substituting this into (1.14) and imposing the orbifold boundary conditions, we obtain the solution

$$\mathrm{d}s^2 = e^{-2|y|/l}\eta_{\mu\nu}\mathrm{d}x^\mu \mathrm{d}x^\nu + \mathrm{d}y^2, \tag{1.17}$$

and find that the branes must have equal and opposite tensions, fine-tuned against

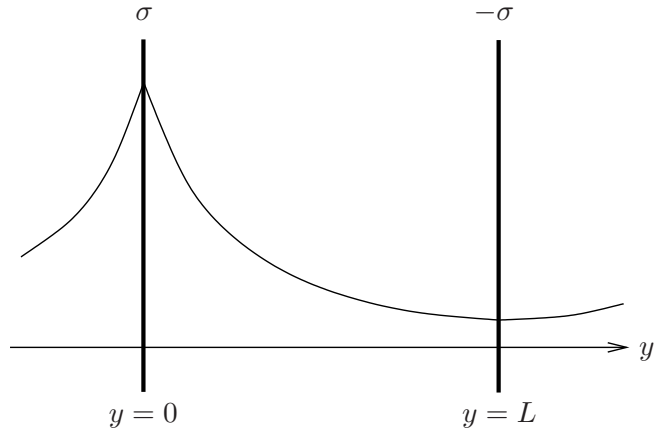

Figure 1.2: The behaviour of the warp factor in the RS1 model.

the bulk cosmological constant:

$$\sigma_1 = -\sigma_2 = \sigma\,; \tag{1.18}$$

$$\Lambda_5 = -\frac{32\pi^2\sigma^2 G_5^2}{3}, \tag{1.19}$$

where $G_5$ is the 5-D Newton's constant. For physical solutions, $\Lambda_5$ must therefore be negative, hence the bulk has to be AdS$_5$. The term $e^{-2|y|/l}$ in (1.17) is called the *warp* factor, which is $\mathbb{Z}_2$-symmetric (by construction), and decays exponentially from the positive tension brane to the negative tension brane (see Figure 1.2).

The primary motivation of the RS1 model was a possible resolution to the hierarchy problem. Assuming we live on the negative tension brane, it can be shown that the effective Planck mass $M_p$ measured on the brane is given by

$$M_p^2 \approx e^{2L/l} M_f^3 \, l, \tag{1.20}$$

where $M_f$ is the fundamental 5-D Planck mass (see [34] for details). Hence the exponential warp factor allows us to generate a large hierarchy between energy scales without introducing a large hierarchy involving the size of the extra dimension; if we wish to take $M_f \sim 1\,\text{TeV}$, we only need $L/l \approx 50$ in order to generate the observed Planck mass $M_p \sim 10^{16}\,\text{TeV}$.

However, there is a subtlety. The inter-brane separation, which corresponds to a scalar degree of freedom called the radion, must be fixed at a particular value for this to work, i.e. the radion must not fluctuate. The stabilisation mechanism for the radion is not thoroughly understood, and introduces its own complications [38–40]. Furthermore, the model requires us to live on a negative tension brane which, as we will see in Chapter 2, results in an unacceptable theory of gravity on the brane.

### 1.3.2 Randall-Sundrum II (RS2)

The RS2 model consists of a single, positive tension brane in an *infinite*, i.e. non-compact, extra dimension (see Figure 1.1). Operationally, the RS2 model can be obtained from the RS1 model by taking the negative tension brane to infinity, thus removing it from the setup. The model no longer addresses the hierarchy problem, but provides an interesting framework for exploring the gravitational effects of an extra dimension. The metric of the RS2 model is the same as in RS1:

$$ds^2 = e^{-2|y|/l}\eta_{\mu\nu}dx^\mu dx^\nu + dy^2, \tag{1.21}$$

where we have retained the $\mathbb{Z}_2$ reflection symmetry across the brane, which resides at $y = 0$. The brane tension is fine-tuned against the bulk cosmological constant as before (see equation (1.19)),

$$\sigma = \frac{3}{4\pi l G_5}, \tag{1.22}$$

(where we have used (1.15)) which ensures the brane has the flat geometry of Minkowski spacetime. Relaxing this condition results in a 4-D effective cosmological constant on the brane, as shown in Section 2.3.

Since the extra dimension is no longer compact, there is a *continuum* of massive KK modes of the 5-D graviton (i.e. the fourier series in (1.7) becomes a fourier integral). However, these KK modes are suppressed near the brane by the curvature of the bulk, with the result that standard 4-D gravity (described by the *massless* zero mode of the graviton) is recovered to leading order on the brane, as shown in detail in Chapter 2. This remarkable property, that 4-D gravity is recovered on the brane even though gravity has access to an infinite extra dimension, has had a huge impact on studies of extra dimensions and spawned a whole new field of research in braneworld gravity.

In some sense, the RS2 model provides the simplest, geometrically appealing model of an extra dimension, and is the one we will be using in this thesis. Henceforth, whenever we refer to braneworlds we are referring specifically to the RS2 braneworld scenario. For braneworld reviews, including discussions of various aspects of braneworlds that will not be considered in this thesis, see [41–44].

## 1.4 Motivations and thesis outline

Browsing the literature on braneworlds the statement that braneworlds are "motivated from" or "inspired by" string theory is often encountered. Indeed, if string theory is correct then braneworlds can be expected to provide useful insights into the full, more complicated theory. However, string theory appears to be in something of a crisis at the moment [45, 46], seemingly needing to resort to the anthropic principle in order to make contact with the observable universe [47]. As a personal comment, I find this unsatisfactory as it would mean string theory is not falsifiable, and therefore not a true scientific theory [48]. Suppose a competing quantum gravity theory was constructed – call it *N theory* – that was able to reproduce cosmology, the standard model etc., but which also contained numerous other possible "vacua". We could then appeal to the anthropic principle to "explain" why the universe appears as we observe it. However, it would then be impossible to distinguish N theory from M theory and determine which of them (if either!) correctly described nature.

While string theory requires extra dimensions, the existence of extra dimensions certainly does not require string theory. Hence, we adopt a slightly unconventional view and regard braneworlds as simply an interesting framework within which to explore the logical possibility of the existence of extra dimensions. Consistent with this viewpoint, we consider only the simplest RS2 model of a single $\mathbb{Z}_2$-symmetric brane in a bulk that is empty except for a cosmological constant (no moduli fields[2] etc.), in order to explore the purely gravitational effects of an extra dimension.

In particular, we focus on the topic of black holes in braneworlds. Black holes are fascinating objects and provide a promising testbed for GR. Black holes in 4-D GR are uniquely characterised by their mass, charge and angular momentum, as described by the Kerr-Newman family of metrics. Black holes in braneworlds are not so well understood and there exist whole classes of possible solutions (although it might be hoped that a unique solution can be found), with some key potential differences from the black holes of GR. Observations of black holes might therefore provide an exciting means of detecting extra dimensions, if they exist.

The thesis is outlined as follows. In Chapter 2 we give a description of braneworld gravity and show that the results of GR are recovered on the brane, but with corrections due to the existence of the extra dimension. In Chapter 3 we discuss static, spherically symmetric braneworld black hole solutions. Using a dynamical systems approach we classify the behaviour of these solutions according to an assumed equation of state for the "Weyl term", which encodes bulk gravitational effects. The gravitational lensing properties of a couple of candidate braneworld black hole metrics are investigated in Chapter 4 using the strong field limit approach. In Chapter 5 we discuss rotating braneworld black holes, and attempt to generate such solutions from known static solutions using the Newman-Janis complexification "trick". We conclude in Chapter 6.

---

[2]In an M-theory realisation of the 5-D braneworld picture, the six extra dimensions required by M-theory must be compactified á la Kaluza-Klein, with their effect on the dynamics being felt through 5-D scalar fields, called *moduli* fields.

# Chapter 2

# Braneworld Gravity

## 2.1 Introduction

A basic requirement of any alternative theory of gravity is that it reproduces Newton's law of gravity in the appropriate limit. Naïvely, braneworld gravity fails at this first hurdle as we might expect the 5-D nature of gravity to be reflected through a $1/r^3$ force law, rather than the familiar 4-D $1/r^2$ behaviour. In overcoming this apparent problem, the *warp factor* in (1.21) is crucial. Essentially, this has the effect of "squeezing" gravity around the brane so that gravity spreads mostly in the four directions parallel to the brane and Newton's law is indeed recovered, despite the fact that gravity has access to an extra dimension. This remarkable property is the key reason for the phenomenological viability of the braneworld scenario.

Of course, the empirical success of standard 4-D GR goes beyond just reproducing Newton's laws in the weak field approximation for a static source and slowly moving test masses. The first big success of GR was the correct prediction of the bending of light by the Sun. Indeed, Eddington's confirmation of this prediction by observing starlight during a solar eclipse in 1919 made Einstein and his theory famous. Other solar system tests confirming the predictions of GR include various radar time-delay experiments and the precession of perihelion for planetary orbits (for the experimental status of GR, see [49]). A further interesting prediction of GR is the "frame dragging" effect produced by a rotating massive body. Although currently unverified, this effect is soon to be tested by the Gravity Probe B satellite [50], launched in 2004, which contains four suspended gyroscopes spinning freely in vacuum. By measuring the tiny gyroscopic precession caused by the Earth's rotation, this experiment will provide an additional important test of GR, with results expected in 2007.

All of the above effects can be treated within the framework of linearised general relativity, where we assume the spacetime can be described by small perturbations around a given background. An exposition of linearised gravity in braneworlds is given in the next section, where it is shown that the general relativistic results are recovered on the brane, with small corrections due to the presence of the extra dimension.

However, not all gravitational phenomena can be treated in the weak field limit, and we clearly require a concrete understanding of the true non-perturbative nature of gravity on the brane. This is provided by the Shiromizu, Maeda, Sasaki approach [51], where the basic idea is to project the 5-D Einstein equations onto the brane to obtain 4-D effective Einstein equations. These equations, which are derived in Section 2.3, closely resemble the standard Einstein equations, but with correction terms arising from bulk gravitational effects.

Having described the foundations of general relativity and presented Einstein's equations, a textbook or lecture course on GR would then proceed to discuss two of the most important applications of GR: cosmology and black holes. The final section of this chapter gives a brief overview of braneworld cosmology, with the remainder of the thesis devoted to a study of braneworld black holes.

## 2.2 Linearised gravity

In the following account we adopt Garriga and Tanaka's classic approach to linearised braneworld gravity [52] (see also [53]). Consider perturbations $h_{\mu\nu}$ about the Randall-Sundrum background (1.21):

$$\mathrm{d}s^2 = \left(e^{-2|y|/l}\eta_{\mu\nu} + h_{\mu\nu}\right)\mathrm{d}x^\mu \mathrm{d}x^\nu + \mathrm{d}y^2. \tag{2.1}$$

Notice that in writing down (2.1) we have chosen hypersurface orthogonal coordinates with $h_{a4} = 0$, which reduces the fifteen functions $h_{ab}$ to the ten functions $h_{\mu\nu}$. This is a choice we can always make in the neighbourhood of the brane (see Appendix A.3). Such coordinates are not unique however, as we can still make coordinate transformations which shift the brane while preserving $h_{a4} = 0$. It turns out that two different gauges,[1] both of them hypersurface orthogonal, are useful for this analysis and we will transform between them as required.

Linearising the bulk Einstein equations (1.14) in the perturbations $h_{\mu\nu}$, we obtain

$$R_{44} + \frac{4}{l^2} = -\frac{1}{2}\partial_y\left(e^{2|y|/l}\partial_y h\right) = 0, \tag{2.2}$$

$$R_{\mu 4} = \frac{1}{2}\partial_y\left[e^{2|y|/l}(\partial_\nu h_\mu{}^\nu - \partial_\mu h)\right] = 0, \tag{2.3}$$

$$R_{\mu\nu} + \frac{4}{l^2}g_{\mu\nu} = \frac{1}{2}e^{2|y|/l}\left(2\partial_\sigma\partial_{(\nu}h_{\mu)}{}^\sigma - \Box^{(4)}h_{\mu\nu} - \partial_\mu\partial_\nu h\right)$$
$$- \frac{1}{2}\partial_y^2 h_{\mu\nu} + \frac{2}{l^2}h_{\mu\nu} + \eta_{\mu\nu}\left(\frac{h}{l^2} + \frac{\partial_y h}{2l}\right) = 0, \tag{2.4}$$

where $h = h_\mu{}^\mu$, $\Box^{(4)} \equiv \eta^{\mu\nu}\partial_\mu\partial_\nu$ is the 4-dimensional Laplacian and 4-dimensional indices are raised and lowered using the Minkowski metric $\eta_{\mu\nu}$. We can choose the transverse, trace-free Randall-Sundrum (RS) gauge [35] everywhere in the bulk,

---

[1] We use the terms "gauge transformation" and "coordinate transformation" interchangeably.

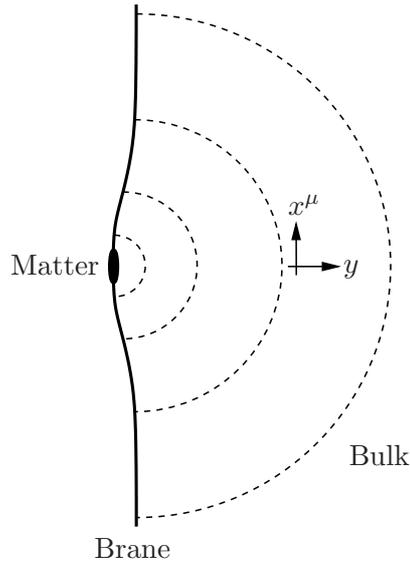

Figure 2.1: Gravitational field of a matter source in the Randall-Sundrum gauge, showing the brane-bending.

defined by

$$h_{a4} = 0, \qquad \partial_\nu h_\mu{}^\nu = h = 0. \tag{2.5}$$

The five polarisations of the 5-D graviton are contained in the five independent components of $h_{\mu\nu}$ in the RS gauge (which has no remaining gauge freedom). In this gauge, equations (2.2) and (2.3) are trivially solved, and equation (2.4) nicely simplifies to give the decoupled field equation for the perturbations:

$$\left(e^{2|y|/l}\Box^{(4)} + \partial_y^2 - \frac{4}{l^2}\right) h_{\mu\nu} = 0. \tag{2.6}$$

Boundary conditions for this equation are given by the Israel junction conditions at the brane, which relate the jump in extrinsic curvature across the brane to the stress-energy tensor of the brane (see Appendix A.4). However, when we choose the RS gauge in the bulk, matter on the brane causes it to bend so that it is no longer located at $y = 0$ in general (see Figure 2.1). It is therefore convenient to transform to Gaussian Normal (GN) coordinates $(\bar{x}^\mu, \bar{y})$ defined by $\bar{h}_{a4} = 0$ with the brane fixed at $\bar{y} = 0$, in which the junction conditions (A.23) can be straightforwardly implemented. Working on the positive side of the brane and assuming even parity under $\bar{y} \to -\bar{y}$ (i.e. $\mathbb{Z}_2$ symmetry), these give:

$$-\frac{2}{l}\eta_{\mu\nu} + \partial_{\bar{y}}\bar{h}_{\mu\nu} = -8\pi G_5 \left(S_{\mu\nu} - \frac{1}{3}q^0_{\mu\nu}S\right), \qquad (\bar{y}=0^+), \qquad (2.7)$$

where $q^0_{\mu\nu} = q_{\mu\nu}(\bar{y}=0) = (\eta_{\mu\nu} + \bar{h}_{\mu\nu})$ is the induced metric on the brane. The brane stress-energy tensor $S_{\mu\nu}$ can be split into the tension $\sigma$ and the energy-momentum tensor $T_{\mu\nu}$ of any matter on the brane:

$$S_{\mu\nu} = -\sigma q^0_{\mu\nu} + T_{\mu\nu}. \qquad (2.8)$$

Substituting this into (2.7) and imposing the fine-tuning condition (1.22) (as required for the background metric to be a solution), we arrive at

$$\left(\frac{2}{l} + \partial_{\bar{y}}\right)\bar{h}_{\mu\nu} = -8\pi G_5 \left(T_{\mu\nu} - \frac{1}{3}\eta_{\mu\nu}T\right), \qquad (\bar{y}=0^+). \qquad (2.9)$$

We now wish to transform this junction condition back into the RS gauge, in order to make it compatible with the field equation (2.6).

Consider the transformation between the two gauges:

$$\bar{y} = y + \xi^4(x^a), \qquad (2.10)$$
$$\bar{x}^\mu = x^\mu + \xi^\mu(x^a), \qquad (2.11)$$

which is small in the sense that $\xi^4$ is small. The requirement that the metric takes the form (2.1) in both gauges, in particular that $\bar{h}_{44} = 0$ and $\bar{h}_{\mu 4} = 0$, gives the conditions

$$\xi^4 = \xi^4(x^\nu), \qquad (2.12)$$
$$\xi^\mu = -\frac{l}{2}e^{2|y|/l}\eta^{\mu\sigma}\partial_\sigma \xi^4(x^\nu) + F^\mu(x^\nu), \qquad (2.13)$$

where the functions $\xi^4(x^\nu)$ and $F^\mu(x^\nu)$ are independent of $y$. Note that there is some residual gauge freedom in (2.13), which we will utilise to simplify later equations,[2] while $\xi^4(x^\nu)$ will be determined self-consistently, as we shall see shortly. It follows that the relation between the perturbations in the two coordinate systems is

$$h_{\mu\nu} = \bar{h}_{\mu\nu} - l\partial_\mu\partial_\nu\xi^4 - \frac{2}{l}e^{-2|y|/l}\eta_{\mu\nu}\xi^4 + e^{-2|y|/l}\partial_{(\mu}\eta_{\nu)\rho}F^\rho. \qquad (2.14)$$

Hence the junction condition (2.9) becomes

$$\left(\frac{2}{l} + \partial_y\right)h_{\mu\nu} = -\Sigma_{\mu\nu}, \qquad (y = 0^+), \qquad (2.15)$$

where

$$\Sigma_{\mu\nu} = 8\pi G_5\left(T_{\mu\nu} - \frac{1}{3}\eta_{\mu\nu}T\right) + 2\partial_\mu\partial_\nu\xi^4 \qquad (2.16)$$

is the effective source term in the RS gauge and includes the effect of the brane bending (recall from (2.10) that the brane is located at $y = -\xi^4$ in the RS gauge). The vanishing of $h_\mu{}^\mu$ in the RS gauge implies $\Sigma_\mu{}^\mu = 0$, and so $\xi^4(x^\nu)$ is determined as the solution to

$$\Box^{(4)}\xi^4 = \frac{4\pi G_5}{3}T, \qquad (2.17)$$

which shows explicitly that the brane is bent by the matter residing on it.

It is possible to combine equations (2.6) and (2.15) into a single equation by including delta functions at the discontinuity, giving us the equations of motion for the perturbations in the RS gauge:[3]

$$\left[e^{2|y|/l}\Box^{(4)} + \partial_y^2 - \frac{4}{l^2} + \frac{4}{l}\delta(y)\right]h_{\mu\nu} = -2\Sigma_{\mu\nu}\delta(y). \qquad (2.18)$$

---

[2]The condition $\bar{y} = 0$ reduces the number of independent components of $\bar{h}_{\mu\nu}$ to nine. However, there should be only five! The four "extra" degrees of freedom are precisely those present in the functions $F^\mu(x^\nu)$, and correspond to the transformations we can perform within the brane while still preserving $\bar{y} = 0$ and the form (2.1). Once we have specified these we will have fixed the gauge completely and will have the correct number of degrees of freedom for the 5-D graviton.

[3]Equation (2.18) then gives the vacuum field equation (2.6) for $y \neq 0$, while integration across $y = 0$ yields the junction condition (2.15).

Defining $G(x, y; x', y')$ to be the 5-D retarded Green's function satisfying

$$\left[e^{2|y|/l}\Box^{(4)} + \partial_y^2 - \frac{4}{l^2} + \frac{4}{l}\delta(y)\right] G(x, y; x', y') = \delta^{(4)}(x - x')\delta(y - y'), \quad (2.19)$$

the solution to (2.18) is then given by

$$\begin{aligned} h_{\mu\nu} &= -2 \int d^4x' dy'\, G(x, y; x', y')\Sigma_{\mu\nu}(x')\delta(y') \\ &= -2 \int d^4x'\, G(x, y; x', 0)\Sigma_{\mu\nu}(x'). \end{aligned} \quad (2.20)$$

To solve (2.19) for the Green's function we require a second boundary condition in addition to the junction condition at the brane. This is taken to be the reasonable condition that the perturbations remain bounded at the AdS horizon $y = \infty$ (for more on this issue, see [54–56]). Then, using standard techniques from Sturm-Liouville theory, the full Green's function can be constructed from a complete set of eigenstates as

$$G(x, y; x', y') = -\int \frac{d^4k}{(2\pi)^4} e^{ik_\mu(x^\mu - x'^\mu)} \left[\frac{e^{-2(|y|+|y'|)/l}l^{-1}}{\mathbf{k}^2 - (\omega+i\epsilon)^2} + \int_0^\infty dm\, \frac{u_m(y)u_m(y')}{m^2 + \mathbf{k}^2 - (\omega+i\epsilon)^2}\right], \quad (2.21)$$

where $k^\mu = (\omega, \mathbf{k})$,

$$u_m(y) = \frac{\sqrt{ml/2}\left[J_1(ml)Y_2(mle^{|y|/l}) - Y_1(ml)J_2(mle^{|y|/l})\right]}{\sqrt{J_1(ml)^2 + Y_1(ml)^2}}, \quad (2.22)$$

and $J_n$, $Y_n$ are Bessel functions of order $n$. The first term in (2.21) corresponds to the graviton zero mode, with the second term representing the continuum of KK modes. We will be primarily interested in the stationary case, for which the Green's function is given by

$$G(\mathbf{x}, y; \mathbf{x}', y') = \int_{-\infty}^\infty dt'\, G(x, y; x', y'). \quad (2.23)$$

The long-distance behaviour of gravity is determined by the Green's function in the limit $|\mathbf{x} - \mathbf{x}'| \gg l$.[4] With suitable expansions of the Bessel functions, and taking both points on the brane, we find

$$G(\mathbf{x}, 0; \mathbf{x}', 0) \approx -\frac{1}{4\pi l r}\left(1 + \frac{l^2}{2r^2} + \dots\right), \qquad (2.24)$$

where $r = |\mathbf{x} - \mathbf{x}'|$. We can see that the zero mode gives the 4-D $1/r$ behaviour, with the KK modes inducing a subleading correction term proportional to $(l/r)^2$. Also, for a source on the wall ($y' = 0$), the leading behaviour for large $r$ and large $y$ is given by [52]

$$G(\mathbf{x}, y; \mathbf{x}', 0) \approx -\frac{e^{-3|y|/l}}{8\pi l}\left[\frac{2e^{-2|y|/l}r^2 + 3l^2}{(e^{-2|y|/l}r^2 + l^2)^{3/2}}\right], \qquad (2.25)$$

showing that the perturbation decays rather steeply off the brane (as depicted in Figure 2.1).

Since we are interested in the perturbation on the brane, it is convenient to perform a final transformation back to GN coordinates. From (2.14) we have

$$\bar{h}_{\mu\nu} = h_{\mu\nu}^{(m)} + \frac{2}{l}e^{-2|y|/l}\eta_{\mu\nu}\xi^4 + h_{\mu\nu}^{(\xi)} + l\partial_\mu\partial_\nu\xi^4 - e^{-2|y|/l}\partial_{(\mu}\eta_{\nu)\rho}F^\rho, \qquad (2.26)$$

where we have split $h_{\mu\nu}$ into matter and brane-bending pieces using (2.20) and (2.16):

$$h_{\mu\nu}^{(m)} = -16\pi G_5 \int d^4x'\, G(x, y; x', 0)\left(T_{\mu\nu} - \frac{1}{3}\eta_{\mu\nu}T\right), \qquad (2.27)$$

$$h_{\mu\nu}^{(\xi)} = -4 \int d^4x'\, G(x, y; x', 0)\, \partial_\mu\partial_\nu\xi^4. \qquad (2.28)$$

Choosing $F^\mu(x^\nu)$ appropriately, we can avail ourselves of the remaining gauge freedom and set the last three terms in (2.26) to zero. Then, setting $y = 0$, we arrive

---

[4] In the opposite limit, $|\mathbf{x} - \mathbf{x}'| \ll l$, we obtain $G(\mathbf{x}, 0; \mathbf{x}', 0) \propto 1/r^2$, which simply reflects the fact that gravity becomes truly 5-dimensional at short distances.

at the relatively simple final expression for the perturbation on the brane:

$$\bar{h}_{\mu\nu} = -16\pi G_5 \int d^4x' \, G(x,0;x',0) \left( T_{\mu\nu} - \frac{1}{3}\eta_{\mu\nu}T \right) + \frac{2}{l}\eta_{\mu\nu}\xi^4, \qquad (2.29)$$

with $G(x,0;x',0)$ given by (2.24) for the stationary far-field case (or (2.21) more generally), and $\xi^4$ determined by (2.17).

### 2.2.1 The graviton propagator

As we will see shortly, equation (2.29) reproduces the results of linearised general relativity for a stationary source on the brane, with small corrections due to the KK modes. Before inputting a particular $T_{\mu\nu}$, however, we will examine equation (2.29) in a little more detail; at issue is the structure of the massless graviton propagator. The KK modes of the Green's function (2.21) are only important at high energies. At low energies, the Green's function is dominated by the zero mode, and so we consider the zero-mode truncation

$$G(x,0;x',0) = \frac{l^{-1}}{\Box^{(4)}}, \qquad (2.30)$$

where

$$\frac{1}{\Box^{(4)}} = -\int \frac{d^4k}{(2\pi)^4} \frac{e^{ik_\mu(x^\mu - x'^\mu)}}{\mathbf{k}^2 - (\omega + i\epsilon)^2} \qquad (2.31)$$

is the massless scalar Green's function for 4-D Minkowski space. The matter part of the metric perturbation on the brane is then given by

$$h_{\mu\nu}^{(m)} = -\frac{16\pi G_5}{l} \int d^4x' \, \frac{1}{\Box^{(4)}} \left( T_{\mu\nu} - \frac{1}{3}\eta_{\mu\nu}T \right). \qquad (2.32)$$

However, this is *not* the propagator for a massless 4-D graviton – we have a factor of $\frac{1}{3}$ instead of the usual $\frac{1}{2}$. This is a manifestation of the van Dam-Veltman-Zakharov (vDVZ) discontinuity [57,58], which states that the $m \to 0$ limit of a massive graviton propagator does not coincide with the massless graviton propagator, since the number of polarisation states of the fields do not match. We have an extra polarisation state, a 4-D scalar field, contained in the 5-D graviton propagator, which persists even in the massless limit (i.e. the zero-mode truncation we are considering here). This could potentially be troublesome, since an incorrect tensor structure for the 4-D propagator results in a prediction for the bending of light that is inconsistent with the value accurately predicted by GR [52, 57].

However, we were careful in the above treatment to include the effect of the brane-bending, which is described by the scalar field $\xi^4(x^\mu)$, given by the solution to equation (2.17):

$$\xi^4 = \frac{4\pi G_5}{3} \int d^4x' \frac{1}{\Box^{(4)}} T. \tag{2.33}$$

Putting $\xi^4$ and $h_{\mu\nu}^{(m)}$ into (2.29) we find that the full metric perturbation on the brane is given by

$$\bar{h}_{\mu\nu} = -\frac{16\pi G_5}{l} \int d^4x' \frac{1}{\Box^{(4)}} \left(T_{\mu\nu} - \frac{1}{2}\eta_{\mu\nu}T\right). \tag{2.34}$$

We can see that the extra polarisation state of the 5-D graviton has been exactly compensated for by the brane-bending effect, resulting in the correct tensor structure for a 4-D massless graviton propagator. Hence, linearised general relativity is indeed recovered on the brane at low energies.

### 2.2.2 The Newtonian potential on the brane

To consider the effects of the KK modes, we now compute the metric perturbation for a point mass $M$ at rest on the brane, with stress-energy tensor

$$T_{\mu\nu} = M\,\delta^{(3)}(\mathbf{x})\delta_\mu^0\delta_\nu^0. \tag{2.35}$$

The brane-bending function $\xi^4$ is determined by equation (2.17):

$$\Box^{(4)}\xi^4 = -\frac{4\pi G_5}{3} M \delta^{(3)}(\mathbf{x}). \tag{2.36}$$

For a time-independent solution the differential operator $\Box^{(4)}$ reduces to the 3-D Laplacian and so we obtain the solution

$$\xi^4 = \frac{G_5 M}{3r}, \tag{2.37}$$

where $r = |\mathbf{x}|$. The matter part of the perturbation is given by (2.27) with the stationary Green's function (2.24):

$$\begin{aligned} h^{(m)}_{\mu\nu} &= 16\pi G_5 \int d^3\mathbf{x}' \frac{1}{4\pi l |\mathbf{x}-\mathbf{x}'|}\left(1 + \frac{l^2}{2|\mathbf{x}-\mathbf{x}'|^2}\right)\left(\delta^0_\mu \delta^0_\nu + \frac{1}{3}\eta_{\mu\nu}\right) M \delta^{(3)}(\mathbf{x}') \\ &= \frac{4G_5 M}{lr}\left(1 + \frac{l^2}{2r^2}\right)\left(\delta^0_\mu \delta^0_\nu + \frac{1}{3}\eta_{\mu\nu}\right). \end{aligned} \tag{2.38}$$

Combining (2.37) and (2.38) in equation (2.29) we find that the metric perturbation on the brane is given by

$$\bar{h}_{\mu\nu} = \frac{2G_5 M}{lr}\left[\left(1 + \frac{l^2}{3r^2}\right)\eta_{\mu\nu} + \left(2 + \frac{l^2}{r^2}\right)\delta^0_\mu \delta^0_\nu\right], \tag{2.39}$$

with components

$$\bar{h}_{00} = \frac{2G_4 M}{r}\left(1 + \frac{2l^2}{3r^2}\right), \qquad \bar{h}_{ij} = \frac{2G_4 M}{r}\left(1 + \frac{l^2}{3r^2}\right)\delta_{ij}, \tag{2.40}$$

where the 4-D and 5-D Newton's constant are related through $G_4 = G_5/l$. Hence we recover the weak field limit of the Schwarzschild metric to leading order, with small corrections due to the KK modes. In particular, the Newtonian potential is given by

$$\phi = \frac{1}{2}\bar{h}_{00} = \frac{G_4 M}{r}\left(1 + \frac{2l^2}{3r^2}\right), \tag{2.41}$$

which does not contradict experimental tests of Newton's inverse square law provided $l \lesssim 0.1$mm (the distance down to which Newton's law has been tested [30]). From equation (1.22) this leads to a lower limit on the brane tension:

$$\sigma \gtrsim (1\,\text{TeV})^4. \tag{2.42}$$

### 2.2.3 The gravitomagnetic potential on the brane

The simplest extension of the static point mass example is to consider a stationary source on the brane that is rotating. Just as a moving electric charge creates a magnetic field, a rotating mass produces a gravitational field that is called, by analogy, the gravitomagnetic field. This is a purely relativistic effect, responsible for the "dragging of inertial frames", and has no Newtonian counterpart (in Newtonian gravity, the field surrounding a rotating body is described solely by the Newtonian potential). The corresponding gravitomagnetic *potential* is given by the off-diagonal component $g_{0i}$ of the metric (using the same analogy, the Newtonian potential $\phi = \frac{1}{2}\bar{h}_{00}$ of the previous section is termed the *gravitoelectric* potential).

The gravitomagnetic potential for a source on the brane was first presented by Nayeri and Reynolds [59], although few details of the calculation were given and the result appears not to be widely known.[5] A point particle can not have a nonzero orbital angular momentum, so we consider a more general pressureless perfect fluid source with $T_{\mu\nu} = \rho\, u_\mu u_\nu$, which we interpret as a rotating solid body. The brane-bending term in (2.29) does not contribute to any off-diagonal component of the metric perturbation, and so the gravitomagnetic potential is given purely by the

---

[5]As illustrated by the fact that [59] has only 2 citations in the SLAC-Spires database at present, as compared with over 400 for Garriga and Tanaka's paper [52].

matter part:

$$\bar{h}_{0i} = -16\pi G_5 \int d^4x' \, G(x,0;x',0) T_{0i}(x') \tag{2.43}$$

$$= 4G_4 \int d^3\mathbf{x}' \, \frac{T_{0i}(\mathbf{x}')}{|\mathbf{x}-\mathbf{x}'|}\left(1 + \frac{l^2}{2\,|\mathbf{x}-\mathbf{x}'|^2}\right), \tag{2.44}$$

where we have used (2.24). Performing the multipole expansion

$$\frac{1}{|\mathbf{x}-\mathbf{x}'|} = \frac{1}{r} + \frac{\mathbf{x}\cdot\mathbf{x}'}{r^3} + \ldots, \tag{2.45}$$

where $r = |\mathbf{x}| \gg |\mathbf{x}'|$, we have

$$\bar{h}_{0i} = 4G_4 \int d^3x' \, T_{0i}(x') \left(\frac{1}{r} + \frac{x^j x'_j}{r^3}\right)\left[1 + \frac{l^2}{2}\left(\frac{1}{r^2} + 2\frac{x^j x'_j}{r^4} + \ldots\right)\right]$$

$$= 4G_4 \left[\left(1 + \frac{l^2}{2r^2}\right)\frac{1}{r}P_i + \left(1 + \frac{3l^2}{2r^2}\right)\frac{x^j}{r^3}J_{ij}\right] + \ldots, \tag{2.46}$$

where

$$P_i = \int d^3x' \, T_{0i}(x'), \qquad J_{ij} = \int d^3x' \, x'_j T_{0i}(x'). \tag{2.47}$$

We will find in Section 2.3 that the standard energy-momentum conservation equation $\nabla_\mu T^{\mu\nu} = 0$ remains valid on the brane, and to linear order this becomes $\partial_\mu T^{\mu\nu} = 0$. For stationary systems this implies $\partial_i T^{i0} = 0$ and therefore, integrating by parts,

$$0 = -\int d^3x \, x_i \frac{\partial}{\partial x^j} T^{j0} = P_i, \tag{2.48}$$

$$0 = -2\int d^3x \, x_i x_j \frac{\partial}{\partial x^k} T^{k0} = J_{ij} + J_{ji}. \tag{2.49}$$

The antisymmetry of $J_{ij}$ allows us to write it as

$$J_{ij} = \frac{1}{2}\varepsilon_{ijk} J^k, \tag{2.50}$$

where
$$J^k = \int \mathrm{d}^3 x\, \varepsilon_{mnk} x^m T^{n0} \tag{2.51}$$

is the system's angular momentum. The metric perturbation (2.46) therefore becomes

$$\bar{h}_{0i} = 2G_4 \frac{\varepsilon_{ijk} x^j J^k}{r^3}\left(1 + \frac{3l^2}{2r^2}\right), \tag{2.52}$$

and so we recover the linearised Kerr metric, with a subleading correction term arising from the KK modes. To make this more concrete, consider the case of just a single non-zero angular momentum component $J^z = Ma$. Then, incorporating the results of Section 2.2.2 and setting $G_4 = 1$, we find the metric

$$\begin{aligned}\mathrm{d}s^2 =\ & -\left[1 - \frac{2M}{r}\left(1 + \frac{2l^2}{3r^2}\right)\right]\mathrm{d}t^2 + \left[1 + \frac{2M}{r}\left(1 + \frac{l^2}{3r^2}\right)\right](\mathrm{d}x^2 + \mathrm{d}y^2 + \mathrm{d}z^2) \\ & - \frac{4Ma}{r^3}\left(1 + \frac{3l^2}{2r^2}\right)(x\,\mathrm{d}y - y\,\mathrm{d}x)\mathrm{d}t,\end{aligned} \tag{2.53}$$

which is recognizable as the linearised Kerr metric with $\sim l^2/r^2$ corrections.

### 2.2.4 Beyond linear order

The recovery of the linearised Schwarzschild and Kerr metrics, to leading order in $(l/r)$, is a key requirement for the observational consistency of braneworld gravity. However, for the precession of perihelion of Mercury, the observational accuracy is up to the first order of *non-linearity* in the metric component $g_{00}$. Indeed, for the PPN-type expansion of the static, spherically symmetric metric

$$\mathrm{d}s^2 = -\left(1 - \frac{2GM}{r} - \zeta\frac{G^2 M^2}{r^2} + \ldots\right)\mathrm{d}t^2 + \left(1 + \frac{2GM}{r} + \ldots\right)\mathrm{d}r^2 + r^2 \mathrm{d}\Omega^2, \tag{2.54}$$

the precession angle per orbit is given by [49, 60]

$$\delta = \frac{\pi GM}{L}(6 + \zeta), \tag{2.55}$$

where $L$ is the semilatus rectum of the orbit.

The above perturbative analysis of braneworld gravity can be extended to second order in $GM$, and the general relativistic result $\zeta = 0$ is recovered to leading order in $(l/r)$ [61, 62]. Thus, the RS2 braneworld scenario is indeed consistent with all solar system tests of gravity.

## 2.3 Non-perturbative gravity

Perhaps paradoxically, a description of non-perturbative gravity on the brane is in many ways more straightforward than the analysis of linearised braneworld gravity, since we do not have to tackle the awkward issue of gauge invariance, nor do we have to perform any expansions. We follow the elegant approach of Shiromizu, Maeda and Sasaki [51], where the idea is to use a Gauss-Codacci approach to project the 5-D Einstein equations onto the brane to obtain 4-D effective Einstein equations. The basic results required for this are derived in the appendix.

We begin with the Gauss equation (A.15), relating the 4-D curvature $^{(4)}R^a{}_{bcd}$ constructed from the induced brane metric $q_{ab}$ to the 5-D curvature $^{(5)}R^a{}_{bcd}$ constructed from the full metric $g_{ab}$, and the extrinsic curvature $K_{ab}$ of the brane:

$$^{(4)}R^a{}_{bcd} = {}^{(5)}R^e{}_{fgh} q_e{}^a q_b{}^f q_c{}^g q_d{}^h + K^a{}_c K_{bd} - K^a{}_d K_{bc}. \tag{2.56}$$

Performing the appropriate contractions gives us the Ricci tensor and Ricci scalar on the brane,

$$^{(4)}R_{bd} = {}^{(5)}R_{fh} q_b{}^f q_d{}^h - {}^{(5)}R^e{}_{fgh} q_b{}^f q_d{}^h n_e n^g + K K_{bd} - K^a{}_d K_{ba}, \tag{2.57}$$

$$^{(4)}R = {}^{(5)}R - 2 {}^{(5)}R_{fh} n^f n^h + {}^{(5)}R^e{}_{fgh} n_e n^f n^g n^h + K^2 - K^{ab} K_{ab}, \tag{2.58}$$

from which we construct the Einstein tensor on the brane,

$$^{(4)}G_{ab} = \left(^{(5)}R_{cd} - \frac{1}{2}{}^{(5)}Rg_{cd}\right)q_a{}^c q_b{}^d + {}^{(5)}R_{cd}n^c n^d q_{ab} - {}^{(5)}R^c{}_{def}n_c n^d q_a{}^e q_b{}^f$$
$$+ KK_{ab} - K^c{}_b K_{ac} - \frac{1}{2}\left(K^2 - K^{cd}K_{cd}\right)q_{ab}. \qquad (2.59)$$

We now use the bulk Einstein equations, assuming the bulk to be empty except for a cosmological constant $\Lambda_5$,

$$^{(5)}R_{ab} - \frac{1}{2}{}^{(5)}Rg_{ab} = -\Lambda_5 g_{ab}, \qquad (2.60)$$

together with the decomposition of the Riemann tensor into the Weyl tensor, Ricci tensor and Ricci scalar;

$$^{(5)}R_{abcd} = {}^{(5)}C_{abcd} + \frac{2}{3}\left(g_{a[c}{}^{(5)}R_{d]b} - g_{b[c}{}^{(5)}R_{d]a}\right) - \frac{1}{6}g_{a[c}g_{d]b}{}^{(5)}R. \qquad (2.61)$$

Equation (2.59) then becomes

$$^{(4)}G_{ab} = -\frac{1}{2}\Lambda_5 q_{ab} + KK_{ab} - K^c{}_b K_{ac} - \frac{1}{2}\left(K^2 - K^{cd}K_{cd}\right)q_{ab} - \mathcal{E}_{ab}, \qquad (2.62)$$

where

$$\mathcal{E}_{ab} = C_{defg}n^d n^e q_a{}^f q_b{}^g \qquad (2.63)$$

is the projection of the bulk Weyl tensor onto the brane. Due to the Weyl tensor symmetries, $\mathcal{E}_{ab}$ is symmetric and trace-free:

$$\mathcal{E}_{[ab]} = \mathcal{E}_a{}^a = 0. \qquad (2.64)$$

For convenience, we now choose Gaussian Normal coordinates $(x^\mu, y)$ such that the hypersurface $y = 0$ coincides with the brane (see Appendix A.3). The field equa-tions on the brane will then be given by evaluating equation (2.62) on the brane.[6] The extrinsic curvature $K_{ab}$ is determined by the Israel junction conditions, which take account of the singular energy-momentum of the brane (see Appendix A.4). Imposing the $\mathbb{Z}_2$ symmetry and working on the '+' side of the brane, we have from (A.23):

$$K_{\mu\nu} = -4\pi G_5 \left( S_{\mu\nu} - \frac{1}{3} q_{\mu\nu} S \right), \qquad (2.65)$$

where $S_{\mu\nu}$ is the brane stress-energy tensor. We can further decompose this into the brane tension $\sigma$ and the energy-momentum $T_{\mu\nu}$ of any matter:

$$S_{\mu\nu} = -\sigma q_{\mu\nu} + T_{\mu\nu}, \qquad (2.66)$$

so that

$$K_{\mu\nu} = -4\pi G_5 \left( T_{\mu\nu} + \frac{1}{3} q_{\mu\nu} (\sigma - T) \right). \qquad (2.67)$$

Inserting this into (2.62) we obtain the 4-D effective Einstein equations on the brane:

$$^{(4)}G_{\mu\nu} = -\Lambda_4 q_{\mu\nu} + 8\pi G_4 T_{\mu\nu} + \frac{48\pi G_4}{\sigma} \Pi_{\mu\nu} - \mathcal{E}_{\mu\nu}, \qquad (2.68)$$

where

$$\Lambda_4 = \frac{1}{2} \left( \Lambda_5 + 8\pi G_4 \sigma \right), \qquad (2.69)$$

$$G_4 = \frac{4\pi G_5^2 \sigma}{3}, \qquad (2.70)$$

$$\Pi_{\mu\nu} = \frac{1}{12} T T_{\mu\nu} - \frac{1}{4} T_{\mu\alpha} T^\alpha{}_\nu + \frac{1}{24} \left( 3 T_{\alpha\beta} T^{\alpha\beta} - T^2 \right). \qquad (2.71)$$

The first two terms on the RHS of (2.68) are familiar from standard GR; the first is a 4-D cosmological constant term arising from the mismatch between the brane tension and the bulk cosmological constant. In the original RS2 model [35], $\Lambda_4$ was tuned to zero – see equation (1.22). The second term is the usual matter source term, with the 4-D Newton's constant determined by the brane tension, $G_4 \propto \sigma$. In particular, we must live on a positive tension brane in order for $G_4$ to have the correct sign, and so the RS1 model [34] is ruled out on these grounds.

---

[6] Strictly, in the limit $y \to \pm 0$, since $\mathcal{E}_{ab}$ is not defined exactly on the brane [51].

The final two terms in (2.68) are correction terms arising from bulk gravitational effects. The first of these, $\Pi_{\mu\nu}(T^2)$, consists of squares of $T_{\mu\nu}$ and hence is a local, high energy correction term. The final term, $\mathcal{E}_{\mu\nu}$, is the projection of the bulk Weyl tensor onto the brane and is *non-local* from the brane point of view, since it carries information of the gravitational field *outside* the brane. In order to determine $\mathcal{E}_{\mu\nu}$ one needs to solve the full 5-D equations, meaning that the system of equations (2.68) is not closed, in general. This is to be expected as it simply reflects our ignorance – as 4-dimensional beings confined to the brane – of the full five dimensional dynamics.

A (1+3)-covariant analysis of the equations (2.68) can be developed from the viewpoint of a brane-bound observer [63]. Consider the general decomposition of $\mathcal{E}_{\mu\nu}$ with respect to a 4-velocity field $u^\mu$:

$$\mathcal{E}_{\mu\nu} = -\left[ U\left(u_\mu u_\nu + \frac{1}{3}h_{\mu\nu}\right) + 2q_{(\mu} u_{\nu)} + P_{\mu\nu} \right], \qquad (2.72)$$

where $h_{\mu\nu} = g_{\mu\nu} + u_\mu u_\nu$ projects orthogonally to $u^\mu$. This allows us to consider $\mathcal{E}_{\mu\nu}$ as an effective "Weyl fluid", with non-local energy density $U$, momentum density $q_\mu$ and anisotropic stress $P_{\mu\nu}$, respectively given by

$$U = -\mathcal{E}_{\mu\nu} u^\mu u^\nu, \qquad (2.73)$$

$$q_\mu = h_\mu{}^\alpha \mathcal{E}_{\alpha\beta} u^\beta, \qquad (2.74)$$

$$P_{\mu\nu} = \left[\frac{1}{3} h_{\mu\nu} h^{\alpha\beta} - h_{(\mu}{}^\alpha h_{\nu)}{}^\beta \right] \mathcal{E}_{\alpha\beta}. \qquad (2.75)$$

Note that $\mathcal{E}_{\mu\nu}$ is in "Planck" units, i.e. there is no preceding $8\pi G_4$ since $\mathcal{E}_{\mu\nu}$ is derived from gravity in the bulk. To compare $U$, $q_\mu$ and $P$ with the *physical* energy-momentum of 4-D matter, we should rescale by $1/8\pi G_4$. It is found that there exist evolution equations for $U$ and $q_\mu$, but not for $P_{\mu\nu}$ [63]. This is because $P_{\mu\nu}$ incorporates gravitational wave modes of the 5-D graviton, which can not be predicted by brane observers. In special cases, including that of a Friedmann-Robertson-Walker (FRW) brane (to be discussed in the next section), the effective anisotropic stress $P_{\mu\nu}$ may vanish by symmetry and thus the equations (2.68) close in terms of data on the brane. In general, however, this will not be the case and the system of equations (2.68) will not be closed.

One final important issue to consider is the conservation of energy-momentum on the brane. Consider the Codacci equation (A.16):

$$D^a K_{ab} - D_b K = {}^{(n)}R_{cd} n^d q^c{}_b, \qquad (2.76)$$

where $D_a$ is the derivative operator on the brane associated with the induced metric $q_{ab}$. Using (2.60),

$$ {}^{(n)}R_{cd} n^d q^c{}_b = \left(\frac{1}{2}{}^{(5)}R - \Lambda_5\right) g_{cd} n^d q^c{}_b = 0 \qquad (\text{since } n_c q^c{}_b = 0). \qquad (2.77)$$

From (2.67) we have

$$D^\mu K_{\mu\nu} - D_\nu K = -4\pi G_5 \, D^\mu T_{\mu\nu}, \qquad (2.78)$$

therefore (2.76) implies the standard 4-D conservation equation

$$D^\mu T_{\mu\nu} = 0. \qquad (2.79)$$

The 4-D Bianchi identity $D^\mu G_{\mu\nu} = 0$ applied to (2.68) then implies

$$D^\mu \mathcal{E}_{\mu\nu} = \frac{48\pi G_4}{\sigma} D^\mu \Pi_{\mu\nu}, \qquad (2.80)$$

which shows qualitatively how bulk KK modes can be sourced by variations in the matter on the brane.

## 2.4 Braneworld cosmology

Armed with a description of non-perturbative gravity on the brane, we now investigate the braneworld generalisation of the FRW universe. Consider the isotropic, homogeneous metric

$$\mathrm{d}s^2 = -\mathrm{d}t^2 + a^2(t)\gamma_{ij}^{(k)}\mathrm{d}x^i\mathrm{d}x^j, \tag{2.81}$$

where $a(t)$ is the scale factor and $\gamma_{ij}^{(k)}$ is the maximally symmetric 3-metric with spatial curvature $k = +1, 0$ or $-1$. We make the usual assumption that the matter content of the universe can be modelled as a perfect fluid with pressure $p$ and energy density $\rho$:

$$T_{\mu\nu} = (\rho + p)u_\mu u_\nu + p\, q_{\mu\nu}, \tag{2.82}$$

where $u_\mu$ is the 4-velocity of the fluid. The braneworld generalisation of the Friedmann equation that follows from equation (2.68) is then found to be [64–66]

$$H^2 = \frac{\dot{a}^2}{a^2} = \frac{8\pi G_4}{3}\rho - \frac{k}{a^2} + \frac{\Lambda_4}{3} + \frac{4\pi G_4}{3\sigma}\rho^2 - \frac{\mathcal{C}}{a^4}, \tag{2.83}$$

where $\dot{a} = \partial a/\partial t$ and $\mathcal{C}$ is an integration constant. From equation (2.79) we obtain the standard conservation equation

$$\dot{\rho} + 3H(\rho + p) = 0. \tag{2.84}$$

The first three terms in (2.83) reproduce the standard Friedmann equation of 4-D GR. The fourth term is a high energy correction term arising from $\Pi_{\mu\nu}$ in (2.68). Although negligible at low energies ($\rho < \sigma$), this term can dominate at high energies, resulting in an unconventional expansion of the universe. Observation requires that such an unconventional phase of expansion takes place *before* nucleosynthesis,

leading to the bound

$$\sigma \gtrsim (1\,\text{MeV})^4, \qquad (2.85)$$

which is, however, much weaker than the limit (2.42) imposed by table-top tests of Newton's law.

The final term in (2.83) is a "dark radiation" term arising from the Weyl term $\mathcal{E}_{\mu\nu}$ in (2.68). Although in general we would have needed to solve the full 5-D equations to determine $\mathcal{E}_{\mu\nu}$, the cosmological symmetries have rendered the 4-D effective equations fully integrable in this case [63]. The Weyl parameter $\mathcal{C}$ is also constrained by the number of additional relativistic degrees of freedom allowed during nucleosynthesis [67–69].

The above derivation was performed in GNC, with the brane fixed at $y = 0$, and while it evolves cosmologically the metric evolves in time according to (2.83). This entails no physical restriction, at least locally, and is very useful for describing cosmology on the brane. However, the nature of the bulk spacetime in this *brane-based* approach is less transparent. An alternative is to consider a *bulk-based* approach in which the brane *moves* through a particular bulk spacetime. The general bulk spacetime that satisfies the 5-D vacuum Einstein equations and is compatible with a FRW brane is Schwarzschild-AdS (the fact that this is also *static* can be viewed as a higher-dimensional generalisation of Birkhoff's theorem [70]). Taking the brane to be an arbitrary boundary between two copies of this spacetime, the Gauss-Codacci equations and Israel junction conditions relate the energy-momentum of the brane to its trajectory. It is found that the effective Friedmann equation (2.83) is reproduced in this picture [70, 71], with $a(t)$ reinterpreted as the radial displacement of the brane and the cosmological evolution on the brane being due to its motion through the bulk. The two viewpoints are therefore completely equivalent, as can be shown explicitly by a coordinate tranformation between them. In the bulk-based approach, the constant $\mathcal{C}$ appearing in (2.83) is understood as the mass of the bulk black hole. If this term vanishes, the bulk is purely AdS.

We have seen that the standard cosmological evolution is recovered on the brane at low energies and late times. However, we do not live in a universe that is perfectly isotropic and homogeneous. Indeed, much of the current research in cosmology is focussed on accurately modelling the formation of structure in the Universe, and making comparisons with the ever-improving observational data sets. It is important that such studies are also performed in braneworld gravity, since they might provide key observational signatures of extra dimensions. Progress in this direction has been made (see [41, 42] and references therein), however a description of the theory of braneworld cosmological perturbations is beyond the scope of this thesis.

# Chapter 3

# Static Braneworld Black Holes

## 3.1 Introduction

The static, spherically symmetric vacuum solution to Einstein's equations was discovered by Schwarzschild in 1916 [72], just one year after the formulation of GR. According to GR, the Schwarzschild metric uniquely describes the spacetime surrounding every non-rotating spherical object in the universe, and forms the basis for the classic observational tests of Einstein's theory. The Schwarzschild metric for an object of mass $M$ is given by

$$\mathrm{d}s^2 = -\left(1 - \frac{2M}{r}\right)\mathrm{d}t^2 + \left(1 - \frac{2M}{r}\right)^{-1}\mathrm{d}r^2 + r^2\left(\mathrm{d}\theta^2 + \sin^2\theta\,\mathrm{d}\phi^2\right), \qquad (3.1)$$

which appears to behave badly at $r = 2M$. In fact, $r = 2M$ is a perfectly well-behaved (although interesting) surface, and the apparently singular behaviour of the metric there is simply a result of our choice of coordinates (sometimes called a *coordinate singularity*).

For most objects, such as stars and planets, the surface $r = 2M$ is of little consequence since it lies deep within the object, where the vacuum solution (3.1) does not apply. However, through gravitational collapse it is possible for an object to form that lies entirely *within* the surface $r = 2M$. Such an object is called a *black* *hole*, and $r\!=\!2M$ then defines the black hole's *event horizon* – a surface from which light can not escape.[1]

There is convincing indirect evidence for the existence of black holes [74], and the case is particularly strong for a supermassive black hole at the centre of our galaxy [75, 76], with the exciting possibility that we will directly see the *shadow* of this black hole in the near future [77]. Hence these once mysterious entities now have an important place in our catalogue of astrophysical objects, allowing us to probe strong gravitational fields and test GR in its full, non-linear form. Black holes also provide an interesting arena for discussing quantum gravity ideas, following Hawking's discovery that black holes *radiate* quantum mechanically [78].

It is clearly important in the braneworld context to find braneworld black hole (BBH) solutions, both to shed light on braneworld gravity itself and, more importantly, because observations of black holes might provide an exciting means of detecting extra dimensions, if they exist.

Ideally, we would like a full 5-D solution describing a black hole localised on the brane. Such a solution would be given by a suitable slicing of a 5-D accelerating black hole metric, known as the *C-metric* in 4-D [79]. The reason for this is that the Poincaré coordinate system used to chart the brane (and in which the brane induced metric is flat) is, from the bulk point of view, an accelerating patch covering a part of AdS space – loosely speaking, the situation is similar to that of Rindler coordinates in Minkowski spacetime. Therefore a BBH must also be accelerating in order to "keep up" with the brane and remain localised on it. This has been explicitly demonstrated in one dimension lower, where the 4-D C-metric can indeed be sliced to construct a (2+1)-brane containing a localised black hole [80]. However, this is a very special case. Unfortunately, the 5-D C-metric has not yet been found, and is not even guaranteed to exist!

---

[1]As an interesting historical note, the concept of an object so dense that not even light can escape its gravitational pull goes back to John Michell in 1784 [73].

In the absence of a 5-D C-metric, attempts to address the full 5-D problem have had only limited success [81–87]. Here, instead, we study the problem from a 4-D perspective, using the effective Einstein equations (2.68) to investigate possible 4-D geometries describing black holes on the brane. The motivation will be to see if we can categorise classes of BBH behaviour, and see if there is any universality to the 4-D braneworld solutions. In principle, the 5-D structure of these solutions can be obtained by numerical integration into the bulk, however this is plagued with difficulties [84–87].

In the next section we discuss the general system of equations for the problem, which is not closed on the brane due to the Weyl term $\mathcal{E}_{\mu\nu}$, as discussed in Section 2.3. Therefore an assumption about $\mathcal{E}_{\mu\nu}$ or $g_{\mu\nu}$ must be made in order to close the system of equations and obtain a BBH solution. Several different solutions, making various assumptions, have been presented in the literature and are discussed in Section 3.3. In Section 3.4 we take a practical approach to finding BBH solutions and assume an equation of state for $\mathcal{E}_{\mu\nu}$, which allows us to classify the general behaviour of BBH solutions. Using holography considerations, we propose a particular solution as a "working metric" for the near-horizon form of the BBH metric, and use simple bounds to constrain the solution close to the horizon.

## 3.2 Basic equations

The general static, spherically symmetric metric on the brane can be written as

$$ds^2 = -A^2(r)dt^2 + B^2(r)dr^2 + C^2(r)d\Omega_{I\!I}^2, \tag{3.2}$$

where $d\Omega_{I\!I}^2 = d\theta^2 + \sin^2\theta\, d\phi^2$ is the metric of the unit 2-sphere. Unlike in standard GR, spherical symmetry itself does not imply staticity, i.e. Birkhoff's theorem does not apply. It should be noted that there is no consensus on whether gravitational collapse on the brane results in the formation of a black hole that is static or time-dependent [88–92]. By assuming the metric to be static, we are taking the point of view that there exists a 5-D solution analogous to the 4-D C-metric, which possesses a timelike Killing vector and therefore can be sliced in such a way as to create a static 4-D BBH.

Another feature to note about the metric (3.2) is that we have not specified it as fully as we might, since we can still choose our radial coordinate, $r$, quite arbitrarily. It might be tempting to make the standard choice $C = r$, which we call the *area gauge* since the area of the 2-spheres surrounding the black hole then behaves in the usual way: $\mathcal{A}(r) = 4\pi r^2$. However, this gauge is only valid for the entire black hole exterior if $\mathcal{A} = 4\pi C^2$ increases monotonically between the horizon and spacelike infinity (so that the mapping $C \mapsto r$ is one-to-one). The second derivative of the area function $C$, i.e. the radial function defined by $\sqrt{\mathcal{A}/4\pi}$, is given by

$$\frac{C''}{C} = \frac{B^2}{2}(G_t^{\ t} - G_r^{\ r}) + \frac{C'}{C}\left(\frac{B'}{B} + \frac{A'}{A}\right). \tag{3.3}$$

Hence for the area function $C(r)$ to be guaranteed to be monotonic and increasing as $r \to \infty$ we must have $G_t^{\ t} - G_r^{\ r} \leq 0$, which is equivalent to the dominant energy condition. While this is generally satisfied in standard Einstein gravity, it need not be in the case of extra dimensions. For example, the Weyl term $\mathcal{E}_{\mu\nu}$ appearing in the braneworld effective Einstein equations arises purely from the gravitational field of the bulk, and so does not necessarily satisfy any kind of energy condition.

Our reason for suspecting that $\mathcal{A}(r)$ might not be monotonic for a BBH lies in the putative higher dimensional C-metric, which would consist of an accelerating black hole being "pulled" by a string. The appropriate higher dimensional metric for a Poincaré invariant string has a turning point in the area function, and the "horizon" is in fact singular [93]. This might therefore also render the BBH horizon singular, and induce a turning point in the area function of the 4-D BBH metric. Such a *wormhole* region in the black hole geometry is a key potential difference from the black holes of standard GR, so it is vital that we do not make the restrictive ansatz $C = r$.

The vacuum brane field equations following from (2.68), with $\Lambda_4$ set equal to zero, are

$$G_{\mu\nu} = -\mathcal{E}_{\mu\nu}, \tag{3.4}$$

where

$$\mathcal{E}_\mu{}^\mu = \nabla_\mu \mathcal{E}^{\mu\nu} = \mathcal{E}_{[\mu\nu]} = 0. \tag{3.5}$$

Although $\mathcal{E}_{\mu\nu}$ is an unknown from the brane point of view, it can be generally decomposed with respect to a 4-velocity field $u^\mu$ as in (2.72),

$$\mathcal{E}_{\mu\nu} = -\left[U\left(u_\mu u_\nu + \frac{1}{3}h_{\mu\nu}\right) + 2q_{(\mu}u_{\nu)} + P_{\mu\nu}\right], \tag{3.6}$$

where $h_{\mu\nu} = g_{\mu\nu} + u_\mu u_\nu$ projects orthogonally to $u^\mu$ and $U$, $q_\mu$ and $P_{\mu\nu}$ are the effective non-local energy density, momentum density and anisotropic stress, respectively. Static, spherical symmetry implies $q_\mu = 0$ and $P_{\mu\nu} = P\left(r_\mu r_\nu - \frac{1}{3}h_{\mu\nu}\right)$, where $r^\mu$ is a unit radial vector, so this further reduces to

$$-\mathcal{E}_{\mu\nu} = U(r)\left(u_\mu u_\nu + \frac{1}{3}h_{\mu\nu}\right) + P(r)\left(r_\mu r_\nu - \frac{1}{3}h_{\mu\nu}\right). \tag{3.7}$$

In an inertial frame with $u^\mu = \frac{1}{A}(1,0,0,0)$, the field equations (3.4) then read:

$$G_t{}^t = -\frac{1}{C^2} + \frac{1}{B^2}\left[2\frac{C''}{C} - 2\frac{B'C'}{BC} + \frac{C'^2}{C^2}\right] = -U, \tag{3.8}$$

$$G_r{}^r = -\frac{1}{C^2} + \frac{1}{B^2}\left[2\frac{A'C'}{AC} + \frac{C'^2}{C^2}\right] = \frac{1}{3}(U + 2P), \tag{3.9}$$

$$G_\theta{}^\theta = G_\phi{}^\phi = \frac{1}{B^2}\left[\frac{A''}{A} + \frac{C''}{C} + \frac{A'C'}{AC} - \frac{A'B'}{AB} - \frac{B'C'}{BC}\right] = \frac{1}{3}(U - P), \tag{3.10}$$

where $x' = \partial x/\partial r$. A useful alternative equation is the Bianchi identity, or conservation of Weyl "energy-momentum":

$$(U + 2P)' + 2\frac{A'}{A}(2U + P) + 6P\frac{C'}{C} = 0. \tag{3.11}$$

The system (3.8)-(3.11) consists of three independent equations in four unknowns ($U$, $P$ and two of $A$, $B$ and $C$), therefore an assumption about $g_{\mu\nu}$ or $\mathcal{E}_{\mu\nu}$ must be made in order to close the system of equations and obtain a BBH solution. Several specific solutions to these equations, as well as more general techniques, have been presented in the literature and are discussed in the next section. In Section 3.4 we assume an equation of state for $\mathcal{E}_{\mu\nu}$, linearly relating $U$ and $P$, which allows us to make a more systematic analysis of BBH solutions.

## 3.3 Black hole solutions: examples

### 3.3.1 The linearised weak field metric

The linearised, weak field form of the metric for a static source on the brane was derived in Section 2.2.2, and in the area gauge is given by

$$ds^2 = -\left(1 - \frac{2M}{r} - \frac{4Ml^2}{3r^3}\right)dt^2 + \left(1 - \frac{2M}{r} - \frac{2Ml^2}{r^3}\right)^{-1}dr^2 + r^2 d\Omega_{I\!I}^2. \tag{3.12}$$

Due to the trace-free property of $\mathcal{E}_{\mu\nu}$, a valid BBH solution must have a vanishing Ricci scalar. The Ricci scalar for the above metric is readily calculated to be

$$R = 2l^2 M^2 \frac{[-42l^2 r^3 - 9r^5 + 4M(8l^4 + 18l^2 r^2 + 3r^4)]}{r^5 (4l^2 M + 6Mr^2 - 3r^3)^2}, \tag{3.13}$$

so $R = 0$ up to $O(M)$, as expected since the derivation of (3.12) is performed in the linearised approximation, which neglects terms of order $M^2$ and higher. However, equation (3.13) shows that the Ricci scalar does *not* vanish beyond linear order, therefore the metric (3.12) can only be taken as a solution to the field equations in the linearised far-field limit, and is clearly not a valid solution for the entire horizon exterior of a BBH. The reason for its inclusion in this section is that the solution is obtained by solving the full 5-D linearised equations, assuming only that the bulk is asymptotically AdS and that the perturbations are bounded at the AdS horizon. Therefore the correct metric for a static, spherically symmetric black hole on the brane should be expected to have this asymptotic form.

The Weyl tensor $\mathcal{E}_{\mu\nu}$ for this metric is given by:

$$\mathcal{E}_t{}^t = -\frac{4Ml^2}{r^5}, \qquad \mathcal{E}_r{}^r = -\frac{2Ml^2}{r^5}, \qquad \mathcal{E}_\theta{}^\theta = \mathcal{E}_\phi{}^\phi = \frac{3Ml^2}{r^5}, \tag{3.14}$$

or, in terms of $U$ and $P$ in the decomposition (3.7),

$$U = -\frac{4}{5}P = -\frac{4Ml^2}{r^5}. \tag{3.15}$$

Here, terms are calculated to linear order in $M$, consistent with the linearised approximation used to derive (3.12). Note that it is the non-zero $\mathcal{E}_{\mu\nu}$ that is responsible for the corrections to the Schwarzschild metric in (3.12), so the correct BBH solution should also have a non-zero $\mathcal{E}_{\mu\nu}$ if it is to agree with the metric (3.12) in the weak field limit.

### 3.3.2 The black string

The canonical form of the Randall-Sundrum braneworld metric (1.21),

$$\mathrm{d}s^2 = e^{-2|y|/l}\eta_{\mu\nu}\mathrm{d}x^\mu\mathrm{d}x^\nu + \mathrm{d}y^2, \tag{3.16}$$

in fact remains a solution to the 5-D field equations (1.14) if we replace $\eta_{\mu\nu}$ with any 4-D Ricci flat metric $g_{\mu\nu}$. Therefore the most obvious attempt at constructing a BBH is to take $g_{\mu\nu}$ to be the Schwarzschild metric, resulting in the black string [94]:

$$\mathrm{d}s^2 = e^{-2|y|/l}\left[-\left(1-\frac{2M}{r}\right)\mathrm{d}t^2 + \left(1-\frac{2M}{r}\right)^{-1}\mathrm{d}r^2 + r^2\mathrm{d}\Omega_{I\!I}^2\right] + \mathrm{d}y^2. \tag{3.17}$$

This solution has $\mathcal{E}_{\mu\nu} = 0$ and so the brane geometry receives no correction from bulk gravitational effects and the weak field limit (3.12) is not satisfied. Each 4-D slice $y = $ constant has the induced geometry of the Schwarzschild metric and there is a line singularity at $r = 0$, extending along all $y$. This is illustrated by the square of the 5-D Riemann tensor,

$$R_{abcd}R^{abcd} = \frac{40}{l^2} + \frac{48M^2}{r^6}e^{4|y|/l}, \tag{3.18}$$

which also demonstrates the unsatisfactory behaviour that the curvature diverges as $y \to \infty$. Furthermore, the black string suffers from the Gregory-Laflamme instability [95, 96]. Clearly, the black string is not a viable candidate for the BBH metric! Even so, it is the only known 5-D non-perturbative solution, which has led to it being used to explore the gravitational wave properties of BBHs, which requires a full 5-D solution since gravity can propagate in the bulk. Introducing a second brane to cut off the black string instability (and ameliorate the problem of the curvature diverging as $y \to \infty$), it has been shown that the back-reaction of bulk gravity waves onto the brane results in observational signatures that are distinctive from those of standard GR [97, 98], even though the induced brane geometry is iden-tically Schwarzschild. Hence gravity wave experiments might provide an exciting opportunity to detect extra dimensions, if they exist.

### 3.3.3  The tidal Reissner-Nordström black hole

Being antisymmetric and trace-free (see equation (3.5)), the Weyl term $\mathcal{E}_{\mu\nu}$ has the same algebraic symmetries as the electromagnetic stress-energy tensor $T^{\text{em}}_{\mu\nu}$ of GR, allowing us to make the formal correspondence $-\mathcal{E}_{\mu\nu} \leftrightarrow T^{\text{em}}_{\mu\nu}$. Therefore Einstein-Maxwell solutions in GR result in *vacuum* braneworld solutions, and it is this observation that led Dadhich *et al.* [99] to the tidal Reissner-Nordström (tidal-RN) BBH solution:

$$\mathrm{d}s^2 = -\left(1 - \frac{2M}{r} + \frac{Q}{r^2}\right)\mathrm{d}t^2 + \left(1 - \frac{2M}{r} + \frac{Q}{r^2}\right)^{-1}\mathrm{d}r^2 + r^2\mathrm{d}\Omega^2_{I\!I}. \qquad (3.19)$$

This metric is also the general solution to the brane field equations under the assumption $A(r) = 1/B(r)$.

The solution (3.19) has the form of the Reissner-Nordström (RN) solution of GR, but there is *no* electric field present on the brane; $Q$ is a *tidal* charge parameter arising from the gravitational field of the bulk. Unlike in the GR case, $Q$ can be both positive *and* negative. For $Q > 0$ the tidal-RN metric has qualitatively the same properties as the standard RN geometry: there are two horizons, both of which lie inside the Schwarzschild horizon, and the singularity at $r = 0$ is timelike. However, we can now have $Q < 0$ in which case there is just one horizon, lying outside Schwarzschild, and the central singularity is spacelike. Indeed, negative $Q$ is the more natural case since it gives a *positive* contribution to the gravitational potential and therefore *strengthens* the gravitational field, as we would intuitively expect since the tidal charge arises from the source mass $M$ on the brane (see [99] for further discussion).

The Weyl tensor for this solution is given by

$$\mathcal{E}_t{}^t = \mathcal{E}_r{}^r = -\mathcal{E}_\theta{}^\theta = -\mathcal{E}_\phi{}^\phi = \frac{Q}{r^4}, \tag{3.20}$$

or, in terms of the decomposition (3.7),

$$U = -\frac{1}{2}P = \frac{Q}{r^4}. \tag{3.21}$$

The tidal-RN metric (3.19) does not satisfy the weak field limit (3.12), and so can not describe the entire spacetime surrounding a BBH. Nevertheless, let us briefly discuss this solution in the context of *small* black holes on the brane. For a BBH with an horizon size much less than the AdS length scale, $r_h \ll l$, the AdS curvature has very little effect on the geometry and so the spacetime will be that of the induced 5-D Schwarzschild-Tangherlini (ST) metric [100]:[2]

$$ds^2 = -\left(1 - \frac{r_h^2}{r^2}\right) dt^2 + \left(1 - \frac{r_h^2}{r^2}\right)^{-1} dr^2 + r^2 d\Omega_{I\!I\!I}^2. \tag{3.22}$$

Therefore the tidal-RN metric (3.19) shows the correct 5-D behaviour of gravity at short distances, where the $Q/r^2$ term dominates and the metric approximates the induced 5-D ST metric (in fact, the 5-D ST metric can be viewed as a special case of the tidal-RN metric, with $M = 0$ and $Q = -r_h^2$). Hence the tidal-RN metric might be a good approximation in the strong field regime for small black holes. This solution will be discussed further in Chapter 4, where we investigate its gravitational lensing properties.

---

[2] Primordial BBHs formed in the early universe with an horizon size $r_h \ll l$ would therefore be described by this metric. It has been shown that such black holes can live much longer than their GR counterparts, possibly even surviving to the present day, with some interesting phenomenological consequences [101, 102].

### 3.3.4 Solutions assuming a particular metric form

The above approaches to finding BBH solutions were guided by the form of the Schwarzschild metric (3.1) in standard GR; the black string takes the brane metric to be *identically* Schwarzschild, while the tidal-RN metric has $A(r) = 1/B(r)$, as in the Schwarzschild case. In the same spirit, the obvious next attempt at finding BBH solutions is to fix one of $A(r)$ or $B(r)$ to take the Schwarzschild form, and solve the field equations for the other. This leads to the solutions of Casadio *et al.* [103]:[3]

$$\mathrm{d}s^2 = -\left[(1+\epsilon)\sqrt{1-\frac{2M}{r}}-\epsilon\right]^2 \mathrm{d}t^2 + \left(1-\frac{2M}{r}\right)^{-1}\mathrm{d}r^2 + r^2\mathrm{d}\Omega_{I\!I}^2, \quad (3.23)$$

$$\mathrm{d}s^2 = -\left(1-\frac{2M}{r}\right)\mathrm{d}t^2 + \frac{\left(1-\frac{3M}{2r}\right)}{\left(1-\frac{2M}{r}\right)\left(1-\frac{r_0}{r}\right)}\mathrm{d}r^2 + r^2\mathrm{d}\Omega_{I\!I}^2. \quad (3.24)$$

The second of these was first derived in [104] as a possible metric outside a star on the brane. The reader is referred to [103] for a detailed discussion of the properties of these solutions. However, once again we note that they do not satisfy the far-field limit (3.12) and so can not be considered satisfactory BBH solutions.

Taking the above approach to finding BBH metrics further, Visser and Wiltshire [105] (see also [106]) presented a more general, algorithmic method that generates a BBH solution for a given radial metric function $B(r)$.

## 3.4 Equation of state method

There is good reason to believe that a BBH might have a turning point in the area function $C(r)$, as discussed in Section 3.2. In all the above examples, however, the area gauge $C=r$ was used, thus precluding the possibility of such a wormhole region in the geometry (although [105] did comment on how to use their method for a non-monotonic $C(r)$). We also mentioned in Section 3.2 that a BBH might possess a

---

[3]which we present in a slightly different form to [103].

*singular* horizon, due to the possible connection between the cosmic *p*-brane [93] and the higher-dimensional C-metric. This is further supported by an argument based on the AdS/CFT correspondence [107], in which a static geometry must give rise to a BBH with a singular horizon [89].

It is clearly of interest to explore these two radical differences from the Schwarzschild black hole of standard GR. To this end, we perform a systematic analysis of BBH solutions, making the assumption of an equation of state for the Weyl term:[4]

$$P = \frac{\gamma - 1}{2} U. \tag{3.25}$$

In so doing, we are able to classify the behaviour of BBH solutions and give definitive answers to the following questions:

- When is the horizon singular?

- When does the area function have a turning point?

- When are the black hole solutions asymptotically flat?

Of course, a priori there is no reason to suppose that the Weyl term should obey an equation of state. However, it is quite possible that it might have certain asymptotic equations of state which may be useful as near-horizon or long-range approximations to the (as yet unknown) exact solution.

A similar approach is used in cosmology, where an equation of state $p = w\rho$ is assumed to model the energy-momentum content of the universe, with $w = 0$ for dust, $\frac{1}{3}$ for radiation, and $-1$ for the cosmological constant. Clearly the actual evolution and matter content of the universe is more detailed and complicated, but this method is useful, accurate, and universally accepted.

---

[4]such an equation of state for $\mathcal{E}_{\mu\nu}$ also follows from the assumption of particular conformal symmetries of the metric [108].

## 3.5  The dynamical system

The most convenient gauge for our analysis is $B = C$, which we call the *dynamical system* (DS) gauge. With the above equation of state relating $U$ and $P$, and defining

$$X = \frac{C'}{C} + 2\frac{A'}{A}, \qquad Y = \frac{C'}{C}, \qquad (3.26)$$

the field equations (3.8)-(3.10) in the DS gauge reduce to the two-dimensional dynamical system

$$2X' + X^2 - 1 = 3UC^2 = \frac{9}{\gamma}(XY - 1), \qquad (3.27)$$

$$1 - 2Y' - Y^2 = UC^2 = \frac{3}{\gamma}(XY - 1), \qquad (3.28)$$

together with the constraint

$$1 - XY = -\frac{C^2}{3}(U + 2P) = -\gamma\frac{UC^2}{3}. \qquad (3.29)$$

The Bianchi identity (3.11) reads:

$$\gamma U' + (3 + \gamma)\frac{A'}{A}U + 3(\gamma - 1)\frac{C'}{C}U = 0. \qquad (3.30)$$

A plot of the $(X, Y)$ phase plane gives us curves which are solutions, $X(\rho), Y(\rho)$, to this dynamical system, where $\rho$ is a radial coordinate, used to distinguish from the coordinate $r$ of the area gauge. Whether or not these trajectories correspond to an actual black hole depends on whether the integrated solutions $A(\rho), C(\rho)$ have the behaviour we expect for a black hole, such as an horizon, asymptotic flatness and so on. It is therefore useful before proceeding with the general analysis of the dynamical system to extract some asymptotic information from the Einstein equations in order to identify general features of the solution on the phase plane.

### 3.5.1 Asymptotic analysis

There are two clear asymptotic regions in which we would like to have some information about the black hole solution: the horizon and infinity. Near infinity we would like spacetime to be asymptotically flat, which means $A \sim 1 - O(r^{-1})$, $B \sim 1 - O(r^{-1})$, $C = r$ in the area gauge, or $A \sim 1 - O(e^{-\rho})$, $B = C \sim e^\rho + O(1)$ in the DS gauge. In the far-field limit we can adopt the area gauge, since any turning point in the area function would be expected only to occur closer to the horizon. In this limit, $A'/A \sim O(r^{-2})$ can be neglected compared with $C'/C \sim O(r^{-1})$ in equation (3.30), so the leading behaviour of the Weyl energy is given by

$$U = \frac{U_0 r^{\frac{3}{\gamma}}}{r^3}, \tag{3.31}$$

and the leading behaviour of the radial metric function is then read off from equation (3.8) as:

$$B^{-2} = 1 - \frac{\mu}{r} - \frac{\gamma u_0 r^{\frac{3}{\gamma}}}{3r}. \tag{3.32}$$

This clearly shows that for $0 < \gamma < 3$ the braneworld metric does not tend to Minkowski spacetime for large $r$. Indeed, even for $\gamma > 3$ the metric is not asymptotically flat, since asymptotic flatness requires corrections to Minkowski spacetime to be $O(1/r)$ (i.e. to fall off sufficiently fast with $r$). Therefore, *an asymptotically flat BBH requires an asymptotic equation of state $\gamma < 0$*.

In the opposite, near-horizon limit we can not assume the area gauge, and integrating the conservation equation (3.30) gives

$$U \propto A^{-\frac{3}{\gamma}-1} C^{\frac{3}{\gamma}-3}. \tag{3.33}$$

Whether or not the horizon corresponds to a singularity in the Weyl energy $U$ depends on the magnitude and sign of $\gamma$, as well as on the behaviour of $C$.

## 3.5.2 A simple analytic solution

To demonstrate the utility of the dynamical systems approach and the DS gauge, we now derive an alternate form of a known analytic solution with $U = 0$. A glance at the equation of state (3.25) shows that this is formally the limit $\gamma \to \pm\infty$. Since letting $\gamma$ become infinite decouples the equation of state from the dynamical system, the phase plane is in fact the same whether $\gamma$ is $+\infty$ or $-\infty$. Our main reason for deriving this solution afresh is that we wish to use it in Section 3.7.3 for a near horizon limit, but clearly, any exact solution is helpful, and our derivation makes the overall structure of the spacetime somewhat clearer than using the area gauge.

For $U = 0$, (3.27) and (3.28) have the solutions

$$X, Y = \tanh \frac{(\rho - \rho_0)}{2} \quad \text{or} \quad \coth \frac{(\rho - \rho_1)}{2}. \tag{3.34}$$

Clearly the former solution is appropriate for $Y$, as this corresponds to

$$C = C_0 \cosh^2 \frac{(\rho - \rho_0)}{2}, \tag{3.35}$$

which, since we expect an asymptotically flat spacetime which has $C \sim e^\rho$, gives $C_0 = 4e^{\rho_0}$. We then have for $A$:

$$2 \ln A + \ln C = 2 \ln \cosh \frac{(\rho - \rho_1)}{2} \quad \text{or} \quad 2 \ln \sinh \frac{(\rho - \rho_1)}{2}, \tag{3.36}$$

i.e.

$$A^2 A_0^{-2} C_0 \cosh^2 \frac{(\rho - \rho_0)}{2} = \cosh^2 \frac{(\rho - \rho_1)}{2} \quad \text{or} \quad \sinh^2 \frac{(\rho - \rho_1)}{2}. \tag{3.37}$$

For an horizon, we expect that $A^2$ will have a zero, hence we choose the second solution, giving

$$A^2 = \frac{A_0^2}{C_0} \left( \frac{\cosh(\rho - \rho_1) - 1}{\cosh(\rho - \rho_0) + 1} \right), \tag{3.38}$$

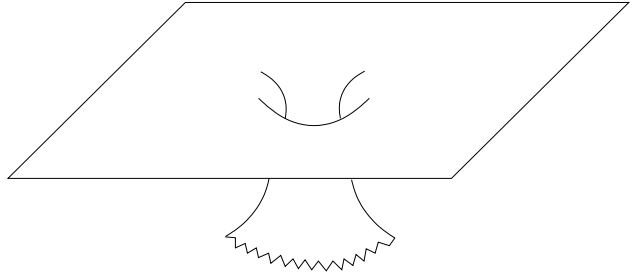

Figure 3.1: Pictorial representation of a constant time slice of the metric (3.40).

with asymptotic flatness requiring $A_0^2 = 4e^{\rho_1}$. Overall the metric is:

$$ds^2 = -e^{\rho_1-\rho_0}\left(\frac{\cosh(\rho-\rho_1)-1}{\cosh(\rho-\rho_0)+1}\right)dt^2 + 16e^{2\rho_0}\cosh^4\frac{(\rho-\rho_0)}{2}\left[d\rho^2+d\Omega_{I\!I}^2\right] \quad (3.39)$$

or, writing $r = e^\rho$:

$$ds^2 = -\frac{(r-r_1)^2}{(r+r_0)^2}dt^2 + \frac{(r+r_0)^4}{r^4}dr^2 + \frac{(r+r_0)^4}{r^2}d\Omega_{I\!I}^2, \quad (3.40)$$

which has the horizon at $r = r_1$ and a turning point in the area function at $r = r_0$. This metric has appeared in the area gauge in Section 3.3.4 (see equation (3.23)) as:

$$ds^2 = -\left[(1+\epsilon)\sqrt{1-\frac{2M}{R}}-\epsilon\right]^2 dt^2 + \left(1-\frac{2M}{R}\right)^{-1}dR^2 + R^2 d\Omega_{I\!I}^2, \quad (3.41)$$

$$\text{where} \quad R = (r+r_0)^2/r, \quad M = 2r_0, \quad M\epsilon = r_1 - r_0. \quad (3.42)$$

While the area gauge gives a familiar spatial part of the metric, note that at the old Schwarzschild "horizon", $R = 2M$, the $tt$-component of the metric does not vanish. For $\epsilon > 0$, $g_{tt}$ will be zero *before* $R = 2M$ and so the area gauge holds outside the black hole (strictly of course, this is not a black hole, as the "horizon" is singular). If $\epsilon < 0$, however, we see from (3.42) that the turning point $r = r_0$ lies *outside* of the horizon $r = r_1$, a situation that is depicted in Figure 3.1. In the area gauge in this case, $g_{tt} = 0$ at $R < 2M$ and $R = 2M$ becomes a *coordinate singularity* accessible by timelike observers, the result of choosing an inappropriate gauge.

## 3.6 General analysis of the dynamical system

In order to investigate the general behaviour of BBH solutions, we now turn to the general analysis of the dynamical system (3.27)-(3.28):

$$X' = \frac{1-X^2}{2} + \frac{9}{2\gamma}(XY-1), \tag{3.43}$$

$$Y' = \frac{1-Y^2}{2} + \frac{3}{2\gamma}(XY-1), \tag{3.44}$$

where $X$ and $Y$ were defined in (3.26). The plot of the $(X, Y)$ phase plane, obtained by solving (3.43)-(3.44) numerically, gives us curves which are solutions, $X(\rho), Y(\rho)$, to this dynamical system. In principle, these in turn can be integrated to find $A(\rho), C(\rho)$.

A selection of representative phase plane plots, for different values of the equation of state parameter $\gamma$, are given in Figure 3.2. These plots show that there are attractors in the phase plane, and a general trajectory flows in from infinity on the phase plane to one of these attractors. Whether these trajectories correspond to an actual black hole depends on whether the integrated solutions $A(\rho), C(\rho)$ have the behaviour we expect for a black hole, in particular whether they are asymptotically flat and possess an horizon.

An horizon corresponds to $A \to 0$, therefore we expect $X$ becomes infinite for the black hole horizon. Flat space on the other hand corresponds to $A' = 0$, $C \propto e^\rho$, hence $X = Y = 1$. A black hole solution must therefore be a trajectory which comes in from large $X$ (possibly large $Y$ as well) and terminates on $(1, 1)$. To find out which equations of state allow this, and whether the area function has any turning points, which correspond to zeros of $Y$, requires a detailed analysis of the phase plane.

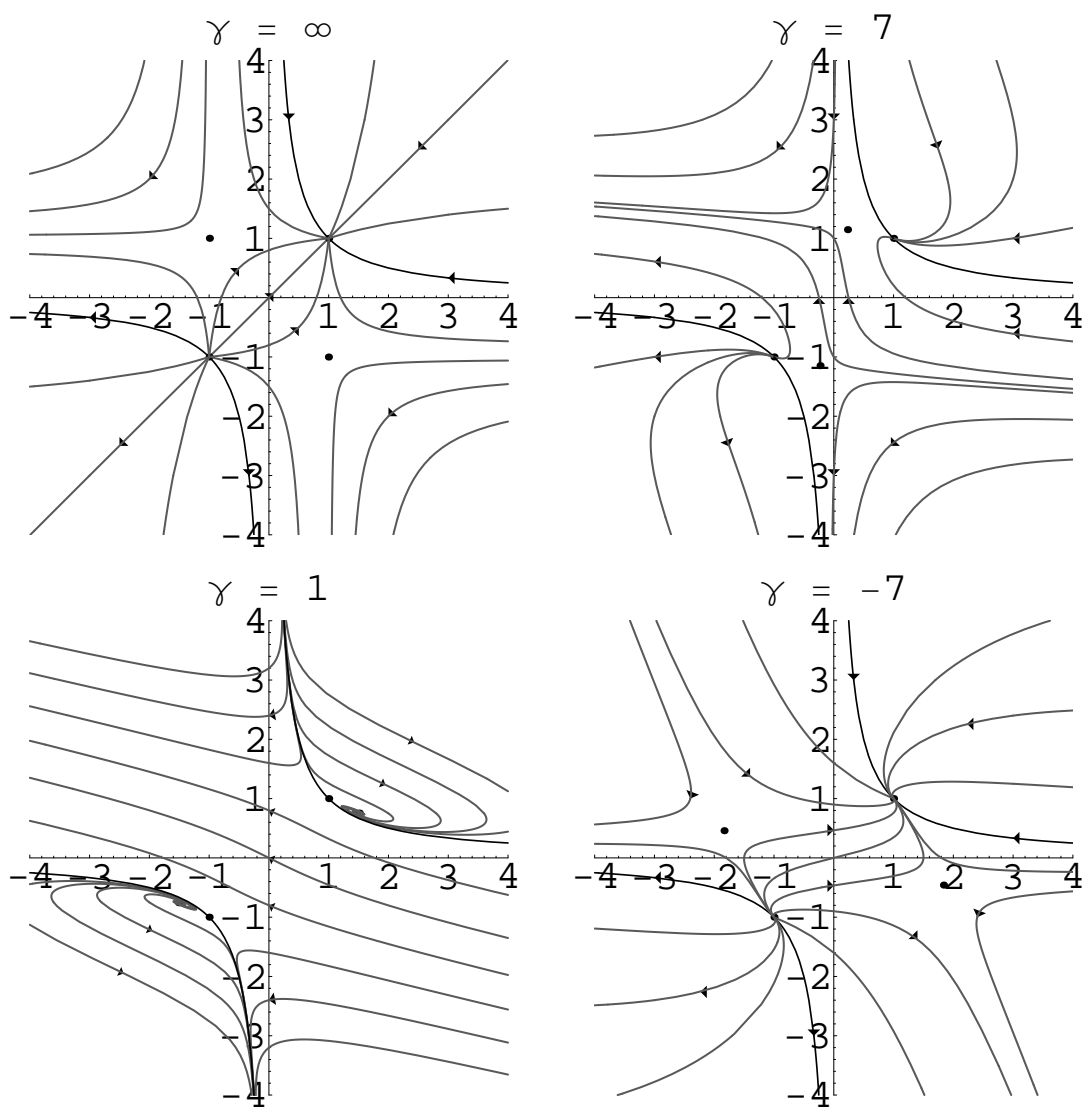

Figure 3.2: Phase planes for a range of equations of state.

### 3.6.1 Features of the phase plane

A crucial part of any dynamical systems analysis is the identification and classification of critical points, invariant submanifolds, and other distinguishing features of the phase plane. In this case, an invariant hyperboloid is easily identified from the constraint (3.29) as $XY = 1$. Along this hyperboloid $(U + 2P) = 0$, therefore – apart from the special case $\gamma = 0$ – this corresponds to a vanishing Weyl term $U = P = 0$, and hence the pure Einstein case (i.e. the Schwarzschild solution, given by (3.35), (3.38) with $\rho_0 = \rho_1$ and $A_0^2 = C_0$). If $\gamma = 0$, in addition to the Schwarzschild solution we have $X = \pm\sqrt{3} = 1/Y$, which is covered by the analysis of the critical points below.

**Critical points**

The system has 4 critical points, defined as points with $X' = Y' = 0$, given by:

$$P_\pm = \pm(1,1), \tag{3.45}$$

$$Q_\pm = \pm(27+\gamma^2)^{-1/2}(9-\gamma, 3+\gamma). \tag{3.46}$$

The $Q_\pm$ critical points move from $Q_\pm = (\mp 1, \pm 1)$ to $Q_\pm = (\pm 1, \mp 1)$ as $\gamma$ runs from $+\infty$ to $-\infty$, tracing out an ellipse in the $(X, Y)$ plane given by $X^2 + 3Y^2 = 4$. For $\gamma = 3$, $P$ and $Q$ are coincident. The nature of the critical points is as follows:

- $3 < \gamma$      $P_+$ is an attractor and $P_-$ a repellor, the $Q$'s are saddle points.

- $0 < \gamma < 3$   $Q_+$ is an attractor and $Q_-$ a repellor, the $P$'s are saddle points.

- $\gamma < 0$      $P_+$ is an attractor and $P_-$ a repellor, the $Q$'s are saddle points.

The critical point $P_+$ corresponds to flat space:

$$X = Y = 1 \;\;\Rightarrow\;\; A' = 0,\; C' = C$$

$$\Rightarrow\;\; C = e^\rho, \tag{3.47}$$

therefore any asymptotically flat solution must terminate on this critical point. While $P_+$ is an attractor this is not a problem, but for the range of $\gamma$ where it is a saddle point, only the invariant hyperboloid can satisfy this, and by definition, this is where we have the exact Schwarzschild solution.

In contrast, the critical point $Q_+$ corresponds to a *non-asymptotically flat* spacetime, which for $\gamma > -3$ in the area gauge is:

$$ds^2 = -r^{\frac{2(3-\gamma)}{(3+\gamma)}} dt^2 + \frac{(27+\gamma^2)}{(3+\gamma)^2} dr^2 + r^2 d\Omega_{II}^2, \tag{3.48}$$

which we do not expect to be appropriate to the metric for an isolated source. This solution can also be used for $\gamma < -3$, provided one remembers that increasing $r$ actually corresponds to moving *towards* the black hole ($g_{tt} \to 0$ as $r \to \infty$). This is a genuine wormhole solution, in that the area $\mathcal{A}(r)$ increases unboundedly towards the event horizon, which is located at infinite proper distance. Unlike the Schwarzschild wormhole however, this inner asymptotic region is not flat, but leads in to a null asymptopia. It is perhaps worth noting that the critical value $\gamma = -3$ corresponds to the marginal case of no wormhole, but an infinite throat:

$$ds^2 = -r^2 dt^2 + \frac{dr^2}{r^2} + r_0^2 d\Omega_{II}^2, \tag{3.49}$$

exactly analogous to the extreme Reissner-Nordstrom black hole throat.

Clearly, trajectories which terminate on the critical point $Q_+$ do not describe desirable black hole solutions, and the general solution of the BBH therefore requires $\gamma > 3$ or $\gamma < 0$ in order to terminate on the critical point $P_+$. Fortunately, this range of $\gamma$ is precisely that for which spacetime asymptotes flat space, although only for $\gamma < 0$ is it actually asymptotically flat (as discussed in Section 3.5.1).

**Asymptotes**

The other asymptotic region of interest, corresponding in general to large values of $X$, is the black hole horizon, for which we can identify the characteristic behaviour.

The line $Y = (\gamma - 3)X/(\gamma + 9)$ is a separatrix for $\gamma < 0$ and $\gamma > 3$, but an asymptote for $\gamma \in [0, 3]$, corresponding to the solution

$$\mathrm{d}s^2 = -(\rho - \rho_0)^{\frac{24\gamma}{\gamma^2+27}} \mathrm{d}t^2 + (\rho - \rho_0)^{\frac{4\gamma(\gamma-3)}{\gamma^2+27}} \left[\mathrm{d}\rho^2 + \mathrm{d}\Omega_{II}^2\right]. \tag{3.50}$$

From equation (3.33) it follows that this has a singular horizon.

For negative $\gamma$ and $\gamma > 3$ the asymptotic solution is

$$X = \frac{2}{\rho - \rho_0}, \qquad Y \propto (\rho - \rho_0)^{-3/\gamma}, \tag{3.51}$$

giving the metric

$$\mathrm{d}s^2 = -(\rho - \rho_0)^2 \mathrm{d}t^2 + C_0^2 \left(1 + c_1(\rho - \rho_0)^{1-3/\gamma}\right) \left[\mathrm{d}\rho^2 + \mathrm{d}\Omega_{II}^2\right]. \tag{3.52}$$

The horizon is singular in this case for $|\gamma| > 3$.

### 3.6.2 Special solutions

Because of the number and nature of the critical points, there can be special solutions which start on one critical point and terminate on another. Depending on the value of $\gamma$, the attractor $P_+$ can have up to three special solutions corresponding to trajectories from each of the other critical points. It is easiest to demonstrate these solutions for the extreme case $\gamma = \infty$ (where we have the analytic solution of the phase plane discussed in Section 3.5.2). In addition, there are special values of $\gamma$ for which there are $Q_- \to Q_+$ trajectories.

- $P_- \to P_+$ solution:

  The solution from $P_-$ to $P_+$ is easy to define in the analytic case $\gamma = \infty$: it is $X = \tanh(\rho - \rho_1)/2$, $Y = \tanh(\rho - \rho_0)/2$, which corresponds to a two parameter family solution of metrics:

  $$\mathrm{d}s^2 = -\frac{(r+r_1)^2}{(r+r_0)^2}\mathrm{d}t^2 + \frac{(r+r_0)^4}{r^4}\mathrm{d}r^2 + \frac{(r+r_0)^4}{r^2}\mathrm{d}\Omega_{I\!I}^2. \qquad (3.53)$$

  For $r_0 = r_1$, this is the Schwarzschild wormhole, and corresponds to the straight line path between $P_-$ and $P_+$ in Figure 3.2. For $r_1 \neq r_0$, the solution corresponds to the other paths between the critical points. The wormhole is supported here by the negative anisotropic stress $P = -6/C^3 = -6r^3/(r+r_0)^6$. See [106] for more detailed discussions of wormhole solutions in braneworlds. Although for finite $\gamma$ the form of this metric will change, the general nature – that of a wormhole connecting two asymptotically flat regions – will not. This solution exists for $\gamma > 9$ and $\gamma < -3$ (when the $Q_+$ critical point lies outside of the top-right quadrant of the phase plane).

- $Q_- \to P_+$ solution:

  This solution only exists in the $\gamma = \infty$ limit. It has $X \equiv 1$ and $Y = \tanh \rho/2$, which integrates to the metric

  $$\mathrm{d}s^2 = -\frac{r^2}{(r+r_0)^2}\mathrm{d}t^2 + \frac{(r+r_0)^4}{r^4}\mathrm{d}r^2 + \frac{(r+r_0)^4}{r^2}\mathrm{d}\Omega_{I\!I}^2. \qquad (3.54)$$

  This can be seen to be a limiting form of the metric (3.40), i.e. a wormhole with an inner asymptotic null asymptopia. For $\gamma < -3$, the $Q_+$ critical point has $Y < 0$ and hence the $Q_+ \to P_+$ solution also assumes this form.

- $Q_+ \to P_+$ solution:

  This solution has $Y \equiv 1$ and $X = \tanh \rho/2$ for the analytic case ($\gamma = \infty$). This integrates to the metric

  $$ds^2 = -\left(1 + \frac{1}{r}\right)^2 dt^2 + dr^2 + r^2 d\Omega_{I\!I}^2, \qquad (3.55)$$

  which is not a wormhole, but a flat 3-space with a distorted Newtonian potential. The solution will have this form for $\gamma > 3$, but for $\gamma < -3$ the $Q_+$ critical point moves below the $X$-axis and the solution joining $Q_+$ to $P_+$ takes the above form (3.54). For $\gamma = -3$ this trajectory is identical to the extremal Reissner-Nordström solution:

  $$ds^2 = -\left(1 - \frac{r_0}{r}\right)^2 dt^2 + \left(1 - \frac{r_0}{r}\right)^{-2} dr^2 + r^2 d\Omega_{I\!I}^2. \qquad (3.56)$$

- $Q_- \to Q_+$ solution:

  For special values of $\gamma$ there are also $Q_- \to Q_+$ trajectories. Specifically, for $\gamma = 9$ the critical points $Q$ lie on the $Y$-axis, giving us the wormhole solution:

  $$ds^2 = -\frac{r}{(r+r_0)^2} dt^2 + \frac{3}{4}\left(\frac{r+r_0}{r}\right)^4 dr^2 + \frac{(r+r_0)^4}{r^2} d\Omega_{I\!I}^2, \qquad (3.57)$$

  and for $\gamma = -3$ they lie on the $X$-axis, giving us a "throat" solution with bouncing Newtonian potential:

  $$ds^2 = -A_0^2 \cosh^2 r \, dt^2 + dr^2 + R_0^2 d\Omega_{I\!I}^2, \qquad (3.58)$$

  or $\mathrm{AdS}_2 \times S^2$.

### 3.6.3 Summary

To summarize: the phase plane shows clearly that flat spacetime is a critical point, which is an attractor for equations of state with $\gamma > 3$ or $\gamma < 0$. Only for this latter range of $\gamma$ is the spacetime asymptotically flat. For these ranges of $\gamma$ the near horizon behaviour is also given by a defined class of metrics (3.52), which are singular for $|\gamma| > 3$. Solutions are allowed both with and without turning points in the area function, though only for $\gamma < -3$ do these correspond to solutions which are asymptotically flat.

In addition, there are special solutions which correspond to trajectories between critical points. The trajectory from $P_-$ to $P_+$ represents a solution with two asymptotic flat regions, and is a genuine braneworld wormhole. The solution from one $Q$ critical point also represents a wormhole spacetime, which in the case of $\gamma = -3$, has a throat.

## 3.7 Physical solutions

In order to decide what a reasonable BBH metric might look like, and also to put the above analysis in context, it is useful to compare with some known results. In particular, the far-field linearised metric, and the small black hole approximation.

### 3.7.1 The weak field limit

One region in which we *do* know the metric of the black hole is at large $r$, where the spacetime should be described by the linearised Garriga-Tanaka metric (3.12):

$$ds^2 = -\left(1 - \frac{2M}{r} - \frac{4Ml^2}{3r^3}\right)dt^2 + \left(1 - \frac{2M}{r} - \frac{2Ml^2}{r^3}\right)^{-1}dr^2 + r^2 d\Omega_{I\!I}^2. \quad (3.59)$$

The asymptotic form of the Weyl tensor for this metric is given by (3.15), which has the equation of state $\gamma = -3/2$. However, as discussed in Section 3.3.1, the metric (3.59) can only be taken as a solution to the field equations in the linearised, far-field limit. Therefore we make the following conjecture:

*The exact BBH solution will interpolate between an asymptotic equation of state $\gamma = -3/2$ at infinity, and some other asymptotic value of $\gamma$ in the near-horizon limit.*

### 3.7.2 The small black hole limit

Another limit in which we can be fairly sure of the form of the BBH metric is for a small black hole with horizon size much smaller than the AdS curvature length. As discussed in Section 3.3.3, such a black hole should be described by the induced 5-D Schwarzschild-Tangherlini metric (3.22):

$$\mathrm{d}s^2 = -\left(1 - \frac{r_h^2}{r^2}\right)\mathrm{d}t^2 + \left(1 - \frac{r_h^2}{r^2}\right)^{-1}\mathrm{d}r^2 + r^2\mathrm{d}\Omega_{\mathbb{II}}^2, \qquad (3.60)$$

and might also be well-approximated by the more general tidal-RN metric (3.19). It can be seen from equation (3.21) that both these solutions have an equation of state $\gamma = -3$.

A plot of the phase plane for $\gamma = -3$ shows that these trajectories form a stable family of solutions terminating on $P_+$ (see Figure 3.3). There are no trajectories crossing the $X$-axis as $Y = 0$ is a solution to the equations of motion (in other words, there are solutions to the dynamical system which represent truly infinite throats with varying Newtonian potentials).

For $\gamma < -3$, the $Q_+$ critical point moves below the $X$-axis and there are once again trajectories which cross the $Y$-axis, and hence have a turning point in the area function, or a wormhole (see, e.g., the phase plane for $\gamma = -7$ in Figure 3.2).

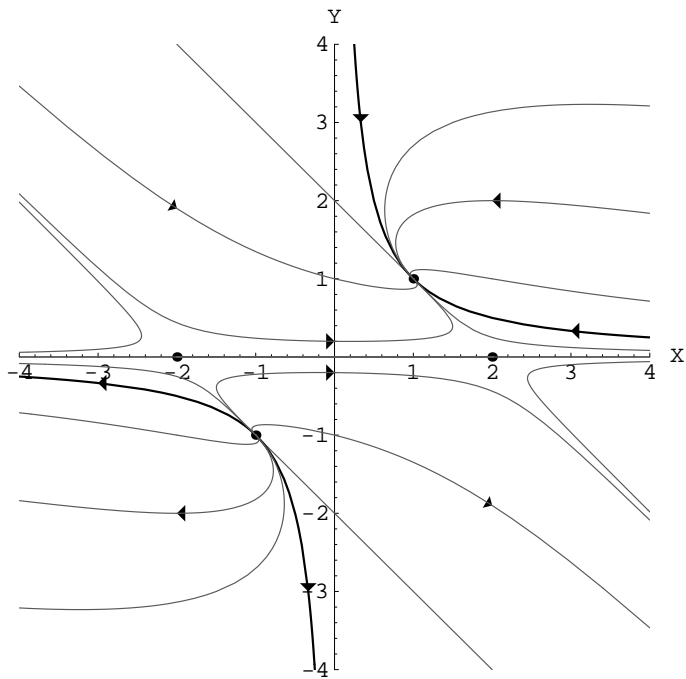

Figure 3.3: Phase plane for $\gamma = -3$.

### 3.7.3 A possible near-horizon solution

Having compared the equation of state method with known asymptotic solutions, it is clear that if an equation of state applies, it is generally a negative one. Moreover, from the intuition gained by looking at very small black holes, $\gamma$ curiously appears to become more negative closer to the horizon.

This interpretation is further supported by heuristic arguments based on the holographic correspondence, according to which classical solutions in 5-D project to quantum corrected solutions in 4-D [89]. This idea works nicely in cosmology, where the projection of the Weyl curvature of the bulk black hole onto the brane gives rise to an effective radiation source, as discussed in Section 2.4. This source can be interpreted holographically as a conformal field theory (CFT) in a thermal state corresponding to the Hawking temperature of the bulk black hole. The isotropy of the brane cosmological metric demands that only a radiation energy Weyl term is allowed, which corresponds from the bulk perspective to every point on the brane being the same distance from the bulk black hole.

One way of imagining a *braneworld* black hole forming is to transport this bulk black hole in towards and onto the brane (for a recent attempt at realising this process, see [83]). In doing this, the symmetry of equidistance from the bulk black hole is broken. From the holographically dual point of view, we therefore introduce an anisotropic stress by shifting the black hole from its equidistant position, and the closer we bring it towards the brane, the more anisotropic the setup. Hence, we might expect that the effective anisotropic stress $P$ on the brane becomes more and more important as we transport the black hole towards the brane, or as we move closer to the event horizon.

A physically reasonable expectation might therefore be that equations of state with large $\gamma$ are relevant near the horizon. Taking this reasoning to its logical extreme, we therefore propose as a "working metric" for the near-horizon solution the $U = 0$ (i.e. $\gamma = \pm\infty$) metric (3.40) of Section 3.5.2:

$$\mathrm{d}s^2 = -\frac{(r-r_1)^2}{(r+r_0)^2}\mathrm{d}t^2 + \frac{(r+r_0)^4}{r^4}\mathrm{d}r^2 + \frac{(r+r_0)^4}{r^2}\mathrm{d}\Omega_{I\!I}^2. \tag{3.61}$$

There is clearly a degree of arbitrariness in this choice, however we believe that the horizon is likely to be singular and that a turning point in the area function is also likely. This metric exhibits both these features, and has the added advantage of being analytic, which means that many properties can be calculated explicitly. In the area gauge, this solution is given by (3.41):

$$\mathrm{d}s^2 = -\left[(1+\epsilon)\sqrt{1-\frac{2GM}{R}}-\epsilon\right]^2\mathrm{d}t^2 + \left(1-\frac{2GM}{R}\right)^{-1}\mathrm{d}R^2 + R^2\mathrm{d}\Omega_{I\!I}^2. \tag{3.62}$$

The first point to note about (3.62) is that the ADM mass $M$ and gravitational mass (defined by $g_{tt}$) are no longer the same. As a result, the weak field tests of light bending and perihelion precession will be modified at the $O(\epsilon)$ level. Indeed, the metric (3.62) can easily be compared with the parametrised post-Newtonian (PPN) formalism, and observational limits on these parameters from solar system tests impose the limit $|\epsilon| \lesssim 10^{-3}$. This would only be applicable, however, if the metric (3.62) covered the entire spacetime, but we are considering this solution only as a possible *near-horizon* limit of some more general metric. The only mathematical limit on this solution as a near-horizon metric is $\epsilon > -\frac{1}{2}$ from the requirement that $r_1$ is positive definite (see equation (3.42)).

Another key feature of this solution is that the horizon is *always* singular, even for apparently negligibly small $\epsilon$ (except for $\epsilon = 0$ which corresponds to $\mathcal{E}_{\mu\nu} = 0$ and so is identically Schwarzschild). This is a true singularity, as the effective energy density from $\mathcal{E}_{\mu\nu}$ becomes infinite at the horizon (see equation (3.33)). An intrepid observer plunging into a supermassive black hole, expecting to sail seamlessly through the horizon and explore spacetime close to the singularity before finally succumbing to tidal forces, is instead crushed out of existence *at* the horizon. Indeed, they are slammed into the horizon at infinite proper speed! There are no timelike or null geodesics which connect anywhere outside the horizon with the standard Schwarzschild singularity at $r = 0$; infalling matter or light simply can not reach this point.

We now explore some simple tests of the metric (3.62) as a possible near-horizon solution, and constrain $\epsilon$ by connection to the observations of accreting black holes in our galaxy. Although this form of the metric is not valid throughout the entire horizon exterior if $\epsilon < 0$, it is sufficient for our purposes since it turns out that there can be no stable orbits within the wormhole region, i.e. the coordinate singularity at $R = 2GM$ is below the radii of interest for light and particle orbits, even when $\epsilon < 0$ (this point is further clarified in Section 4.3.1). The luminosity and temperature of the blackbody emission from the accretion disc can be measured from X-ray spectra, giving a direct estimate of the emitting area. There are uncertainties in this approach, but some confidence can be derived from the fact that changes in the luminosity give rise to the $L \propto T^4$ behaviour expected from a constant emitting area [109]. Assuming that the uncertainties are less than a factor of 2 then the data

| $\epsilon$ | -0.5 | -0.1 | 0 | 0.1 | 0.5 | 1 | 2 |
|---|---|---|---|---|---|---|---|
| $R_h$ | $\infty^*$ | $2.02^*$ | 2 | 2.01 | 2.25 | 2.67 | 3.6 |
| $R_{ms}$ | 4.26 | 5.63 | 6 | 6.37 | 7.88 | 9.82 | 13.75 |
| $h^2$ | 4.15 | 10.11 | 12 | 14.03 | 23.78 | 39.54 | 83.06 |
| $R_{ph}$ | 2.25 | 2.83 | 3 | 3.17 | 3.91 | 4.86 | 6.82 |

Table 3.1: Orbital parameters for the metric (3.62), in units of $GM$: $R_h$ is the radius of the horizon, $R_{ms}$ is the radius of the minimum stable orbit for matter, with associated angular momentum $h^2$, and $R_{ph}$ is the radius of the unstable photon orbit. Note that for negative $\epsilon$, the star indicates that the horizon is on the lower Kruskal branch, and therefore *inside* $R = 2GM$.

strongly constrain the minimum stable orbit to be $< 12GM$.

We assume that the metric (3.62) applies on these size scales and solve the Euler-Lagrange equations for motion in the equatorial plane to find the innermost stable particle orbit, and its associated angular momentum $h = |g_{\phi\phi}|\dot{\phi}$. Table 3.1 shows these as a function of $\epsilon$, together with the horizon position ($g_{tt} = 0$) and the radius at which light can (unstably) orbit around the black hole. The observations of accretion disc size in X-ray binary systems therefore give an upper limit to $\epsilon$ of about 2, and are easily consistent with our expectation of $\epsilon < 0$. Note that there is an extra degree of uncertainty in this conclusion since we have taken $M$ to be the ADM mass of the near-horizon metric (3.62), but in the absence of a complete solution for all $R$ it is not clear how this is related to the mass of the black hole that is operationally measured at larger distances.

A more detailed study of possible observational signatures of a black hole with this geometry is performed in the next chapter, where we investigate its gravitational lensing properties.

# Chapter 4

# Gravitational Lensing by Braneworld Black Holes

## 4.1 Introduction

The theory of gravitational lensing has been mostly developed in the weak field approximation, where it has been successful in explaining all observations and has become a valuable tool in astrophysics and cosmology [110–112]. However, one of the most spectacular consequences of the strong gravitational field surrounding a black hole is the large bending of light that can result for a light ray passing through this region. In particular, light rays passing close enough to the black hole's photon sphere may loop around the black hole one or more times before reaching the observer and forming so-called *relativistic images*, in addition to the two classical weak field images. A full description of these images requires heavy numerical integrations, although a surprisingly simple formula for the deflection angle in the strong field limit (SFL) was proposed by Darwin in 1958 [113], and revived in [114–116].

The study of strong gravitational lensing has received renewed interest recently. This is partly due to the fact that advances in technology have made observations of these relativistic images more viable [117], and partly because of the exciting oppor-tunity afforded by such observations to discriminate amongst alternative theories of gravity. Virbhadra and Ellis [118] studied lensing by the galactic supermassive black hole, in an asymptotically flat background, while Frittelli, Kling and Newman [119] found an exact lens equation without reference to a background metric and compared their results with those of Virbhadra and Ellis. In [116] Bozza *et al.* used the SFL to investigate Schwarzschild black hole lensing analytically. This technique was then applied to Reissner-Nordström black holes [120] and various other black hole (and wormhole) geometries [121–126], and was generalised to an arbitrary static, spherically symmetric metric by Bozza [127]. Alternative formalisms for treating gravitational lensing beyond the weak field approximation have also been devised [128–130], and the SFL method has recently been extended to include Kerr black hole lensing [131]. In this chapter we utilise the SFL method to investigate the gravitational lensing properties of a couple of candidate BBH metrics, to see how braneworld effects might manifest themselves in observations of black holes.

Similar studies have been performed for a BBH with the induced geometry of the 5-D Schwarzschild-Tangherlini solution: $-g_{tt} = g_{rr}^{-1} = 1 - r_h^2/r^2$. Both weak field lensing [132] and strong field lensing [133] for this geometry have been studied. However, this metric is only appropriate for black holes with a horizon size smaller than the AdS length scale of the extra dimension: $r_h \ll l \lesssim 0.1$mm. Hence, this metric is not appropriate for investigating the phenomenology of massive astrophysical black holes. Weak gravitational lensing for the linearised Garriga-Tanaka metric (3.12) has been studied by Kar and Sinha [134], and in more detail by Keeton and Petters [135]. As expected, braneworld effects are negligible in traditional lensing scenarios, although [135] raised the interesting possibility of observing primordial BBHs through their "attolensing" properties.

## 4.2 Lensing setup and equations

The lensing setup is shown in Figure 4.1. Light emitted by the source $S$ is deflected by the lens $L$ and reaches the observer $O$ at an angle $\theta$ to the optic axis $OL$, instead of $\beta$. The spacetime is asymptotically flat, and both observer and source are located in the flat region. By simple trigonometry, the lens equation can be written down:

$$\tan\beta = \tan\theta - \frac{D_{ls}}{D_{os}}\left[\tan\theta + \tan(\alpha - \theta)\right], \tag{4.1}$$

where $D_{ls}$, $D_{os}$ and the deflection angle $\alpha$ are defined by Figure 4.1. The general static, spherically symmetric metric on the brane can be written as:

$$\mathrm{d}s^2 = -A^2(r)\mathrm{d}t^2 + B^2(r)\mathrm{d}r^2 + C^2(r)\mathrm{d}\Omega_{I\!I}^2. \tag{4.2}$$

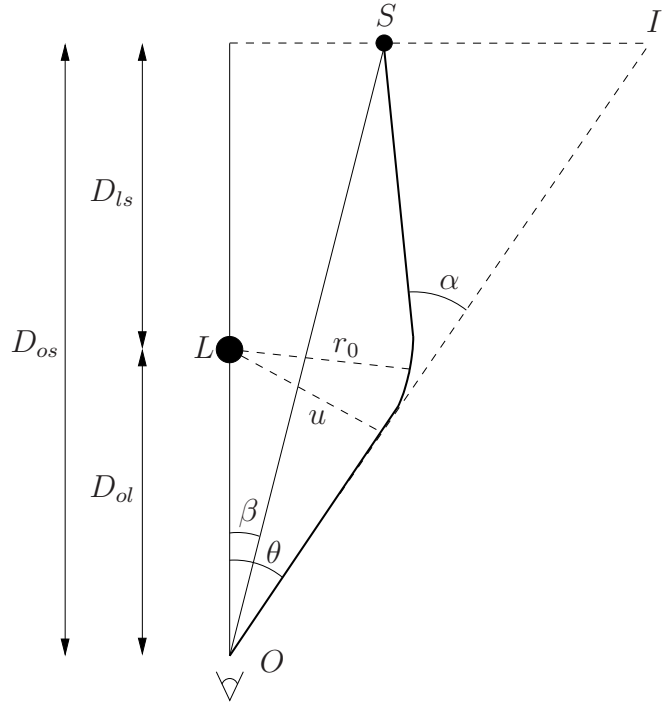

Figure 4.1: Gravitational lensing diagram.

From the null geodesic equations it is straightforward to show that the angular deflection of light as a function of radial distance from the lens is

$$\frac{\mathrm{d}\phi}{\mathrm{d}r} = \frac{B}{C\sqrt{\frac{C^2}{u^2 A^2} - 1}}. \tag{4.3}$$

By conservation of angular momentum, the impact parameter $u$ is given by

$$u = \frac{C_0}{A_0}, \tag{4.4}$$

where the subscript 0 indicates that the function is evaluated at the closest approach distance $r_0$. Hence, the deflection angle is given by

$$\begin{aligned} \alpha(r_0) &= I(r_0) - \pi \\ &= \int_{r_0}^{\infty} \frac{2B}{C} \left( \frac{C^2}{C_0^2} \frac{A_0^2}{A^2} - 1 \right)^{-\frac{1}{2}} \mathrm{d}r - \pi. \end{aligned} \tag{4.5}$$

Equations (4.1) and (4.5) are the basic equations of gravitational lensing. In principle, the deflection angle $\alpha$ for a given metric can be calculated from (4.5). This can then be plugged into the lens equation (4.1), and the image position $\theta$ for a given source position $\beta$ can be found.

The theory of gravitational lensing has been developed mostly in the weak field limit, where several simplifying assumptions can be made. The angles in (4.1) are taken to be small, so that $\tan x$ can be replaced by $x$ for $x = \beta, \theta, \alpha$, and the integrand in (4.5) is expanded to first order in the gravitational potential. For the Schwarzschild geometry, and setting $\beta = 0$, this leads to the well-known result:

$$\theta_E = \sqrt{\frac{4GM}{c^2} \frac{D_{ls}}{D_{os} D_{ol}}}, \tag{4.6}$$

where $\theta_E$ is the Einstein radius. In this formulation, GR has been successful in explaining all observations (see [110–112] for detailed reviews). However, it is important that gravitational lensing is not conceived of as a purely weak field phe-nomenon. Indeed, gravitational lensing in strong fields is one of the most promising tools for testing GR in its full, non-linear form.

### 4.2.1 Strong field limit

As the impact parameter $u$ of a light ray decreases, the deflection angle $\alpha$ increases as shown in Figure 4.2. At some point, the deflection angle exceeds $2\pi$ and the photon performs a complete loop around the black hole before emerging. The images thus formed are termed relativistic images and a theoretically infinite number of such images are formed on either side of the lens, corresponding to successive winding numbers around the black hole. The photon sphere is the radius $r_p$ at which a photon can unstably orbit the black hole, and is defined as the largest solution to the equation

$$\frac{A'(r)}{A(r)} = \frac{C'(r)}{C(r)}. \tag{4.7}$$

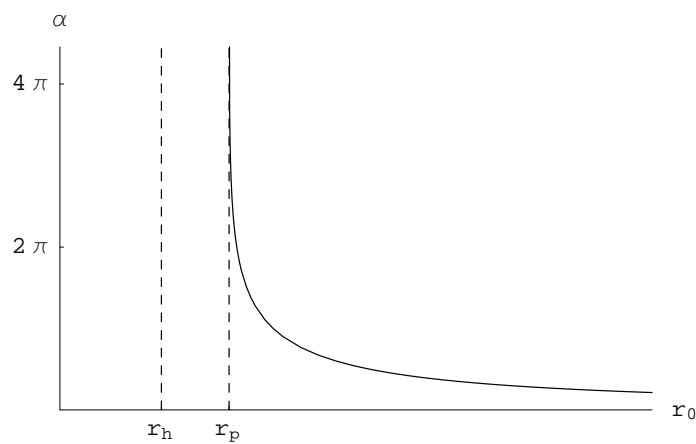

Figure 4.2: General behaviour of the deflection angle as a function of $r_0$. As $r_0$ decreases, $\alpha$ increases, and each time it reaches a multiple of $2\pi$ the photon completes a loop around the black hole.

As $r_0$ approaches $r_p$, with corresponding impact parameter

$$u_p = \frac{C_p}{A_p}, \tag{4.8}$$

the deflection angle diverges and for $r_0 < r_p$ the photon is captured by the black hole.

Bozza [127] has shown that this divergence is logarithmic for all spherically symmetric black hole metrics of the form (4.2). Hence the deflection angle can be expanded close to the divergence in the form

$$\alpha(r_0) = -a \ln\left(\frac{r_0}{r_p} - 1\right) + b + O(r_0 - r_p), \qquad (4.9)$$

or in terms of the angular position of the image, $\theta = u/D_{ol}$,

$$\alpha(\theta) = -\bar{a} \ln\left(\frac{\theta D_{ol}}{u_p} - 1\right) + \bar{b} + O(u - u_p), \qquad (4.10)$$

where the *strong field limit* (SFL) coefficients $\bar{a}$ and $\bar{b}$ depend on the metric functions evaluated at $r_p$. This formula allows a simple, analytic description of the relativistic images and their properties, rather than having to use the exact deflection angle calculated numerically from equation (4.5).

Equation (4.10) can be derived from (4.5) by splitting the integral into a divergent and non-divergent piece, and performing some expansions. Defining the new variable

$$z = \frac{A^2 - A_0^2}{1 - A_0^2}, \qquad (4.11)$$

the integral in equation (4.5) becomes

$$I(r_0) = \int_0^1 R(z, r_0) f(z, r_0) \mathrm{d}z, \qquad (4.12)$$

where

$$R(z, r_0) = \frac{BC_0}{C^2 A'}(1 - A_0^2), \qquad (4.13)$$

$$f(z, r_0) = \left(A_0^2 - [(1 - A_0^2)z + A_0^2]\frac{C_0^2}{C^2}\right)^{-\frac{1}{2}}. \qquad (4.14)$$

The function $R(z, r_0)$ is regular for all values of $z$ and $r_0$, but $f(z, r_0)$ diverges for $z \to 0$. Expanding the argument of the square root in $f(z, r_0)$ to second order in $z$:

$$f(z, r_0) \sim f_0(z, r_0) = \left(m(r_0) z + n(r_0) z^2\right)^{-1/2}, \tag{4.15}$$

$$m(r_0) = \frac{A_0(1 - A_0^2)}{A_0'} \left(\frac{C_0'}{C_0} - \frac{A_0'}{A_0}\right), \tag{4.16}$$

$$n(r_0) = \frac{(1 - A_0^2)^2}{4 A_0 A_0' C_0} \left[ 3 C_0' \left(1 - \frac{A_0 C_0'}{A_0' C_0}\right) + \frac{A_0}{A_0'} \left(C_0'' - \frac{C_0' A_0''}{A_0'}\right) \right], \tag{4.17}$$

it is clear why the deflection angle diverges logarithmically for $r_0 = r_p$: with $r_p$ given by equation (4.7), $m(r_p)$ vanishes. Hence for $r_0 = r_p$, $f_0 \propto 1/z$ and the integral (4.12) diverges logarithmically.

Proceeding to split the integral (4.12) into a divergent and a regular piece, and performing further expansions (see [127] for the detailed derivation[1]), the SFL coefficients are obtained:

$$\bar{a} = \frac{R(0, r_p)}{2\sqrt{n_p}}, \tag{4.18}$$

$$\bar{b} = \bar{a} \ln \frac{2 n_p}{A_p^2} + b_R - \pi, \tag{4.19}$$

where

$$n_p = n(r_p) = \frac{(1 - A_p^2)^2}{4 C_p A_p'^3} (C_p'' A_p' - C_p' A_p'') \tag{4.20}$$

and

$$b_R = \int_0^1 [R(z, r_p) f(z, r_p) - R(0, r_p) f_0(z, r_p)] \, dz, \tag{4.21}$$

is the integral $I(r_p)$ with the divergence subtracted.

---

[1] Note: The expressions in [127] look slightly different to ours since we define the metric as $g_{\mu\nu} = \text{diag}(-A^2, B^2, C^2)$ as opposed to $g_{\mu\nu} = \text{diag}(-A, B, C)$ in [127].

## 4.3 Braneworld black hole lensing

In this section we apply the method outlined in the previous section to calculate the deflection angle in the strong field limit for two candidate BBH metrics. It is important to emphasise that we envisage these only as possible near-horizon asymptotes of a more general metric, which has yet to be found. Neither metric satisfies the long distance $1/r^3$ correction to the gravitational potential, and both would be constrained in the weak field limit by the Solar System PPN observations.

### 4.3.1 $U = 0$ metric

The first metric we will consider is the $U = 0$ (i.e. $\gamma = \pm\infty$) solution (3.40) of Section 3.5.2:

$$\mathrm{d}s^2 = -\frac{(r - r_h)^2}{(r + r_t)^2}\,\mathrm{d}t^2 + \frac{(r + r_t)^4}{r^4}\,\mathrm{d}r^2 + \frac{(r + r_t)^4}{r^2}\,\mathrm{d}\Omega_{I\!I}^2, \qquad (4.22)$$

which has a turning point in the area function at $r = r_t$, and the horizon at $r = r_h$ is singular (except for the special case $r_h = r_t$, which is just the standard Schwarzschild solution in isotropic coordinates). Although this choice of metric is somewhat arbitrary, we believe that the horizon is likely to be singular and that a turning point in the area function is also likely, and this metric exhibits both these features. It has the added advantage of being analytic, and so seems a good choice for exploring these radical differences to the standard Schwarzschild geometry.

Normalising the distances to $4r_t$ (which corresponds to a distance $2M$ where $M$ is the ADM mass – see equation (3.42)), the metric functions become

$$A^2(r) = \frac{(r - r_h)^2}{(r + 1/4)^2}\,; \quad B^2(r) = \frac{(r + 1/4)^4}{r^4}\,; \quad C^2(r) = \frac{(r + 1/4)^4}{r^2}. \qquad (4.23)$$

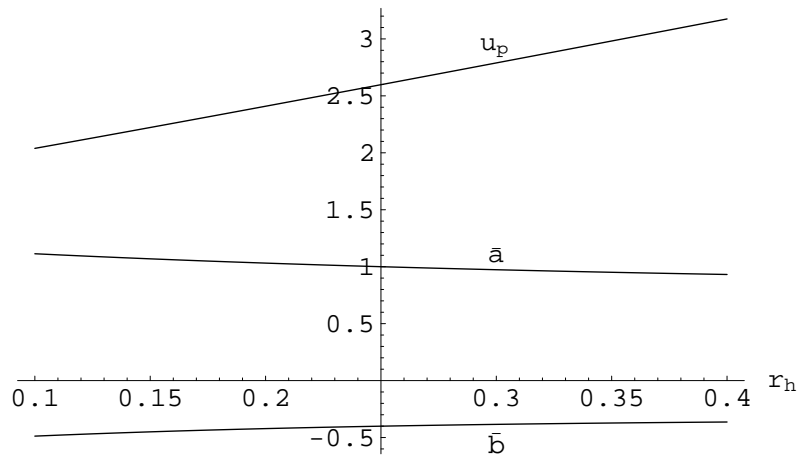

Figure 4.3: SFL coefficients for the $U = 0$ metric (4.23), as functions of $r_h$. The standard Schwarzschild case is given by $r_h = 1/4$.

The radius of the photon sphere is given by

$$r_p = \frac{1}{4}\left(1 + 4r_h + \sqrt{1 + 4r_h + 16r_h^2}\right). \quad (4.24)$$

The turning point in the area function occurs at $r = 1/4$, however note from (4.24) that $r_p > 1/4$ for all values of $r_h$. Therefore the turning point always lies *inside* of the photon sphere and light rays that are not captured by the black hole can never probe the wormhole region. Hence, perhaps surprisingly, the presence of a wormhole region exterior to the horizon (which occurs for $r_h < 1/4$) does not have a dramatic effect on the lensing properties of the black hole.

The SFL coefficients $\bar{a}, \bar{b}$ and $u_p$, calculated from equations (4.18), (4.19) and (4.8), are shown in Figure 4.3. It can be seen that the biggest deviation from standard Schwarzschild lensing is for the minimum impact parameter $u_p$. This is because as the horizon is shifted inwards/outwards relative to the Schwarzschild case, the photon sphere is pulled/pushed along with it.

We can check the accuracy of the SFL approximation by comparing the exact deflection angle $\alpha_{\text{exact}}$ calculated from (4.5) with the SFL $\alpha_{\text{SFL}}$ from (4.10). The outermost relativistic image appears where $\alpha \simeq 2\pi$, which occurs for an impact parameter $u_1 = u_p + x$, where $x \sim 0.003$ depends on $r_h$. The discrepancy between $\alpha_{\text{exact}}(u_1)$ and $\alpha_{\text{SFL}}(u_1)$ is less than 0.13% for all values of $r_h$ we consider. Hence, the SFL of the deflection angle is very accurate and can be reliably used to obtain accurate results for the properties of the relativistic images.

### 4.3.2 Tidal Reissner-Nordström metric

The second metric we consider is the tidal-RN solution discussed in Section 3.3.3:

$$\mathrm{d}s^2 = -\left(1 - \frac{2GM}{r} + \frac{Q}{r^2}\right)\mathrm{d}t^2 + \left(1 - \frac{2GM}{r} + \frac{Q}{r^2}\right)^{-1}\mathrm{d}r^2 + r^2 \mathrm{d}\Omega_{I\!I}^2. \qquad (4.25)$$

This has the correct 5-D ($\sim 1/r^2$) short distance behaviour and so could be a good approximation in the strong field regime for small black holes. Taking this argument further, we can postulate this solution as a near-horizon metric for any size BBH: we might expect braneworld effects to be more prominent as we approach the event horizon, and so the extra dimension might begin to make its presence felt through the $1/r^2$ correction to the potential that is characteristic of 5-D gravity.

Normalising the distances to $2GM$, with $q = Q/(2GM)^2$, the metric (4.25) becomes

$$A^2(r) = 1 - \frac{1}{r} + \frac{q}{r^2}; \quad B^2(r) = \left(1 - \frac{1}{r} + \frac{q}{r^2}\right)^{-1}; \quad C^2(r) = r^2. \qquad (4.26)$$

The radius of the photon sphere is given by

$$r_p = \frac{1}{4}\left(3 + \sqrt{9 - 32q}\right). \qquad (4.27)$$

The SFL coefficients are shown in Figure 4.4. These results reproduce those of Eiroa *et al.* [120] for $q > 0$ (see also [127]), but we have extended the results to negative $q$. We emphasise that there is *no* electric charge present for the tidal RN solution – $q$ is a tidal charge parameter arising from the bulk Weyl tensor. It is also worth noting that any electric charge acquired by an astrophysical black hole is expected to be neutralised by plasma in its vicinity, and so discussions of astrophysical black hole phenomena are commonly restricted to the Kerr family of black holes. This argument does not apply in the braneworld case, where an effective Reissner-Nordström metric emerges *geometrically*.

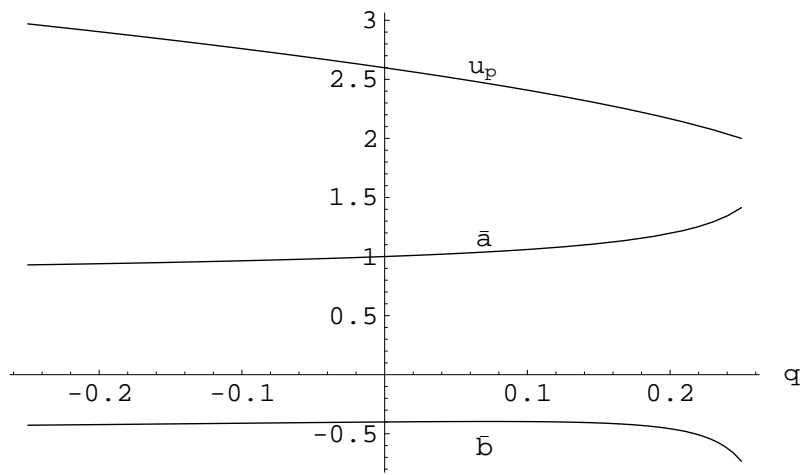

Figure 4.4: SFL coefficients for the tidal Reissner-Nordström metric (4.26), as functions of $q$. The standard Schwarzschild case given by $q = 0$.

Again, comparing $\alpha_{\text{exact}}$ with $\alpha_{\text{SFL}}$ for an impact parameter corresponding to the outermost image, it is found that the discrepancy is less than 0.5% for the values of $q$ considered here.

## 4.4 Observables

In Section 4.2.1 it was shown how to calculate the deflection angle in the strong field limit and in Section 4.3 this method was applied to the candidate BBH metrics. In this section we put the SFL of the deflection angle into the lens equation to obtain analytic formulae for the properties of the relativistic images in terms of the SFL coefficients $\bar{a}, \bar{b}$ and $u_p$.

As expected, the relativistic images formed by light rays winding around the black hole are greatly de-magnified compared to the standard weak field images, and are most prominent when the source, lens and observer are highly aligned [118]. Hence, we restrict our attention to the case where $\beta$ and $\theta$ are small (see [117] for the general case where this assumption is relaxed). Although we can not assume $\alpha$ is small, if a light ray is going to reach the observer after winding around the black hole, $\alpha$ must be very close to a multiple of $2\pi$. Writing $\alpha = 2n\pi + \Delta\alpha_n$, $n \in \mathbb{Z}$, the lens equation (4.1) becomes

$$\beta = \theta - \frac{D_{ls}}{D_{os}} \Delta\alpha_n. \tag{4.28}$$

Firstly, we have to find the values $\theta_n^0$ such that $\alpha(\theta_n^0) = 2n\pi$. With $\alpha$ given by (4.10) we find

$$\theta_n^0 = \frac{u_p}{D_{ol}}(1 + e_n), \tag{4.29}$$

where

$$e_n = e^{(\bar{b} - 2n\pi)/\bar{a}}. \tag{4.30}$$

Thus the position of the $n^{\text{th}}$ relativistic image can be approximated by [127]

$$\theta_n = \theta_n^0 + \frac{u_p e_n (\beta - \theta_n^0) D_{os}}{\bar{a} D_{ls} D_{ol}}, \tag{4.31}$$

where the correction to $\theta_n^0$ is much smaller than $\theta_n^0$. Approximating the position of the images by $\theta_n^0$, the magnification of the $n^{\text{th}}$ relativistic image is given by

$$\mu_n = \frac{1}{(\beta/\theta)\partial\beta/\partial\theta}\bigg|_{\theta_n^0} \simeq \frac{u_p^2 e_n (1 + e_n) D_{os}}{\bar{a} \beta D_{ol}^2 D_{ls}}. \tag{4.32}$$

Equations (4.31) and (4.32) relate the position and magnification of the relativistic images to the SFL coefficients. We now focus on the simplest situation, depicted in Figure 4.5, where only the outermost image $\theta_1$ is resolved as a single image, with the remaining images packed together at $\theta_\infty = u_p/D_{ol}$. Therefore we define the

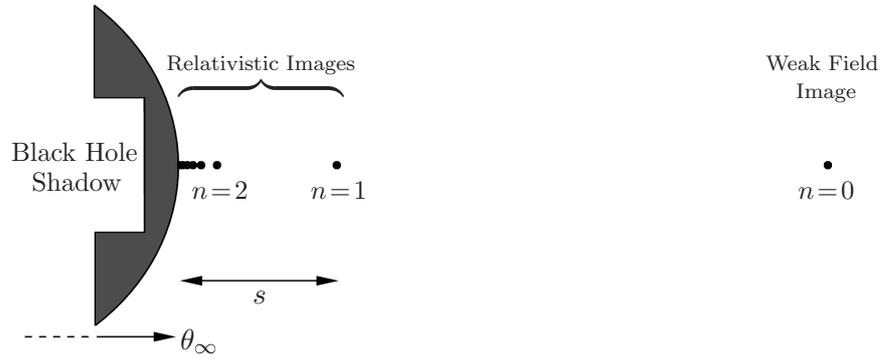

Figure 4.5: Diagrammatic representation of the observables defined in Section 4.4, showing the images formed by photons that wind around the black hole $n$ times.

observables

$$s = \theta_1 - \theta_\infty, \tag{4.33}$$

$$f = \frac{\mu_1}{\sum_{n=2}^{\infty} \mu_n}, \tag{4.34}$$

which are respectively the separation between the outermost image and the others, and the flux ratio between the outermost image and all the others. It is found that these simplify to [127]

$$s = \theta_\infty e^{(\bar{b}-2\pi)/\bar{a}}, \tag{4.35}$$

$$f = e^{2\pi/\bar{a}}. \tag{4.36}$$

These equations are easily inverted to give $\bar{a}, \bar{b}$ and so if an observation were able to measure $s, f$ and $\theta_\infty$ the SFL coefficients could be determined and the nature of the lensing black hole identified.

### 4.4.1 An example: the galactic supermassive black hole

It is believed that the centre of our galaxy harbours a black hole of mass $M = 2.8 \times 10^6 M_\odot$ [75, 76]. Taking $D_{ol} = 8.5\,\text{kpc}$, Virbhadra and Ellis [118] studied the lensing of a background source by this black hole (assuming it to be Schwarzschild)

|  | Sch. | $U=0$ | | | | Tidal RN | | | |
|---|---|---|---|---|---|---|---|---|---|
|  |  | $r_h\!=\!0.1$ | $r_h\!=\!0.2$ | $r_h\!=\!0.3$ | $r_h\!=\!0.4$ | $q\!=\!-0.2$ | $q\!=\!-0.1$ | $q\!=\!0.1$ | $q\!=\!0.2$ |
| $\theta_\infty$ | 16.87 | 13.24 | 15.65 | 18.11 | 20.62 | 18.85 | 17.92 | 15.64 | 14.07 |
| $s$ | 0.0211 | 0.0303 | 0.0235 | 0.0192 | 0.0164 | 0.0150 | 0.0173 | 0.0286 | 0.0502 |
| $f_m$ | 6.82 | 6.13 | 6.61 | 7.01 | 7.32 | 7.26 | 7.08 | 6.44 | 5.70 |

Table 4.1: Estimates for the lensing observables for the central black hole of our galaxy. $\theta_\infty$ and $s$ are defined in Section 4.4 and given in $\mu$ arc seconds, and $f_m = 2.5 \log f$ is $f$ converted to magnitudes.

and found that the relativistic images are formed at about $17\,\mu$ arc sec. from the optic axis.

In Table 4.1 we estimate the observables $\theta_\infty, s, f$ defined in the previous section for the $U = 0$ and tidal-RN BBH metrics, as well as the standard Schwarzschild metric. It is clear that the easiest observable to resolve is $\theta_\infty$, since a microarcsecond resolution is in principle attainable by very long baseline interferometry (VLBI) projects such as MAXIM [136] and ARISE [137]. However, the disturbances inherent in such observations would make the identification of the faint relativistic images very difficult [118], although not impossible [117].

If a measurement of $\theta_\infty$ was made, it would be immediately capable of distinguishing between Schwarzschild and other types of geometry. However, to determine all the SFL coefficients and thus unambiguously identify the nature of the lensing black hole, it is necessary to also measure $s$ and $f$. This would require the resolution of two extremely faint images separated by $\sim 0.02\,\mu$ arc sec. Such an observation in a realistic astrophysical environment is certainly not feasible in the near future, although if such an observation were ever possible, it would provide an excellent test of gravity in a strong field.

Nevertheless, Table 4.1 clearly shows that BBHs could have significantly different observational signatures than the standard Schwarzschild black hole, and this encourages the investigation of more realistically observable situations. An interesting possibility in this direction is the study of the accretion discs surrounding black holes.

The observed disc emission depends on several factors that could be modified by braneworld effects. A key factor is the radius of the innermost stable circular orbit, since emitting material at this radius sets the maximum temperature for the disc emission [109, 138–140]. Just as the radius of the photon sphere can be shifted inwards or outwards relative to the Schwarzschild case for a BBH, so too can the radius of the innermost stable orbit for matter (see Section 3.7.3). In addition, the observed disc emission, and in particular the iron fluorescence line profile, is affected by relativistic effects [141, 142] (doppler shift, gravitational redshift), which would be modified if the metric in the emitting region was not that of standard general relativity, but was a modified braneworld metric. Also, the light rays are gravitationally lensed by the central black hole as they escape the disc, and we have shown here that such lensing can be different from the Schwarzschild case for BBHs. In light of the results found here, it is not unreasonable to anticipate that these effects could result in distinctive observational signatures for accretion discs.

# Chapter 5

# Rotating Braneworld Black Holes

## 5.1 Introduction

Astrophysically, the Kerr metric [143] is undoubtedly the most important solution in GR, since it is believed to describe the spacetime surrounding every planet, star and black hole in the observable Universe. It is therefore important in the braneworld context to find rotating black hole solutions, both for completeness of the model and, more importantly, because observations of such objects might provide an exciting opportunity to detect extra dimensions, if they exist.

However, due to the reduced symmetry as compared to the static, spherically symmetric case, the derivation of rotating solutions is a much more complicated problem. This is highlighted by the fact that the Kerr metric [143] was not discovered until 1963, nearly five decades after the discovery of the Schwarzschild solution [72].

Ideally, we would like a full, 5-D solution that describes a rotating black hole localised on the brane. Such a solution would be given by a suitable slicing of the 5-D rotating C-metric (see discussion in Section 3.1), however the 5-D C-metric has not even been found in the static case, let alone with rotation included! Gravity in one dimension lower is much simpler, however, and it has been demonstrated that the 4-D rotating C-metric can indeed be sliced to construct a (2+1)-brane containing a localised rotating black hole [144]. Here, as before, we will have to content ourselves with a study of the 4-D-projected effective Einstein equations (2.68), in order to investigate possible 4-D geometries describing rotating black holes on the brane.

In the next section we will elucidate the general system of equations for the problem. The few currently known rotating braneworld solutions are described in Section 5.3: the linearised weak field metric, the rotating black string and the "tidal Kerr-Newman" solution. These solutions are the rotating analogues of the static solutions presented in Sections 3.3.1–3.3.3. In Section 5.4 we will attempt to generate rotating solutions using the complexification "trick" of Newman and Janis [145]. Unfortunately this is only partially successful and we are not able to generate any new solutions via this method.

## 5.2 Basic equations

A spacetime is said to be stationary if there exists a timelike Killing vector $k = \partial/\partial t$, and axisymmetric if there exists a spacelike Killing vector $m = \partial/\partial \phi$ whose integral curves are closed, where $t$ and $\phi$ are coordinates adapted to $k$ and $m$, respectively. We also restrict attention to *circular* spacetimes: a stress-energy tensor $T$ is circular if it satisfies

$$\begin{aligned} k^\delta T_{\delta[\alpha} k_\beta m_{\gamma]} &= 0, \\ m^\delta T_{\delta[\alpha} k_\beta m_{\gamma]} &= 0. \end{aligned} \quad (5.1)$$

These conditions are equivalent to the absence of momentum currents in the meridional planes orthogonal to both $k$ and $m$. For a fluid, this means that there is no convective motion but only circular motion about the axis of symmetry. For such a spacetime, it can be shown [146, 147] that there exists a family of two-surfaces everywhere orthogonal to the plane defined by $k$ and $m$. Hence we may choose coordinates $(t, x_1, x_2, \phi)$, where $(x_1, x_2)$ span the two-surfaces orthogonal to $k$ and $m$, such that $g_{01} = g_{02} = g_{31} = g_{32} = 0$.[1] Hence the metric can be written in block diagonal form:

$$\mathrm{d}s^2 = (g_{tt}\mathrm{d}t^2 + 2g_{t\phi}\mathrm{d}t\mathrm{d}\phi + g_{\phi\phi}\mathrm{d}\phi^2) + [g_{11}\mathrm{d}x_1^2 + 2g_{12}\mathrm{d}x_1 x_2 + g_{22}\mathrm{d}x_2^2], \qquad (5.2)$$

where the metric functions must be independent of $t$ and $\phi$ in order to respect the symmetries. Now, any two-dimensional space with a positive- or negative-definite signature $(+,+$ or $-,-)$ is conformally flat [148]. Hence the part in square brackets in (5.2) can be written $\pm f(x_1, x_2)[\mathrm{d}x_1^2 + \mathrm{d}x_2^2]$. However, it is often convenient for problem solving to retain some coordinate freedom in the metric and write it in a slightly less specific form. Labelling the coordinates $(x_1, x_2) \to (r, \theta)$, we can therefore write the general stationary, axisymmetric, circular metric in the form

$$\mathrm{d}s^2 = -A^2\mathrm{d}t^2 + B^2(\mathrm{d}\phi - \omega\mathrm{d}t)^2 + C^2\mathrm{d}r^2 + D^2\mathrm{d}\theta^2, \qquad (5.3)$$

where $A, B, C, D, \omega$ are functions of $(r, \theta)$ only, and we are free to impose a convenient coordinate condition relating $C$ and $D$. Now, consider an orthonormal tetrad for this metric:

$$\begin{aligned}
\Theta^{(0)} &= A\,\mathrm{d}t, \\
\Theta^{(1)} &= C\,\mathrm{d}r, \\
\Theta^{(2)} &= D\,\mathrm{d}\theta, \\
\Theta^{(3)} &= B(\mathrm{d}\phi - \omega\,\mathrm{d}t),
\end{aligned} \qquad (5.4)$$

$$\mathrm{d}s^2 = -\left(\Theta^{(0)}\right)^2 + \left(\Theta^{(1)}\right)^2 + \left(\Theta^{(2)}\right)^2 + \left(\Theta^{(3)}\right)^2. \qquad (5.5)$$

A point in spacetime that is at rest in this locally inertial frame,

$$u^{(a)} = (1, 0, 0, 0),$$

---
[1] An alternative statement of circularity is that the spacetime is invariant under the simultaneous inversion $t \to -t$ and $\phi \to -\phi$. Then $g_{01} = g_{02} = g_{31} = g_{32} = 0$ since these terms would change sign under this transformation.

will be assigned a 4-velocity

$$u^\mu = u^t(1, 0, 0, \omega)$$

in the coordinate frame. This makes clear the interpretation of $\omega$ as the angular velocity, as measured at infinity, of a zero angular momentum (or *locally inertial*) observer.

The vacuum braneworld effective Einstein equations we wish to solve are:

$$G_{\mu\nu} = -\mathcal{E}_{\mu\nu}, \tag{5.6}$$

where

$$\mathcal{E}_\mu{}^\mu = \nabla_\mu \mathcal{E}^{\mu\nu} = \mathcal{E}_{[\mu\nu]} = 0. \tag{5.7}$$

The trace-free property of $\mathcal{E}_{\mu\nu}$ implies $R = 0$, so the field equations reduce to

$$R_{\mu\nu} = -\mathcal{E}_{\mu\nu}. \tag{5.8}$$

As before (see equation (2.72)), $\mathcal{E}_{\mu\nu}$ can be decomposed with respect to a 4-velocity field $u^\mu$ as follows:

$$\mathcal{E}_{\mu\nu} = -\left[U\left(u_\mu u_\nu + \frac{1}{3}h_{\mu\nu}\right) + 2q_{(\mu}u_{\nu)} + P_{\mu\nu}\right], \tag{5.9}$$

where $h_{\mu\nu} = g_{\mu\nu} + u_\mu u_\nu$ projects orthogonally to $u^\mu$, and $U$, $q_\mu$ and $P_{\mu\nu}$ are the effective non-local energy density, momentum density and anisotropic stress, respectively. In order to respect stationarity and axisymmetry, the 4-velocity must take the form $u^\mu = \frac{1}{N}(1, 0, 0, \Omega)$, and the most "natural" choice is to take $\Omega = \omega$ so that the Weyl "fluid" is co-rotating with the black hole:[2]

$$u^\mu = \frac{1}{N}(1, 0, 0, \omega), \tag{5.10}$$

---

[2]i.e. in the locally inertial frame (5.4), the Weyl "fluid" is at rest: $u^{(a)} = (1, 0, 0, 0)$.

where $N$ is determined by the normalisation condition $u_\mu u^\mu = -1$. The symmetries of the problem suggest that we further decompose $P_{\mu\nu}$ as[3]

$$P_{\mu\nu} = P(r,\theta) \left( b(r,\theta) r_\mu r_\nu - (b(r,\theta) - 1) \theta_\mu \theta_\nu + f(r,\theta) r_\mu \theta_\nu - \frac{1}{3} h_{\mu\nu} \right), \quad (5.11)$$

where $r^\mu$ is a unit radial vector, $\theta^\mu$ is a unit vector in the $\theta$-direction, and $P(r,\theta)$ and $b(r,\theta)$ essentially parameterise the different pressures in the $r$-, $\theta$- and $\phi$-directions. It seems reasonable also to assume $f(r,\theta) = 0$, so we take $\mathcal{E}_{\mu\nu}$ to be

$$-\mathcal{E}_{\mu\nu} = U \left( u_\mu u_\nu + \frac{1}{3} h_{\mu\nu} \right) + 2 q_{(\mu} u_{\nu)} + P \left( b\, r_\mu r_\nu - (b-1) \theta_\mu \theta_\nu - \frac{1}{3} h_{\mu\nu} \right). \quad (5.12)$$

The static, spherically symmetric case (3.7) is recovered for $b(r,\theta) = 1$, with a non-zero $(b-1)$ representing the breaking of symmetry between the $\theta$- and $\phi$-directions.

---

[3] Compare to the static case (equation (3.7)) for which $P_{\mu\nu} = P(r) \left( r_\mu r_\nu - \frac{1}{3} h_{\mu\nu} \right)$.

The field equations (5.8) are most compactly written in the tetrad frame (5.4). With the above form for $\mathcal{E}_{\mu\nu}$ and $u_\mu$ given by (5.10), we find:

$$R_{00} = \frac{1}{C^2}\left[\frac{A''}{A} + \frac{A'}{A}\left(\frac{B'}{B} + \frac{D'}{D} - \frac{C'}{C}\right) - \frac{B^2\omega'^2}{2A^2}\right] \quad (5.13)$$

$$+ \frac{1}{D^2}\left[\frac{\ddot{A}}{A} + \frac{\dot{A}}{A}\left(\frac{\dot{B}}{B} + \frac{\dot{C}}{C} - \frac{\dot{D}}{D}\right) - \frac{B^2\dot{\omega}^2}{2A^2}\right] = U$$

$$R_{11} = -\frac{1}{C^2}\left[\frac{A''}{A} + \frac{B''}{B} + \frac{D''}{D} - \frac{C'}{C}\left(\frac{A'}{A} + \frac{B'}{B} + \frac{D'}{D}\right) - \frac{B^2\omega'^2}{2A^2}\right] \quad (5.14)$$

$$- \frac{1}{D^2}\left[\frac{\ddot{C}}{C} + \frac{\dot{C}}{C}\left(\frac{\dot{A}}{A} + \frac{\dot{B}}{B} - \frac{\dot{D}}{D}\right)\right] = \frac{1}{3}[U + P(3b-1)]$$

$$R_{22} = -\frac{1}{D^2}\left[\frac{\ddot{A}}{A} + \frac{\ddot{B}}{B} + \frac{\ddot{C}}{C} - \frac{\dot{D}}{D}\left(\frac{\dot{A}}{A} + \frac{\dot{B}}{B} + \frac{\dot{C}}{C}\right) - \frac{B^2\dot{\omega}^2}{2A^2}\right] \quad (5.15)$$

$$- \frac{1}{C^2}\left[\frac{D''}{D} + \frac{D'}{D}\left(\frac{A'}{A} + \frac{B'}{B} - \frac{C'}{C}\right)\right] = \frac{1}{3}[U - P(3b-2)]$$

$$R_{33} = -\frac{1}{C^2}\left[\frac{B''}{B} + \frac{B'}{B}\left(\frac{A'}{A} - \frac{C'}{C} + \frac{D'}{D}\right) + \frac{B^2\omega'^2}{2A^2}\right] \quad (5.16)$$

$$- \frac{1}{D^2}\left[\frac{\ddot{B}}{B} + \frac{\dot{B}}{B}\left(\frac{\dot{A}}{A} + \frac{\dot{C}}{C} - \frac{\dot{D}}{D}\right) + \frac{B^2\dot{\omega}^2}{2A^2}\right] = \frac{1}{3}(U - P)$$

$$R_{03} = \frac{B}{2AC^2}\left[\omega'' + \omega'\left(-\frac{A'}{A} + 3\frac{B'}{B} - \frac{C'}{C} + \frac{D'}{D}\right)\right] \quad (5.17)$$

$$+ \frac{B}{2AD^2}\left[\ddot{\omega} + \dot{\omega}\left(-\frac{\dot{A}}{A} + 3\frac{\dot{B}}{B} + \frac{\dot{C}}{C} - \frac{\dot{D}}{D}\right)\right] = q_3$$

$$R_{12} = \frac{1}{CD}\left[\frac{D'}{D}\left(\frac{\dot{A}}{A} + \frac{\dot{B}}{B}\right) + \frac{\dot{C}}{C}\left(\frac{A'}{A} + \frac{B'}{B}\right) - \frac{\dot{A}'}{A} - \frac{\dot{B}'}{B} + \frac{B^2\dot{\omega}\omega'}{2A^2}\right] = 0, \quad (5.18)$$

where $x' = \partial x/\partial r$, $\dot{x} = \partial x/\partial \theta$ and $R_{01} = R_{02} = 0$ implies $q_1 = q_2 = 0$. Solving this full system of equations presents a formidable challenge. At the time of writing only a few rotating BBH solutions are known (to be described in the next section) but, as with the static, spherically symmetric case, braneworld gravity should allow whole classes of stationary, axisymmetric black hole solutions on the brane.

## 5.3 Rotating solutions: examples

### 5.3.1 The linearised weak field metric

The linearised, weak field form of the metric for a rotating source on the brane was derived in Section 2.2.3, and in the form (5.3) is given by:

$$ds^2 = -\left[1 - \frac{2M}{r}\left(1 + \frac{2}{3}\frac{l^2}{r^2}\right)\right]dt^2 + r^2\sin^2\theta\left[d\phi - \frac{2Ma}{r^3}\left(1 + \frac{3}{2}\frac{l^2}{r^2}\right)dt\right]^2$$
$$+ r^2 d\theta^2 + \left(1 - \frac{2M}{r} - \frac{2Ml^2}{r^3}\right)^{-1} dr^2, \tag{5.19}$$

so the angular velocity of a zero angular momentum observer is

$$\omega = \frac{2Ma}{r^3}\left(1 + \frac{3}{2}\frac{l^2}{r^2}\right). \tag{5.20}$$

As in the static case (see Section 3.3.1), this metric only satisfies $R = 0$ to linear order in the far field limit, and so is not a valid solution for describing the entire horizon exterior of a rotating BBH. Nevertheless, this solution is obtained by solving the full 5-D linearised equations, assuming only that the bulk is asymptotically AdS and that the perturbations are bounded at the AdS horizon, and so the correct metric for a rotating black hole on the brane should be expected to have this asymptotic form. The Weyl tensor $\mathcal{E}_{\mu\nu}$ for this metric is given by:

$$\mathcal{E}_t{}^t = -\frac{4Ml^2}{r^5}, \tag{5.21}$$

$$\mathcal{E}_r{}^r = -\frac{2Ml^2}{r^5}, \tag{5.22}$$

$$\mathcal{E}_\theta{}^\theta = \mathcal{E}_\phi{}^\phi = \frac{3Ml^2}{r^5}, \tag{5.23}$$

$$\mathcal{E}_\phi{}^t = -r^2\sin^2\theta\,\mathcal{E}_t{}^\phi = \frac{15Ml^2}{r^5}a\sin^2\theta. \tag{5.24}$$

Here, terms are calculated to linear order, i.e. terms of order $M^2$ and higher are neglected, which is consistent with the linearised approximation used to derive (5.19).

Assuming $\mathcal{E}_{\mu\nu}$ can be decomposed as in (5.12), we find[4]

$$U = -\frac{4Ml^2}{r^5}, \qquad (5.25)$$

$$P = \frac{5Ml^2}{r^5}, \qquad (5.26)$$

$$q^\phi = -\frac{15Ml^2 a}{r^7}, \qquad (5.27)$$

$$b = 1 + \frac{Ma^2 \sin^2\theta}{10l^2 r}\left[44 + \frac{3l^2}{r^2}\left(68 + 81\frac{l^2}{r^2}\right)\right]. \qquad (5.28)$$

The expressions for $U$ and $P$ are the same as in the static case; in particular they satisfy the same equation of state

$$U = -\frac{4}{5}P. \qquad (5.29)$$

The effects of the rotation are to introduce a small non-zero effective momentum density $q^\phi$, and a value for $b(r,\theta)$ that is slightly different from 1.

### 5.3.2 The rotating black string

As mentioned in Section 3.3.2, the canonical form of the Randall-Sundrum braneworld metric (1.21),

$$ds^2 = e^{-2|y|/l}\eta_{\mu\nu}dx^\mu dx^\nu + dy^2,$$

remains a solution to the 5-D field equations (1.14) if we replace $\eta_{\mu\nu}$ with any 4-D vacuum Einstein solution $g_{\mu\nu}$. Therefore the most obvious attempt at constructing a rotating braneworld black hole is to take $g_{\mu\nu}$ to be the Kerr metric, resulting in

---

[4]In terms of the more general decomposition (5.11), $f(r,\theta) \sim O(M^2)$ and so vanishes in the linear approximation, as expected.

the rotating black string [149]:

$$\begin{aligned}ds^2 &= e^{-2|y|/l}\left[-\left(1-\frac{2Mr}{\Sigma}\right)dt^2 - \frac{4Mra\sin^2\theta}{\Sigma}dtd\phi \right.\\ &\left. + \frac{\Sigma}{\Delta}dr^2 + \Sigma d\theta^2 + \left(r^2+a^2+\frac{2Mra^2\sin^2\theta}{\Sigma}\right)\sin^2\theta d\phi^2\right] + dy^2,\end{aligned} \quad (5.30)$$

where

$$\Delta = r^2 + a^2 - 2Mr,$$

$$\Sigma = r^2 + a^2\cos^2\theta.$$

This solution has $\mathcal{E}_{\mu\nu} = 0$ and so the brane geometry receives no correction from bulk gravitational effects and the weak field limit (5.19) is not satisfied. The square of the 5-D curvature tensor is given by

$$R_{abcd}R^{abcd} = \frac{40}{l^2} + \frac{48M^2}{\Sigma^6}e^{4|y|/l}(r^2-a^2\cos^2\theta)(r^4-14a^2r^2\cos^2\theta+a^4\cos^4\theta), \quad (5.31)$$

which demonstrates the usual ring singularity of the Kerr metric ($\Sigma = 0$) extending along all $y$, together with the unsatisfactory behaviour that the curvature diverges as $y \to \infty$. Furthermore, as in the static case, the solution is unstable to large scale perturbations [150].

### 5.3.3 The tidal Kerr-Newman black hole

As discussed in Section 3.3.3, Einstein-Maxwell solutions in general relativity lead to vacuum braneworld solutions, because of the correspondence between the trace-free electromagnetic stress-energy tensor $T_{\mu\nu}^{\text{em}}$ and the Weyl term $\mathcal{E}_{\mu\nu}$. This allows us to immediately write down the *tidal Kerr-Newman* metric, which was first derived in [151] by considering a Kerr-Schild form for the metric and solving the braneworld effective Einstein equations;

$$ds^2 = -\left(1 - \frac{2Mr - Q}{\Sigma}\right)dt^2 - 2\frac{a(2Mr - Q)\sin^2\theta}{\Sigma}dt d\phi + \frac{\Sigma}{\Delta}dr^2 + \Sigma d\theta^2$$
$$+ \left(r^2 + a^2 + \frac{2Mr - Q}{\Sigma}a^2\sin^2\theta\right)\sin^2\theta d\phi^2, \tag{5.32}$$

where

$$\Delta = r^2 + a^2 - 2Mr + Q, \tag{5.33}$$

$$\Sigma = r^2 + a^2\cos^2\theta. \tag{5.34}$$

This is essentially just the Kerr-Newman metric, but with the electric charge $q^2$ replaced by the *tidal charge* $Q$. Unlike in the general relativity case, $Q$ can be both positive and negative, with some interesting consequences (see below). Indeed, negative $Q$ is the more natural since intuitively we would expect the tidal charge to *strengthen* the gravitational field, as it arises from the source mass $M$ on the brane (see [99] for further discussion).

The Weyl tensor $\mathcal{E}_{\mu\nu}$ for this solution is given by:

$$\mathcal{E}_t{}^t = -\mathcal{E}_\phi{}^\phi = -\frac{Q}{\Sigma^3}\left(\Sigma - 2(r^2 + a^2)\right),$$

$$\mathcal{E}_r{}^r = -\mathcal{E}_\theta{}^\theta = \frac{Q}{\Sigma^2}, \tag{5.35}$$

$$\mathcal{E}_\phi{}^t = -(r^2 + a^2)\sin^2\theta\,\mathcal{E}_t{}^\phi = -\frac{2Qa}{\Sigma^3}(r^2 + a^2)\sin^2\theta.$$

Written in the form (5.3), this metric has

$$\omega = \frac{a(2Mr - Q)}{\Sigma(r^2 + a^2) + (2Mr - Q)a^2\sin^2\theta}, \tag{5.36}$$

and assuming the decomposition (5.12) for $\mathcal{E}_{\mu\nu}$ we find[5]

$$U = -\frac{1}{2}P = \frac{N^2 Q}{\Sigma^3 \Delta}\left[\Delta a^2 \sin^2\theta + (a^2 + r^2)^2\right], \tag{5.37}$$

$$q^\phi = -\frac{2NaQ}{\Sigma^2}(r^2 + a^2), \tag{5.38}$$

$$b = \frac{(r^2 + a^2)^2}{(r^2 + a^2)^2 + a^2 \Delta \sin^2\theta}. \tag{5.39}$$

Notice that the Weyl term satisfies the same equation of state $U = -\frac{1}{2}P$ as the static (tidal-RN) case (3.21).

In addition to their singularity structure, rotating black hole solutions possess two major features: the event horizon structure and the existence of a static limit surface. For a positive tidal charge $Q$, these properties are identical to the Kerr-Newman case in GR. In the following we will focus on the more natural, and interesting, case of negative $Q$. An event horizon is a null surface determined by the equation $\Delta = 0$. This has solutions

$$r_\pm = M \pm \sqrt{M^2 - a^2 - Q}, \tag{5.40}$$

representing the outer $(+)$ and inner $(-)$ event horizons. The event horizon $r_+$ exists provided that

$$M^2 \geq a^2 + Q, \tag{5.41}$$

with the equality corresponding to the extremal case. It can be seen that with a negative tidal charge, it is possible for the black hole's rotation parameter $a$ to exceed its mass $M$, a situation that is impossible in GR! An interesting consequence of this is that a particle in orbit about such a black hole, at the radius of the minimum stable orbit, can have a binding energy *greater* than the 42% binding energy for such an orbit about a maximally rotating Kerr black hole in GR (see [151] for details). Hence the efficiency of an accretion disc around a rotating BBH could be greater than in GR.

---

[5]In terms of the more general decomposition (5.11), $f(r,\theta) = 0$, as expected.

The static limit surface is defined by the equation $g_{tt} = 0$, the largest root of which determines the radius of the ergosphere around the black hole:

$$r_s = M + \sqrt{M^2 - a^2 \cos^2\theta - Q}. \tag{5.42}$$

For the extremal case, $a^2 = M^2 - Q$, the ergosphere lies in the region

$$M < r < M + \sin\theta\sqrt{M^2 - Q}. \tag{5.43}$$

This shows that, compared with the GR case, a rotating BBH with negative tidal charge is a more energetic object in terms of the extraction of rotational energy from its ergosphere [151].

Unfortunately, the tidal Kerr-Newman metric (5.32) does not satisfy the far-field limit (5.19), and so can not correctly describe the entire spacetime surrounding a rotating black hole on the brane. However, if we accept as fact that black holes radiate, then quantum gravity and/or extra dimensional effects might become more prominent in the vicinity of the horizon. The tidal Kerr-Newman metric is able to nicely realise this possibility, since it displays the $\sim 1/r^2$ behaviour of the potential that is characteristic of five dimensional gravity. Hence, the tidal Kerr-Newman metric can be considered a plausible "candidate" metric for the near-horizon geometry of a rotating black hole on the brane. With this in mind, it would be interesting to investigate gravitational lensing for this metric, in the same spirit as the presentation in Chapter 4.

## 5.4 The Newman-Janis "trick"

Shortly after the discovery of the Kerr metric [143], it was shown by Newman and Janis (NJ) that this solution could be "derived" by performing a simple complex coordinate transformation on the Schwarzschild metric [145]. The same method was then applied to the Reissner-Nordström solution to generate a new solution to the Einstein-Maxwell equations, now know as the Kerr-Newman metric [152]. Using a different complex coordinate transformation, the technique was also used to obtain a Kerr-NUT solution from the Schwarzschild metric [153]. Demiański [154] applied a more general complex coordinate transformation to a static, spherically symmetric seed metric and proved that the Taub-NUT metric with $\Lambda$ is the most general solution to Einstein's (vacuum) equations with $\Lambda$ that can be so generated. In particular, this shows that Carter's Kerr-de Sitter metric [155] can not be obtained using the NJ method (see also [156]).

At the time of publication, there was no reason as to why this method should work. Indeed, up to the present there is no full, clear explanation of why the method works and exactly what conditions must be satisfied in order for it to be successful in generating a rotating solution from a static, spherically symmetric seed solution. Partial explanations were given by Talbot [157] and Schiffer *et al.* [158] by examining Einstein's vacuum field equations (Talbot by following the Newman-Penrose approach and Schiffer *et al.* for metrics in Kerr-Schild (KS) form), and Finkelstein [159] extended the analysis of Schiffer *et al.* to include the charged (Kerr-Newman) case. Hints at a deeper explanation are provided by Gürses and Gürsey [160] (see also [161]), who showed that

> "complex translation is allowed in general relativity whenever we can find a coordinate system in which the pseudo energy-momentum tensor vanishes or the Einstein equations are linear in this coordinate system ...[this] happens to be true in the algebraically special Kerr-Schild geometry."

Interestingly, Kerr had apparently already shown that the NJ operation works for metrics in KS form (as mentioned in [145]), although his calculation appears not to have been published. However, this can not be completely correct since the Schwarzschild-de Sitter metric *can* be put into KS form, and yet the NJ method does not work for this case. Hence, a KS metric form clearly is not a sufficient condition for the NJ method to be successful. There is also no proof that a KS metric form is a *necessary* condition for the method to be successful. For example, it might be possible that there exist non-KS metrics for which the Einstein equations are linear, and for which complex transformations would thus be allowed according to Gürses and Gürsey (or it might even be possible that the method could work for certain spacetimes for which the Einstein equations are not linear).

Quevedo [162] considered general complex transformations of the "curvature tensor" and showed that

  i) such transformations are allowed when the seed energy-momentum tensor is diagonal and trace-free,[6] and

  ii) the Petrov type of a given curvature tensor is not changed by such a transformation.

The term "curvature tensor" is in quotation marks because it represents only a tensor possessing the same symmetry properties as those of the curvature tensor – this tensor only becomes a true curvature tensor if it is derivable from a metric. Hence the relevence of these results for the applicability of the NJ method is not clear.

Attempts have been made at using the NJ method to generate interior Kerr solutions [163–166] (an approach first mentioned in [160]), however these have been unable to find a solution that is both physically reasonable and can be matched smoothly to the Kerr metric. Lombardo [167] applied the NJ algorithm to the Born-Infeld monopole, but showed that the resulting metric does not coincide with the correct Born-Infeld monopole with rotation. Mallett [168] claimed to have obtained the rotating radiating charged metric in de Sitter space via the NJ transformation, however Xu [169] showed that this solution is incorrect.

---

[6]conditions which are satisfied by the electromagnetic field strength tensor, and the Weyl tensor $\mathcal{E}_{\mu\nu}$...

Examples for which the NJ method has been successfully applied to generate a rotating solution from the corresponding static, spherically symmetric solution include the dilaton-axion black hole [170], the BTZ black hole [171] and the $D$-dimensional (Myers-Perry) Kerr black hole [172]. Drake and Szekeres [173] considered the application of the NJ algorithm to a more general metric and proved some general results concerning the solutions that can be generated using this method.

### 5.4.1 Complex coordinate transformations

Consider the general static, spherically symmetric seed metric in the $r^2$ area gauge (the method can also be applied to metrics that are not in this gauge, as shown by the dilaton-axion example [170]):

$$\mathrm{d}s^2 = -e^{2\phi(r)}\mathrm{d}t^2 + e^{2\lambda(r)}\mathrm{d}r^2 + r^2\mathrm{d}\Omega_{I\!I}^2. \tag{5.44}$$

Now change to Eddington-Finkelstein coordinates:

$$\mathrm{d}t = \mathrm{d}u + e^{\lambda-\phi}\mathrm{d}r,$$

$$\mathrm{d}s^2 = -e^{2\phi(r)}\mathrm{d}u^2 - 2e^{\lambda(r)+\phi(r)}\mathrm{d}u\mathrm{d}r + r^2\mathrm{d}\Omega_{I\!I}^2. \tag{5.45}$$

In contravariant form this is

$$g^{\mu\nu} = \begin{pmatrix} 0 & -e^{-\lambda-\phi} & 0 & 0 \\ -e^{-\lambda-\phi} & e^{-2\lambda} & 0 & 0 \\ 0 & 0 & 1/r^2 & 0 \\ 0 & 0 & 0 & 1/(r^2\sin^2\theta) \end{pmatrix}. \tag{5.46}$$

A null tetrad is a set of vectors $l^\mu, n^\mu, m^\mu, \bar{m}^\mu$ satisfying

$$g^{\mu\nu} = -l^\mu n^\nu - l^\nu n^\mu + m^\mu \bar{m}^\nu + m^\nu \bar{m}^\mu, \tag{5.47}$$

where

$$l^\mu l_\mu = m^\mu m_\mu = n^\mu n_\mu = 0, \quad l_\mu n^\mu = -m_\mu \bar{m}^\mu = -1, \quad l_\mu m^\mu = n_\mu m^\mu = 0.$$

A suitable null tetrad for the metric (5.46) is:

$$\begin{aligned} l^\mu &= \delta_1^\mu, \\ n^\mu &= e^{-\lambda-\phi}\delta_0^\mu - \frac{1}{2}e^{-2\lambda}\delta_1^\mu, \\ m^\mu &= \frac{1}{\sqrt{2}r}\left(\delta_2^\mu + \frac{i}{\sin\theta}\delta_3^\mu\right). \end{aligned} \tag{5.48}$$

The coordinate $r$ is now allowed to take on complex values, and the tetrad (5.48) is "complexified" subject to the conditions:

i) $l^\mu$, $n^\mu$ are kept real, and reduce to those in (5.48) when $r = \bar{r}$.

ii) $\bar{m}^\mu$ is still the complex conjugate of $m^\mu$.

The resulting tetrad is given by

$$\begin{aligned} l^\mu &= \delta_1^\mu, \\ n^\mu &= e^{-\lambda(r,\bar{r})-\phi(r,\bar{r})}\delta_0^\mu - \frac{1}{2}e^{-2\lambda(r,\bar{r})}\delta_1^\mu, \\ m^\mu &= \frac{1}{\sqrt{2}\bar{r}}\left(\delta_2^\mu + \frac{i}{\sin\theta}\delta_3^\mu\right). \end{aligned} \quad (5.49)$$

It is at this stage in the procedure that there is an element of ambiguity, since the new functions $\lambda(r,\bar{r})$ and $\phi(r,\bar{r})$ are not uniquely determined by the above conditions. For example, the Reissner-Nordström metric is given by (5.48) with $e^{2\phi} = e^{-2\lambda} = 1 - 2M/r - Q^2/r^2$ and is complexified as:

$$\begin{aligned} l^\mu &= \delta_1^\mu, \\ n^\mu &= \delta_0^\mu - \frac{1}{2}\left(1 - M\left(\frac{1}{r}+\frac{1}{\bar{r}}\right) - \frac{Q^2}{r\bar{r}}\right)\delta_1^\mu, \\ m^\mu &= \frac{1}{\sqrt{2}\bar{r}}\left(\delta_2^\mu + \frac{i}{\sin\theta}\delta_3^\mu\right). \end{aligned} \quad (5.50)$$

It can be seen that the term $2M/r$ is complexified differently from the $Q^2/r^2$ term. There appears to be no justification for complexifying these terms in this particular way, and yet complexifying them in any other way does not yield the correct Kerr-Newman metric.

Now perform the following coordinate transformation on the metric (5.49):

$$\begin{aligned} u' &= u + iF(\theta), \\ r' &= r + iG(\theta), \\ \theta' &= \theta, \\ \phi' &= \phi, \end{aligned} \quad (5.51)$$

and restrict $u', r', \theta', \phi'$ to be real. This transformation is similar to that considered by Demiański [154], although he restricted the form of the metric to $e^\phi = e^{-\lambda}$. The original NJ transformation given in [145] and used in most subsequent applications of the method is given by $-F = G = a\cos\theta$. The tetrad now becomes

$$\begin{aligned} l^{\mu'} &= \delta_1^{\mu'}, \\ n^{\mu'} &= e^{-\lambda(r',\theta')-\phi(r',\theta')}\delta_0^{\mu'} - \frac{1}{2}e^{-2\lambda(r',\theta')}\delta_1^{\mu'}, \\ m^{\mu'} &= \frac{1}{\sqrt{2}(r'+iG)}\left(iF'\delta_0^{\mu'} + iG'\delta_1^{\mu'} + \delta_2^{\mu'} + \frac{i}{\sin\theta'}\delta_3^{\mu'}\right), \end{aligned} \quad (5.52)$$

where $F' = \frac{\partial F}{\partial \theta}$, $G' = \frac{\partial G}{\partial \theta}$, and the functions $\lambda, \phi$ are now different from those appearing in (5.49). Henceforth, drop the primes on the transformed coordinates. The metric associated with this tetrad can now be read off from (5.47):

$$g^{\mu\nu} = \begin{pmatrix} \frac{F'^2}{\Sigma} & -e^{-\lambda-\phi} + \frac{F'G'}{\Sigma} & 0 & \frac{F'}{\Sigma\sin\theta} \\ \cdot & e^{-2\lambda} + \frac{G'^2}{\Sigma} & 0 & \frac{G'}{\Sigma\sin\theta} \\ \cdot & \cdot & \frac{1}{\Sigma} & 0 \\ \cdot & \cdot & \cdot & \frac{1}{\Sigma\sin^2\theta} \end{pmatrix},$$

where

$$\Sigma = r^2 + G(\theta)^2, \quad (5.53)$$

and the "$\cdot$" is used to indicate $g^{\mu\nu} = g^{\nu\mu}$. In covariant form this is:

$$g_{\mu\nu} = \begin{pmatrix} -e^{2\phi} & -e^{(\lambda+\phi)} & 0 & e^{\phi}\sin\theta(e^{\phi}F' + e^{\lambda}G') \\ \cdot & 0 & 0 & e^{(\lambda+\phi)}\sin\theta F' \\ \cdot & \cdot & \Sigma & 0 \\ \cdot & \cdot & \cdot & \sin^2\theta\left[\Sigma - e^{\phi}F'(e^{\phi}F' + 2e^{\lambda}G')\right] \end{pmatrix}.$$

We now want to transform to "Boyer-Lindquist" coordinates, i.e. with only one off-diagonal term. Perform the following transformation:

$$\{u, r, \theta, \phi\} \to \{t, r, \theta, \psi\};$$
$$\mathrm{d}u = \mathrm{d}t + g\,\mathrm{d}r, \tag{5.54}$$
$$\mathrm{d}\phi = \mathrm{d}\psi + h\,\mathrm{d}r.$$

Then demanding that the metric in the $\{t, r, \theta, \psi\}$ coordinates has no off-diagonal terms except $\mathrm{d}t\,\mathrm{d}\psi$ requires

$$g = \frac{e^{2\lambda}G'F' - \Sigma e^{(\lambda-\phi)}}{\Sigma + e^{2\lambda}G'^2},$$

$$h = \frac{e^{2\lambda}G'}{\sin\theta\,(\Sigma + e^{2\lambda}G'^2)},$$

and the metric becomes

$$g_{\mu\nu} = \begin{pmatrix} -e^{2\phi} & 0 & 0 & e^{\phi}\sin\theta(e^{\phi}F' + e^{\lambda}G') \\ \cdot & \dfrac{\Sigma}{(\Sigma e^{-2\lambda} + G'^2)} & 0 & 0 \\ \cdot & \cdot & \Sigma & 0 \\ \cdot & \cdot & \cdot & \sin^2\theta\left[\Sigma - e^{\phi}F'(e^{\phi}F' + 2e^{\lambda}G')\right] \end{pmatrix}. \tag{5.55}$$

This is the general form of the rotating metric generated by the NJ method from the seed metric (5.44) with the complex transformation (5.51), without specifying the forms of $e^{\lambda(r,\theta)}$ and $e^{\phi(r,\theta)}$ produced by the "complexification" step (5.49) followed by the transformation (5.51). The Kerr-Newman metric is given by (5.55) with the original NJ transformation

$$-F = G = a\cos\theta, \tag{5.56}$$

and

$$e^{2\phi} = e^{-2\lambda} = 1 + \frac{Q^2 - 2Mr}{\Sigma}. \tag{5.57}$$

Drake and Szekeres [173] obtained the same metric (5.55) but for the specific transformation (5.56), and proved the following results:

1. "The only perfect fluid generated by the NJ algorithm is the vacuum" (i.e. the Kerr metric).[7]

2. "The only algebraically special spacetimes generated by the NJ algorithm are Petrov type $D$."

3. "The only Petrov type $D$ spacetime generated by the NJ algorithm with a vanishing Ricci scalar is the Kerr-Newman spacetime."

It would be interesting to repeat their analysis for the more general metric (5.55) and see how the above conclusions are affected.

### 5.4.2 Application to braneworld black holes

When the NJ "trick" is successful in generating a rotating solution from a static, spherically symmetric one, it clearly circumvents a significant amount of hard work involved with solving the field equations directly. Given the methods success at generating a variety of solutions from their non-rotating counterparts [145, 152, 153, 170–172], it seems reasonable to apply the procedure to some of the known BBH solutions in an attempt to generate new rotating BBH solutions.

The NJ method will be applied to three static, spherically symmetric seed metrics, all in the form (5.44):

$$ds^2 = -e^{2\phi(r)}dt^2 + e^{2\lambda(r)}dr^2 + r^2 d\Omega_{II}^2,$$

---

[7]If the NJ procedure is only valid for KS spacetimes, this result would follow as a corollary since non-vacuum perfect fluids can not be represented in KS form (except a $\Lambda$-fluid).

given by

$$\text{Metric 1:} \quad e^{2\phi} = \left[(1+\epsilon)\sqrt{1-\frac{2M}{r}} - \epsilon\right]^2, \quad e^{2\lambda} = \left(1-\frac{2M}{r}\right)^{-1}.$$

$$\text{Metric 2:} \quad e^{2\phi} = \left(1-\frac{2M}{r}\right), \qquad e^{2\lambda} = \frac{\left(1-\frac{3M}{2r}\right)}{\left(1-\frac{2M}{r}\right)\left(1-\frac{r_0}{r}\right)}.$$

$$\text{Metric 3:} \quad e^{2\phi} = \left(1-\frac{2M}{r}-\frac{4}{3}\frac{Ml^2}{r^3}\right), \quad e^{2\lambda} = \left(1-\frac{2M}{r}-\frac{2Ml^2}{r^3}\right)^{-1}.$$

The first two metrics were discussed in Section 3.3.4, and are derived by assuming the Schwarzschild form for $g_{rr}$ or $g_{tt}$. Metric 1 is also the $U = 0$ solution of Section 3.5.2, written in the $r^2$ area gauge. Metric 3 is the linearised Garriga-Tanaka metric (3.12) discussed in Section 3.3.1, and is the expected weak field form of the BBH solution. As well as the mass, each metric contains an additional parameter which can be thought of as encoding the effect of the extra dimension. All three reduce to the Schwarzschild metric in the appropriate limit:

$$\begin{aligned} \text{Metric 1:} & \quad \epsilon \to 0. \\ \text{Metric 2:} & \quad r_0 \to 3M/2. \\ \text{Metric 3:} & \quad l \to 0. \end{aligned} \tag{5.58}$$

Suitable tetrads for these metrics are given by (5.48) with the above forms for $e^\lambda$, $e^\phi$. We now complexify these tetrads in the most "natural" way, subject to the conditions i) and ii) on page 96, with the additional requirement that the resulting tetrads continue to reduce to the standard GR case in the appropriate limit (5.58).

These complexified tetrads are given by:

Metric 1:

$$l^\mu = \delta_1^\mu,$$

$$n^\mu = \left[1 - M\left(\frac{1}{r} + \frac{1}{\bar{r}}\right)\right]^{\frac{1}{2}} \left[(1+\epsilon)\sqrt{1 - M\left(\frac{1}{r} + \frac{1}{\bar{r}}\right)} - \epsilon\right]^{-1} \delta_0^\mu$$
$$- \frac{1}{2}\left[1 - M\left(\frac{1}{r} + \frac{1}{\bar{r}}\right)\right]\delta_1^\mu,$$

$$m^\mu = \frac{1}{\sqrt{2}\bar{r}}\left(\delta_2^\mu + \frac{i}{\sin\theta}\delta_3^\mu\right).$$

Metric 2:

$$l^\mu = \delta_1^\mu,$$

$$n^\mu = \left[\frac{1 - \frac{r_0}{2}\left(\frac{1}{r} + \frac{1}{\bar{r}}\right)}{1 - \frac{3M}{4}\left(\frac{1}{r} + \frac{1}{\bar{r}}\right)}\right]^{\frac{1}{2}} \delta_0^\mu - \frac{1}{2}\frac{\left[1 - M\left(\frac{1}{r} + \frac{1}{\bar{r}}\right)\right]\left[1 - \frac{r_0}{2}\left(\frac{1}{r} + \frac{1}{\bar{r}}\right)\right]}{\left[1 - \frac{3M}{4}\left(\frac{1}{r} + \frac{1}{\bar{r}}\right)\right]}\delta_1^\mu,$$

$$m^\mu = \frac{1}{\sqrt{2}\bar{r}}\left(\delta_2^\mu + \frac{i}{\sin\theta}\delta_3^\mu\right).$$

Metric 3:

$$l^\mu = \delta_1^\mu,$$

$$n^\mu = \left[\frac{1 - M\left(\frac{1}{r} + \frac{1}{\bar{r}}\right) - \frac{Ml^2}{r\bar{r}}\left(\frac{1}{r} + \frac{1}{\bar{r}}\right)}{1 - M\left(\frac{1}{r} + \frac{1}{\bar{r}}\right) - \frac{2}{3}\frac{Ml^2}{r\bar{r}}\left(\frac{1}{r} + \frac{1}{\bar{r}}\right)}\right]^{\frac{1}{2}} \delta_0^\mu$$
$$- \frac{1}{2}\left[1 - M\left(\frac{1}{r} + \frac{1}{\bar{r}}\right) - \frac{Ml^2}{r\bar{r}}\left(\frac{1}{r} + \frac{1}{\bar{r}}\right)\right]\delta_1^\mu,$$

$$m^\mu = \frac{1}{\sqrt{2}\bar{r}}\left(\delta_2^\mu + \frac{i}{\sin\theta}\delta_3^\mu\right).$$

Next, perform the complex coordinate transformation (5.51). For simplicity, con-sider the specific form of the transformation (5.56) originally used by NJ:

$$\begin{aligned} u' &= u - ia\cos\theta, \\ r' &= r + ia\cos\theta, \\ \theta' &= \theta, \\ \phi' &= \phi. \end{aligned} \tag{5.59}$$

After the additional transformation (5.54) to bring the metrics into Boyer-Lindquist form, the resulting metrics are given by (5.55) with $-F = G = a\cos\theta$:

$$\begin{aligned} ds^2 &= -e^{2\phi}dt^2 - 2a\sin^2\theta e^\phi(e^\lambda - e^\phi)dtd\psi + \sin^2\theta\left[\Sigma + a^2\sin^2\theta e^\phi(2e^\lambda - e^\phi)\right]d\psi^2 \\ &+ \frac{\Sigma}{\Delta}dr^2 + \Sigma d\theta^2, \end{aligned} \tag{5.60}$$

where

$$\Sigma = r^2 + a^2\cos^2\theta, \tag{5.61}$$

$$\Delta = \Sigma e^{-2\lambda} + a^2\sin^2\theta, \tag{5.62}$$

with the following forms for $e^\phi$, $e^\lambda$:

Metric 1: $\quad e^\phi = (1+\epsilon)\sqrt{1 - \frac{2Mr}{\Sigma}} - \epsilon, \quad e^\lambda = \left(1 - \frac{2Mr}{\Sigma}\right)^{-\frac{1}{2}}.$ (5.63)

Metric 2: $\quad e^\phi = \left(1 - \frac{2Mr}{\Sigma}\right)^{\frac{1}{2}}, \quad e^\lambda = \frac{\left(1 - \frac{3Mr}{2\Sigma}\right)^{\frac{1}{2}}}{\left(1 - \frac{2Mr}{\Sigma}\right)^{\frac{1}{2}}\left(1 - \frac{r_0 r}{\Sigma}\right)^{\frac{1}{2}}}.$ (5.64)

Metric 3: $\quad e^\phi = \left(1 - \frac{2Mr}{\Sigma} - \frac{4}{3}\frac{Ml^2r}{\Sigma^2}\right)^{\frac{1}{2}}, \quad e^\lambda = \left(1 - \frac{2Mr}{\Sigma} - \frac{2Ml^2r}{\Sigma^2}\right)^{-\frac{1}{2}}.$

(5.65)

All three metrics reduce to the Kerr metric in the appropriate 4-D GR limit (5.58) and in the limit $a \to 0$ (by construction). However, evaluation of the Ricci scalar for these metrics shows that they do not satisfy $R = 0$ in general, as required for them to be valid braneworld solutions. For metric 3 this is to be expected, since the seed metric only satisfies $R = 0$ to linear order, in the far field limit, as discussed in Section 3.3.1. Interestingly, taking the slow rotation approximation (i.e. neglecting terms of order $a^2$) and expanding the square roots in (5.65) in powers of $1/r$, we arrive at the correct linearised metric for a rotating source on the brane, equation (5.19), so the NJ method is at least partially successful in this case.

Of course, the NJ method can also be trivially applied to the tidal-RN BBH metric (3.19) to generate the corresponding tidal Kerr-Newman black hole (5.32), since this precisely mirrors the original derivation of the Kerr-Newman solution from the Reissner-Nordström metric [152]. Unfortunately it appears that the NJ method, at least in the form presented here, is unable to generate more general rotating BBH solutions from static, spherically symmetric seed black hole solutions.

Further work is required to determine why the NJ method is unsuccessful in these cases, and if the method can be modified in any way to make it successful.

# Chapter 6

# Discussion and conclusions

Contrary to the Kaluza-Klein picture of extra dimensions, braneworlds allow extra dimensions which are large, or even infinite in size, with gravity being effectively localised to four dimensions by the curvature of the bulk, rather than straightforward compactification. In this thesis we have investigated black holes in the Randall-Sundrum braneworld scenario. We have considered the simplest RS2 model throughout, with a single positive tension, $\mathbb{Z}_2$-symmetric brane living in a single, infinite extra dimension. The bulk has been assumed to be empty except for a cosmological constant, with gravity being described by the standard Einstein equations in 5-D. Any of these assumptions can be dropped, leading to a generalisation of the model, and in the eight years since the original model was proposed numerous such studies have been performed (see the reviews [41–44] and references therein). As we have seen, however, the simplest RS2 model already provides a sufficiently rich framework for discussing the phenomenology of BBHs, with the appearance of some key differences from the black holes of standard 4-D GR.

We began with a description of braneworld gravity. It was shown that the results of linearised general relativity are recovered on the brane, with small corrections due to the presence of an extra dimension. Hence, braneworld gravity is consistent with all solar system and table-top tests of gravity. We then moved on to a discussion of non-perturbative gravity, which is relevant for a description of cosmology and black holes. Following a Gauss-Codacci approach, the 5-D Einstein equations are projected onto the brane to give 4-D effective Einstein equations. These closely resemble the standard 4-D Einstein equations, but with additional terms arising from bulk gravitational effects. The most troublesome of these is the non-local "Weyl term". Since this term is not given in terms of data on the brane the effective 4-D equations are not closed, in general, causing difficulties if we attempt to take a purely brane-based approach to problems.

It was shown that the Friedmann equation is reproduced on the brane at low energies and late times, and so braneworld gravity appears also to be consistent with cosmology. We did not discuss cosmological perturbations, however, and it is worth emphasising that observational signatures in the cosmic microwave background radiation might provide a key test of braneworld gravity.

Next, we proceeded to discuss static, spherically symmetric BBH solutions. Ideally, we would like a full 5-D solution describing a black hole localised on the brane. Such a solution would be given by a suitable slicing of the 5-D C-metric, however this has not yet been found. Indeed, it is not clear if a 5-D C-metric even exists! Instead, we focus on solutions to the 4-D effective Einstein equations in order to investigate possible 4-D geometries describing black holes on the brane. To obtain such a solution, an assumption must be made in order to close the system of equations on the brane. Various solutions, making different assumptions, have appeared in the literature and were discussed in Section 3.3.

Although the Weyl term is a complete unknown from the brane point of view, the symmetry of the problem allows it to be decomposed into an effective energy density $U$ and anisotropic stress $P$. We adopted a pragmatic approach to BBHs and assumed an equation of state for the Weyl term: $2P = (\gamma - 1)U$. This allowed us to make a systematic analysis of BBH solutions and classify their behaviour according to the equation of state parameter $\gamma$. It was found that asymptotically flat solutions require an equation of state $\gamma < 0$, and for $|\gamma| > 3$ the solutions have a singular horizon, and are allowed both with and without turning points in the area function (i.e. wormhole regions). This clearly demonstrates that BBHs could have radically different properties to those of standard 4-D GR!

Using the strong field limit approach, the gravitational lensing properties of two candidate BBH solutions were investigated. Taking as a concrete example the supermassive black hole at the centre of our galaxy, it is found that BBHs could have significantly different observational signatures to the Schwarzschild black hole of standard GR. For the particular setup considered here the resolutions required for such observations are beyond reach of current technological capabilities. An interesting extension of this study would therefore be to investigate more realistically observable situations, such as the emission from accretion discs surrounding black holes.

Most objects in the Universe are rotating and so it is important to find braneworld generalisations of the Kerr metric. However, relatively few rotating BBH solutions are currently known. Using the Newman-Janis complexification "trick", an attempt was made to generate rotating solutions from known static solutions. Unfortunately this is only partially successful and we are unable to generate any new solutions via this method. Further work is required to determine why this is and, in the wider context, to explain precisely how the method works and under what conditions it can be applied successfully. It is also worth investigating whether rotating BBH metrics can be found via alternative solution generating techniques, such as those described in [174]. Another possible method for finding rotating BBH solutions is to follow the approach discussed in Section 3.3.4 for the static case, that is to fix some of the metric functions to take the standard Kerr form, and solve the constraint $R = 0$ for the remaining function. Preliminary attempts in this direction have been unsuccessful, however.

Of course, as with the static case, the "holy grail" would be a full 5-D solution describing a rotating black hole localised on the brane. Indeed, finding such a solution - either rotating or static - is probably the biggest unsolved problem in braneworld gravity.

In contemporary physics, extra dimensions are most commonly discussed with regard to potential theories of quantum gravity and unification of the forces; in particular, string theory. In this context, braneworlds might provide useful insights to these more fundamental theories, and open up the exciting prospect of subjecting them to astrophysical and cosmological testing. However, the existence of extra dimensions is a tantalising possibility in its own right and, if ever confirmed, would surely be one of the greatest discoveries in physics.

# Appendix A

# Hypersurface Basics

The "hypersurface formalism" is most commonly encountered in the context of the initial-value formulation of GR (see, e.g., [3,7]). However, the formalism is also central to braneworld constructions, in which our observable Universe is conceived of as a submanifold (the brane) embedded in a higher-dimensional spacetime (the bulk). This appendix introduces some basic aspects of hypersurfaces that are relevant for braneworlds.

## A.1 Induced metric and Extrinsic curvature

A general $m$-dimensional submanifold $\Sigma$ of an $n$-dimensional manifold $\mathcal{M}$ can be defined via the $n$ parametric equations

$$x^a = x^a(\sigma^1, \sigma^2, \ldots, \sigma^m),$$

where the $\sigma^\mu$ are the internal coordinates of the submanifold. With braneworlds in mind, we consider the case where $m=(n-1)$, i.e. the submanifold is a co-dimension one, timelike hypersurface $\Sigma$. The $(n-1)$ internal coordinates $\sigma^\mu$ can then be eliminated from the $n$ parametric equations to give one constraint equation defining

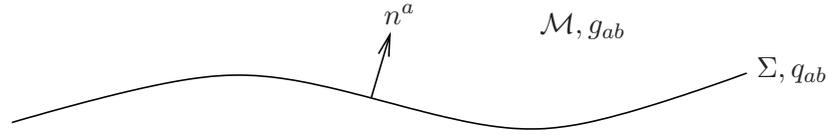

Figure A.1: A brane $\Sigma$ embedded in a bulk spacetime $\mathcal{M}$.

the hypersurface:

$$\Phi(x^a) = c, \tag{A.1}$$

where $c$ is a constant. This allows a unit normal vector to be defined on all points of $\Sigma$ by

$$n_a = \frac{\nabla_a \Phi}{|g^{ab}\nabla_a\Phi\nabla_b\Phi|^{\frac{1}{2}}}, \tag{A.2}$$

(see Figure A.1). The induced metric on $\Sigma$ is given by

$$q_{ab} = g_{ab} - n_a n_b. \tag{A.3}$$

To clarify this, consider two nearby events on $\Sigma$. Since they lie on $\Sigma$, their coordinate separation must satisfy $n_a \mathrm{d}x^a = 0$ and so the interval between them is given by

$$\mathrm{d}s^2 = g_{ab}\mathrm{d}x^a\mathrm{d}x^b = q_{ab}\mathrm{d}x^a\mathrm{d}x^b.$$

Note that $q_{ab}$ is a tensor of $\mathcal{M}$ which is defined only upon $\Sigma$. As well as being the induced metric, $q_{ab}$ has the important property of acting as a projection operator tangent to $\Sigma$. To demonstrate this, consider an arbitrary vector $v^a$ of $\mathcal{M}$ defined at a point $p$ of $\Sigma$, and decompose it into parts tangent and perpendicular to $\Sigma$:

$$v^a = v_\parallel^a + v_\perp n^a,$$

where $v_\parallel^a n_a = 0$. Now, act on this vector with $q^a{}_b$ to find

$$q^a{}_b v^b = (\delta^a{}_b - n^a n_b)(v_\parallel^b + v_\perp n^b) = v_\parallel^a. \tag{A.4}$$

Clearly, a complementary orthogonal projection operator $\perp^a{}_b$ can also be defined:

$$\perp^a{}_b = n^a n_b; \qquad \perp^a{}_b v^b = v_\perp n^a. \tag{A.5}$$

These properties extend to tensors of arbitrary rank, so that a tensor $T^{a_1 \cdots a_k}{}_{b_1 \cdots b_l}$ of $\mathcal{M}$ at a point $p$ of $\Sigma$ is a tensor over the tangent space to $\Sigma$ at $p$ if

$$T^{a_1 \cdots a_k}{}_{b_1 \cdots b_l} = q^{a_1}{}_{c_1} \cdots q^{a_k}{}_{c_k} q_{b_1}{}^{d_1} \cdots q_{b_l}{}^{d_l} T^{c_1 \cdots c_k}{}_{d_1 \cdots d_l}.$$

Hence $q^a{}_b$ quite generally plays the role of a projection operator from the tangent space to $\mathcal{M}$ at $p$ to the tangent space to $\Sigma$ at $p$. This allows us to define a derivative operator $D_a$ on $\Sigma$, simply by projecting all indices onto $\Sigma$ using $q^a{}_b$:

$$D_c T^{a_1 \cdots a_k}{}_{b_1 \cdots b_l} = q^{a_1}{}_{d_1} \cdots q^{a_k}{}_{d_k} q_{b_1}{}^{e_1} \cdots q_{b_l}{}^{e_l} q_c{}^f \nabla_f T^{d_1 \cdots d_k}{}_{e_1 \cdots e_l}, \tag{A.6}$$

where $\nabla_a$ is the derivative operator associated with $g_{ab}$. Furthermore,

$$D_a q_{bc} = q_a{}^d q_b{}^e q_c{}^f \nabla_d (g_{ef} + n_e n_f) = 0, \tag{A.7}$$

since $\nabla_d g_{ef} = 0$ and $q_a{}^b n_b = 0$, thus proving that $D_a$ is the unique derivative operator associated with $q_{ab}$.

The extrinsic curvature of $\Sigma$ is defined by

$$K_{ab} = D_a n_b = q_a{}^c q_b{}^d \nabla_c n_d \tag{A.8}$$

$$= q_a{}^c \nabla_c n_b, \tag{A.9}$$

where the second line follows from the fact that $n_d$ has unit length. Note that it is symmetric, $K_{ab} = K_{(ab)}$. The extrinsic curvature tensor gives the derivative of the normal vector in a direction tangential to $\Sigma$, and so can be interpreted intuitively as describing the "bending" of $\Sigma$ in $\mathcal{M}$ (see Figure A.2).

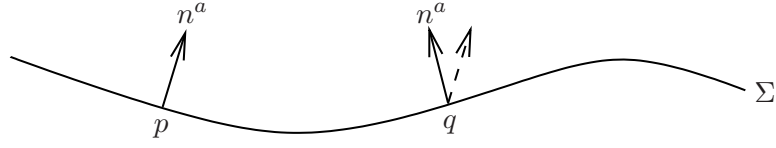

Figure A.2: The normal vector at $p$ is parallel transported to $q$. The failure of this vector to coincide with the normal at $q$ is due to the bending of $\Sigma$ in $\mathcal{M}$, and is encoded in the extrinsic curvature tensor $K_{ab}$.

## A.2  Gauss-Codacci equations

The Gauss-Codacci equations relate $n$-dimensional quantities constructed from the full metric $g_{ab}$ to $(n-1)$-dimensional quantities constructed from the induced metric $q_{ab}$. By definition, the curvature $^{(n-1)}R_{abc}{}^d$ of $\Sigma$ is given by

$$^{(n-1)}R_{abc}{}^d \omega_d = D_a D_b \omega_c - D_b D_a \omega_c, \tag{A.10}$$

where $\omega_a$ is a dual vector field on $\Sigma$. Using the definition (A.6),

$$\begin{aligned}
D_a D_b \omega_c &= q_a{}^f q_b{}^g q_c{}^h \nabla_f \left( q_g{}^e q_h{}^d \nabla_e \omega_d \right) \\
&= q_a{}^f q_b{}^e q_c{}^d \nabla_f \nabla_e \omega_d + q_a{}^f q_b{}^e q_c{}^h \nabla_f q_h{}^d \nabla_e \omega_d + q_a{}^f q_b{}^g q_c{}^d \nabla_f q_g{}^e \nabla_e \omega_d \\
&= q_a{}^f q_b{}^e q_c{}^d \nabla_f \nabla_e \omega_d - q_b{}^e K_{ac} n^d \nabla_e \omega_d - q_c{}^d K_{ab} n^e \nabla_e \omega_d \\
&= q_a{}^f q_b{}^e q_c{}^d \nabla_f \nabla_e \omega_d + K_{ac} K_b{}^d \omega_d + K_{ab} K_c{}^d \omega_d,
\end{aligned} \tag{A.11}$$

where the third equality follows from

$$q_a{}^f q_c{}^h \nabla_f q_h{}^d = q_a{}^f q_c{}^h \nabla_f \left( g_h{}^d - n_h n^d \right) = -K_{ac} n^d, \tag{A.12}$$

and the fourth from

$$q_b{}^e n^d \nabla_e \omega_d = q_b{}^e \nabla_e (n^d \omega_d) - \omega_d q_b{}^e \nabla_e n^d = -\omega_d K_b{}^d. \tag{A.13}$$

Now, antisymmetrising on $a$ and $b$, the final term in (A.11) vanishes since $K_{ab}$ is symmetric, and so (A.10) becomes

$$\begin{aligned} {}^{(n-1)}R_{abc}{}^d \omega_d &= 2q_{[a}{}^f q_{b]}{}^e q_c{}^d \nabla_f \nabla_e \omega_d + 2K_{c[a}K_{b]}{}^d \omega_d \\ &= 2q_a{}^f q_b{}^e q_c{}^k \nabla_{[f}\nabla_{e]}\omega_k + 2K_{c[a}K_{b]}{}^d \omega_d \\ &= q_a{}^f q_b{}^e q_c{}^k q_g{}^d \, {}^{(n)}R_{fek}{}^g \omega_d + K_{ac}K_b{}^d \omega_d - K_{bc}K_a{}^d \omega_d, \end{aligned}$$

where we have used the definition of the curvature ${}^{(n)}R_{fek}{}^g$ of $\mathcal{M}$,

$$2\nabla_{[f}\nabla_{e]}\omega_k = {}^{(n)}R_{fek}{}^g \omega_g. \tag{A.14}$$

Since $\omega_d$ is arbitrary, we obtain the Gauss equation:

$$ {}^{(n-1)}R_{abc}{}^d = q_a{}^e q_b{}^f q_c{}^g q_h{}^d \, {}^{(n)}R_{efg}{}^h + K_{ac}K_b{}^d - K_{bc}K_a{}^d. \tag{A.15}$$

The Codacci equation is derived similarly. Using the definitions (A.6) and (A.9),

$$\begin{aligned} D_a K_{bc} &= q_a{}^d q_b{}^e q_c{}^f \nabla_d (q_e{}^g \nabla_g n_f) \\ &= q_a{}^d q_b{}^g q_c{}^f \nabla_d \nabla_g n_f + q_a{}^d q_b{}^e q_c{}^f \nabla_d q_e{}^g \nabla_g n_f \\ &= q_a{}^d q_b{}^g q_c{}^f \nabla_d \nabla_g n_f - K_{ab} n^g q_c{}^f \nabla_g n_f \quad \text{(using (A.12))}. \end{aligned}$$

Now, antisymmetrise on $a$ and $b$ to find

$$D_{[a}K_{b]c} = q_{[a}{}^d q_{b]}{}^g q_c{}^f \nabla_d \nabla_g n_f = q_a{}^d q_b{}^g q_c{}^f \nabla_{[d} \nabla_{g]} n_f$$

$$\Rightarrow \quad D_a K_{bc} - D_b K_{ac} = q_a{}^d q_b{}^g q^f{}_c \, {}^{(n)}R_{dgfe} n^e \quad \text{(from (A.14))}.$$

Raising the $c$ index and contracting it with $a$ yields the Codacci equation:

$$D_a K^a{}_b - D_b K = {}^{(n)}R_{cd} n^d q^c{}_b. \tag{A.16}$$

## A.3 Gaussian Normal coordinates

A coordinate system that is particularly well adapted to the consideration of hypersurfaces is that of Gaussian Normal Coordinates (GNC), which are defined as follows. Construct the unique geodesic with tangent vector $n^a$ through each point $p$ of $\Sigma$. Choose arbitrary coordinates $(x_1, \ldots, x_{n-1})$ on a portion of $\Sigma$ and label each point in a neighbourhood of $\Sigma$ by the parameter $y$ along the geodesic on which it lies and the coordinates $(x_1, \ldots, x_{n-1})$ of the point $p \in \Sigma$ from which the geodesic emanated. Then $(x_1, \ldots, x_{n-1}, y)$ defines our GNC system (also known as hypersurface orthogonal coordinates). We then have $n_a \mathrm{d}x^a = \mathrm{d}y$ and the metric takes the form

$$\mathrm{d}s^2 = g_{ab}\mathrm{d}x^a\mathrm{d}x^b = q_{\mu\nu}\mathrm{d}x^\mu\mathrm{d}x^\nu + \mathrm{d}y^2. \tag{A.17}$$

Without loss of generality, we can choose the hypersurface to be located at $y = 0$ in GNC. The geodesics originating from $\Sigma$ may eventually cross or run into singularities, but at least in a neighbourhood of the surface GNC are well-defined.

## A.4 Junction conditions

Any physical situation with boundary surfaces requires the formulation of proper junction conditions to deal with the discontinuities across the surface. Examples in GR include the matching of interior and exterior Schwarzschild metrics at the surface of a star, or matching a collapsing interior solution with an exterior vacuum metric in Oppenheimer-Snyder collapse. A geometric, covariant derivation of junction conditions for treating singular hypersurfaces and thin shells in GR was given by Israel [175], building on earlier work by Darmois [176], Lanczos [177] and Sen [178]. These junction conditions are relevant for braneworlds, where our 4-D Universe is represented as a distributional source in the 5-D Einstein equations:

$$^{(5)}G_{ab} = 8\pi G_5 \, S_{ab} \, \delta(y). \tag{A.18}$$

Heuristically, we can understand the junction conditions as follows: $^{(5)}G_{ab}$ consists of the metric and its derivatives up to second order. In order for us to have a well-defined geometry, the metric itself must be continuous across the surface $y = 0$, therefore integrating (A.18) across $y = 0$ we expect to find a relation between the jump in the first derivative of the metric in the $y$-direction, and the energy-momentum $S_{ab}$ of the surface.

We will perform the derivation in Gaussian normal coordinates (GNC), in which it can be shown (from the Gauss-Codacci equations) that $^{(5)}G_{yy}$ and $^{(5)}G_{\mu y}$ are continuous across $y = 0$. Therefore $S_{yy} = S_{\mu y} = 0$ and the discontinuity in (A.18) is only present in the $\mu\nu$-component, which we re-write as

$$^{(5)}R_{\mu\nu} = 8\pi G_5 \left( S_{\mu\nu} - \frac{1}{3} q_{\mu\nu} S \right) \delta(y). \tag{A.19}$$

In GNC, we have

$$K_{\mu\nu} = \frac{1}{2} \partial_y q_{\mu\nu}, \tag{A.20}$$

$$^{(5)}R_{\mu\nu} = -\partial_y K_{\mu\nu} + \underbrace{^{(4)}R_{\mu\nu} + 2 K_{\mu\lambda} K^\lambda{}_\nu - K K_{\mu\nu}}_{Z_{\mu\nu}}, \tag{A.21}$$

where $Z_{\mu\nu}$ is bounded on the brane. Therefore, performing a "pill-box" integration of (A.19) across $y = 0$ we find

$$\lim_{\epsilon \to 0} \int_{-\epsilon}^{\epsilon} dy \left( -\partial_y K_{\mu\nu} + Z_{\mu\nu} \right) = 8\pi G_5 \lim_{\epsilon \to 0} \int_{-\epsilon}^{\epsilon} dy \left( S_{\mu\nu} - \frac{1}{3} q_{\mu\nu} S \right) \delta(y)$$

$$\Rightarrow \quad [K_{\mu\nu}] = -8\pi G_5 \left( S_{\mu\nu} - \frac{1}{3} q_{\mu\nu} S \right), \tag{A.22}$$

where $[K_{\mu\nu}] = K^+_{\mu\nu} - K^-_{\mu\nu}$ is the jump in the extrinsic curvature across the surface. If we impose $\mathbb{Z}_2$ symmetry across the surface we have $K^+_{\mu\nu} = -K^-_{\mu\nu}$ (the minus sign arises because the normal points in opposite directions on either side of the surface), and the junction condition (A.22) becomes

$$2K^+_{\mu\nu} = -8\pi G_5 \left( S_{\mu\nu} - \frac{1}{3} q_{\mu\nu} S \right). \tag{A.23}$$